\documentstyle[epsfig,natbib2,natbibmnfix]{mn}

\newcommand{\be}{\begin{equation}}
\newcommand{\ee}{\end{equation}}
\newcommand{\bea}{\begin{eqnarray}}
\newcommand{\eea}{\end{eqnarray}}
\newcommand{\bc}{\begin{center}}
\newcommand{\ec}{\end{center}}

\renewcommand{\thefootnote}{\fnsymbol{footnote}}

\title{The history of star formation in a $\Lambda$-CDM universe}

\author[V.~Springel and L.~Hernquist] {\parbox{18cm}{Volker
Springel$^{1}$\footnotemark[1] and Lars
Hernquist$^2$\footnotemark[3]}\vspace{0.3cm}\\ $^1$Max-Planck-Institut
f\"{u}r Astrophysik, Karl-Schwarzschild-Stra\ss{}e 1, 85740 Garching
bei M\"{u}nchen, Germany\\ $^2$Harvard-Smithsonian Center for
Astrophysics, 60 Garden Street, Cambridge, MA 02138, USA}

\begin{document}

\maketitle
\begin{abstract}

Employing hydrodynamic simulations of structure formation in a
$\Lambda$CDM cosmology, we study the history of cosmic star formation
from the ``dark ages'' at redshift $z \sim 20$ to the present.  In
addition to gravity and ordinary hydrodynamics, our model includes
radiative heating and cooling of gas, star formation, supernova
feedback, and galactic winds.  By making use of a comprehensive set of
simulations on interlocking scales and epochs, we demonstrate
numerical convergence of our results on all relevant halo mass scales,
ranging from $10^8$ to $10^{15}\, h^{-1}{\rm M}_\odot$.

The predicted density of cosmic star formation, $\dot \rho_\star(z)$,
is broadly consistent with measurements, given observational
uncertainty.  From the present epoch, $\dot \rho_\star(z)$ gradually
rises by about a factor of ten to a peak at $z\sim 5-6$, which is
beyond the redshift range where it has been estimated observationally.
In our model, fully 50\% of the stars are predicted to have formed by
redshift $z\simeq 2.14$, and are thus older than 10.4 Gyr, while only
25\% form at redshifts lower than $z\simeq 1$.  The mean age of all
stars at the present is about 9 Gyr.  Our model predicts a total
stellar density at $z=0$ of $\Omega_\star= 0.004$, corresponding to
about 10\% of all baryons being locked up in long-lived stars, in
agreement with recent determinations of the luminosity density of the
Universe.

We determine the ``multiplicity function of cosmic star formation'' as
a function of redshift; i.e.~the distribution of star formation with
respect to halo mass.  At redshifts around $z\simeq 10$, star
formation occurs preferentially in halos of mass $10^{8}-10^{10}\,
h^{-1}{\rm M}_\odot$, while at lower redshifts, the dominant
contribution to $\dot \rho_\star(z)$ comes from progressively more
massive halos.  Integrating over time, we find that about 50\% of all
stars formed in halos less massive than $10^{11.5}\,h^{-1}{\rm
M}_\odot$, with nearly equal contributions per logarithmic mass
interval in the range $10^{10}-10^{13.5}\, h^{-1}{\rm M}_\odot$, making
up $\sim 70\%$ of the total.

We also briefly examine possible implications of our predicted star
formation history for reionisation of hydrogen in the Universe.
According to our model, the stellar contribution to the ionising
background is expected to rise for redshifts $z>3$, at least up to
redshift $z\sim 5$, in accord with estimates from simultaneous
measurements of the H and He opacities of the Lyman-$\alpha$ forest.
This suggests that the UV background will be dominated by stars for
$z>4$, provided that there are not significantly more quasars at
high-z than are presently known.  We measure the clumping factor of
the gas from the simulations and estimate the growth of cosmic
H{\small II} regions, assuming a range of escape fractions for
ionising photons.  We find that the star formation rate predicted by
the simulations is sufficient to account for hydrogen reionisation by
$z\sim 6$, but only if a high escape fraction close to unity is
assumed.

\end{abstract}
\begin{keywords}
galaxies: evolution -- galaxies: starburst -- methods: numerical.
\end{keywords}

\section{Introduction}

\renewcommand{\thefootnote}{\fnsymbol{footnote}}
\footnotetext[1]{E-mail: volker@mpa-garching.mpg.de}
\footnotetext[3]{\hspace{0.03cm}E-mail: lars@cfa.harvard.edu}

Hierarchical galaxy formation \citep{Whi78} within a $\Lambda$CDM
cosmogony is currently the most successful paradigm for understanding
the distribution of matter in the Universe.  In this scenario,
structure grows via gravitational instability from small perturbations
seeded in an early inflationary epoch.  The dominant mass component is
(unidentified) collisionless cold dark matter, which also determines
the dynamics of the baryons on large scales, where hydrodynamic forces
are unimportant compared to gravity.  The successes of the
$\Lambda$CDM model are impressively numerous, ranging from a detailed
picture of the primary anisotropies of the CMB at high redshift to the
clustering properties of galaxies in the local Universe.

However, while the dynamics of collisionless dark matter is quite well
understood, the same cannot be said for the baryonic processes that
are ultimately responsible for lighting up the Universe with stars.
Hydrodynamic simulations have been shown to produce reliable results
for gas of low to moderate overdensity, allowing for detailed
theoretical studies of e.g. the Lyman-alpha forest
\citep{CMOR94,Zh95,He96} and the intergalactic medium
\citep[e.g.][]{Dave2001,Cr00,Keshet2002}, but at gas densities
sufficiently high for star formation to occur, our knowledge is much
less certain.  Despite numerous attempts to include star formation in
cosmological simulation \citep[e.g.][]{Ce93,Ce00,Ka96,Kat99,
Ye97,We97,Wein2000,St95,Bl99,Pea99,Pea2000,Th98,Th2000}, progress has
been comparatively slow with this approach, primarily because of the
physical complexity of star formation and feedback, but also because
of the computational difficulties inherent in the task of embedding
these processes into the framework of hierarchical galaxy formation.

For this reason, the most popular approach for describing galaxy
evolution in hierarchical universes has been the so-called
semi-analytic technique
\citep[e.g.][]{Wh91,Col91,Lac91,Lac93b,Kau93a,Kau94,Col94}. This
method combines our firm knowledge of the dynamics and growth of dark
matter halos with simplified parameterisations of the baryonic physics
essential to galaxy formation.  In this manner, technical limitations
of direct hydrodynamical simulations can be overcome, allowing some of
the consequences of the assumed physics to be analysed.  On the other
hand, the validity of the assumptions underlying the semi-analytic
approach it is not always clear and must be confirmed by direct
hydrodynamic simulations.

Moreover, while semi-analytic techniques are computationally much less
expensive than hydrodynamical simulations, they are quite limited in
their ability to relate galaxies to the surrounding IGM.  One strength
of hydrodynamical simulations is that they can explore this
relationship {\em from first principles}, making it possible to
constrain galaxy formation and evolution using the rich body of
observational data on the IGM, as probed by quasar absorbers.
Motivated by our desire to understand the connection between galaxies
and their environments in detail, we here attempt to refine the
methodology of direct simulations of galaxy formation, using a novel
approach to describe star formation and feedback.

In this paper, we focus on one aspect of galaxy evolution; namely the
global history of star formation in the Universe.  This is currently
one of the most fundamental quantities in observational and
theoretical cosmology, and it provides a crucial test for any theory
of galaxy formation.  Over the last decade, observational results for
the redshift evolution of the cosmic star formation rate density,
$\dot \rho_\star(z)$, have become available, at both low and high
redshift
\citep[e.g.][]{Gal95,Mad96,Mad98,Lil96,Cow96,Cow99,Con97,Hug98,Trey98,
Tresse98,Pas98,Steid99,Flo99,Gron99,Bald02,Lan2002}.  While these
measurements are still fraught with large observational uncertainty,
they are nevertheless beginning to constrain the epoch of galaxy
formation.  In recent years, therefore, a number of studies have
attempted to compute $\dot \rho_\star(z)$ theoretically and to compare
it with data, using either semi-analytic models \citep{Bau98,Som00},
or numerical simulations \citep{Pea2000,Nag00,Nag01,Asc02}. However,
because of the complexity of the relevant physics and the large range
in scales involved, the theoretical predictions have remained quite
uncertain, a situation we try to improve on in this study.

Besides gravity, ordinary hydrodynamics, and collisionless dynamics of
dark matter, our numerical simulations include radiative cooling and
heating in the presence of a UV background radiation field, star
formation, and associated feedback processes.  Cooling gas settles
into the centres of dark matter potential wells, where it becomes
dense enough for star formation to occur.  We regulate the dynamics of
the gas in this dense interstellar medium (ISM) by an effective
multi-phase model described by \citet{SprHerMultiPhase}. In this
manner, we are able to treat star formation and supernova feedback in
a physically plausible and numerically well-controlled manner,
although there are clearly uncertainties remaining with respect to the
validity and accuracy of our description.

In principle, the cooling-rate of gas in halos can be computed
accurately within simulations, and it is this rate that ultimately
governs the global efficiency of star formation, provided that strong
feedback processes are unimportant.  However, it is clear \citep{Wh91}
that cooling by itself is so efficient that it yields a collapse
fraction of baryons that is considerably higher than that implied by
the measured luminosity density of the Universe \citep[e.g.][]{Bal01}.
Simulations and semi-analytic models agree in this respect
\citep{Yoshida2001}.  Feedback mechanisms related to star formation
are commonly invoked to reduce the efficiency of star formation.  In
our model, we include such a strong feedback process in the form of
galactic winds emanating from star-forming regions. There is
observational evidence for the ubiquitous presence of such winds in
star forming galaxies both locally and at high redshift
\citep[e.g.][]{Heck95,Bland95,
Lehn96,Dahlem97,Mar99,Heck00,Frye01,Pett00,Pett01}.  It is believed
that these galactic outflows are powered by feedback energy from
supernovae and stellar winds in the ISM, but their detailed formation
mechanism is not entirely clear.  Since we also lack the ability to
spatially resolve the interactions of supernova blast waves and
stellar winds within the ISM, we invoke a phenomenological model for
the generation of galactic ``superwinds''.  Note, however, that the
hydrodynamic interaction of the wind with infalling gas in the halos
and with the IGM is treated correctly by our numerical code.  In
particular, we are thus able to investigate how winds associated with
star formation influence galaxy formation, how they disperse and
transport metals, and how they heat the intergalactic medium.

We base our work on a large set of numerical simulations that cover a
vast range of mass and spatial scales.  For example, the masses of
resolved halos in our study spans more than a factor of $10^9$. Our
simulation programme was designed to examine star formation on
essentially all relevant cosmological scales, enabling us to arrive at
a reliable prediction for the star formation density over its full
history, ranging from the present epoch far into the ``dark ages'' at
very high redshift.  As an integral part of our simulation set, we
also carried out extensive convergence tests, allowing us to cleanly
quantify the reliability of our results.

As part of our analysis, we introduce a ``multiplicity function of
star formation'' which gives the cumulative star formation density per
logarithmic interval of halo mass at a given epoch.  Using this
quantity, the global star formation density can be decomposed into the
number density of halos of a given mass scale (i.e. the halo ``mass
function''), and the average star formation efficiency of halos of a
given mass.  Because the cosmological mass function has been reliably
determined using large collisionless simulations \citep{Jen01}, this
decomposition allows us to nearly eliminate the dependence of our
results on cosmic variance.

We also briefly investigate the potential relevance of our results for
the reionisation of hydrogen in the Universe. It is a long-standing
question which sources dominate the ionising UV flux as a function of
redshift.  While the ratio of He{\small II} and H{\small I} optical
depths observed at $z\sim 2.4$ indicates that quasars are the most
significant source of ionising radiation at this low redshift, it has
been suggested that massive stars might dominate at higher redshift
\citep[e.g.][]{Haehn2001,Sok2002b}.  Using our results, we test the
possibility that high-$z$ star formation by itself could have been
responsible for reionisation of hydrogen at a redshift of around
$z\sim 6$.

This paper is organised as follows. In Section~\ref{SecMethod}, we
describe our simulations, and the analysis applied to them.  We then
move on to discuss tests for numerical convergence in
Section~\ref{SecConv}.  In Section~\ref{SecMulti}, we introduce the
concept of a multiplicity function for cosmic star formation, which we
use in Section~\ref{SecSFR} to derive our composite result for the
cosmic star formation history, and compare it to observational
constraints.  In Section~\ref{SecWhenWhere}, we analyse consequences
of our prediction for the age distribution of stars and then discuss
possible implications for the reionisation of the Universe in
Section~\ref{SecReion}.  Finally, we summarise our findings in
Section~\ref{SecDisc}.

\section{Methodology} \label{SecMethod}

\begin{figure*}
\bc
\resizebox{14.0cm}{!}{\includegraphics{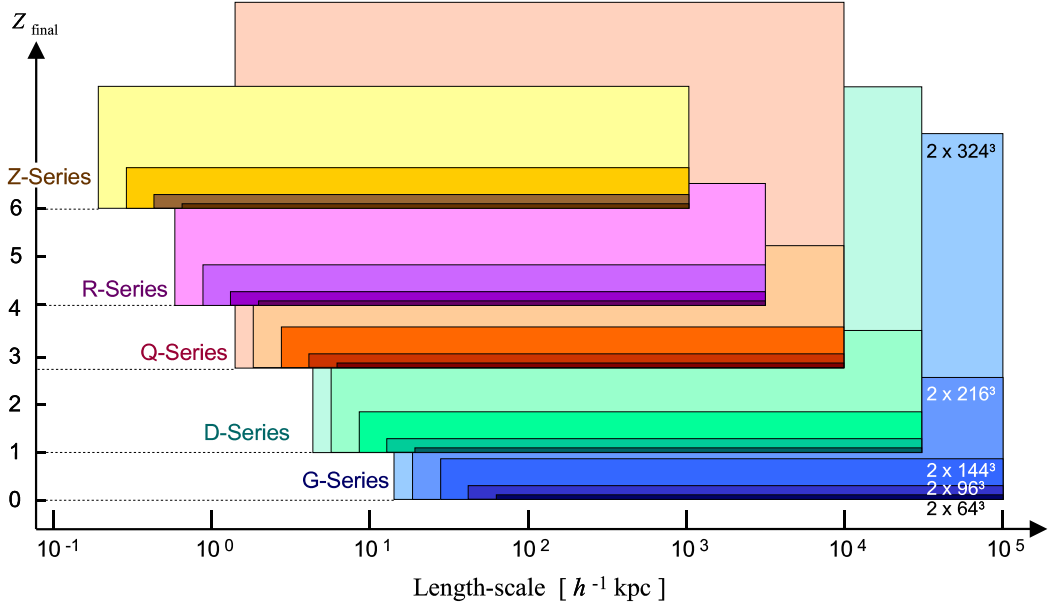}}%
\caption{Illustration of the attributes of the simulations analysed in
this study.  We consider five different box sizes, ranging from
$100\,h^{-1}{\rm Mpc}$ to $1\,h^{-1}{\rm Mpc}$. For each box size, a
number of simulations have been run, with particle number between
$2\times 64^3$ and $2\times 324^3$. In the diagram, each simulation is
represented by a rectangle whose area is proportional to the particle
number. The horizontal width of each rectangle gives the spatial
dynamic range of the corresponding run, from the gravitational
softening scale (left edges) to the box size (right edges), while the
lower edge indicates the ending redshift of each run.  Numerical
parameters of the simulations are listed in
Table~\ref{tabSimulations}.
\label{figSimSet}}
\ec
\end{figure*}

\begin{table*}
\bc
\begin{tabular}{ccrccccc}
\hline
Simulation & $L$ [$\,h^{-1}{\rm Mpc}\,$] & Resolution & $m_{\rm DM}$ [$\,h^{-1}{\rm M}_\odot\,$] & $m_{\rm gas}$  [$\,h^{-1}{\rm M}_\odot\,$] & $z_{\rm start}$ & $z_{\rm end}$ & $\epsilon$ [$\,h^{-1}{\rm kpc}\,$] \\
\hline
Z1 & $1.000$ & $2\times 64^3$   & $2.75\times 10^5$ & $4.24\times 10^4$ & 199 & 6 & $0.63$ \\
Z2 & $1.000$ & $2\times 96^3$   & $8.16\times 10^4$ & $1.25\times 10^4$ & 199 & 6 & $0.42$\\
Z3 & $1.000$ & $2\times 144^3$  & $2.42\times 10^4$ & $3.72\times 10^3$ & 199 & 6 & $0.28$ \\
Z4 & $1.000$ & $2\times 216^3$  & $7.16\times 10^3$ & $1.10\times 10^3$ & 199 & 6 & $0.19$\\
\hline
R1 & $3.375$ & $2\times 64^3$ & $1.06\times 10^7$ & $1.63\times 10^6$ & 199 & 4 & $2.11$\\
R2 & $3.375$ & $2\times 96^3$ & $3.14\times 10^6$ & $4.84\times 10^5$ & 199 & 4 & $1.41$\\
R3 & $3.375$ & $2\times 144^3$& $9.29\times 10^5$ & $1.43\times 10^5$  & 199 & 4 & $0.94$\\
R4 & $3.375$ & $2\times 216^3$& $2.75\times 10^5$ & $4.24\times 10^4$ & 199 & 4 & $0.63$\\
\hline
Q1 & $10.00$ &$2\times 64^3$& $2.75\times 10^8$ & $4.24\times 10^7$ & 159 & 2.75 & $6.25$\\
Q2 & $10.00$ &$2\times 96^3$& $8.16\times 10^7$ & $1.25\times 10^7$ & 159 & 2.75  & $4.17$\\
Q3 & $10.00$ & $2\times 144^3$& $2.42\times 10^7$ & $3.72\times 10^6$ & 159 & 2.75 & $2.78$\\
Q4 & $10.00$ & $2\times 216^3$& $7.16\times 10^6$ & $1.10\times 10^6$ & 159 & 2.75 & $1.85$\\
Q5 & $10.00$ & $2\times 324^3$& $2.12\times 10^6$ & $3.26\times 10^5$ & 159 & 2.75 & $1.23$\\
\hline
D3 & $33.75$ &$2\times 144^3$& $9.29\times 10^8$ & $1.43\times 10^8$ & 159 & 1 & $9.38$\\
D4 & $33.75$ & $2\times 216^3$& $2.75\times 10^8$ & $4.24\times 10^7$ & 159 & 1 &$6.25$\\
D5 & $33.75$ & $2\times 324^3$& $8.15\times 10^7$ & $1.26\times 10^7$ & 159 & 1 &$4.17$\\
\hline
G3 & $100.0$ &$2\times 144^3$& $\;\,2.42\times 10^{10}$ & $3.72\times 10^9$ & 79 & 0  & $18.0$ \\
G4 & $100.0$ & $2\times 216^3$& $7.16\times 10^9$ & $1.10\times 10^9$ & 79 & 0  & $12.0$\\
G5 & $100.0$ & $2\times 324^3$& $2.12\times 10^9$ & $3.26\times 10^8$ & 79 & 0  & $8.00$\\
\hline
\end{tabular}
\caption{Numerical simulations analysed in this study.  Identical
cosmological parameters were employed in all the runs, and the same
simulation technique for treating star formation and feedback
processes was used throughout.  For each of our five different
box-sizes, we carried out a number of simulations, differing only in
mass and spatial resolution. In the table, we list the names of the
runs, their particle and mass resolutions, their starting and ending
redshifts, and the gravitational softening length $\epsilon$.
Simulations with particle resolutions of $2\times 64^3$ carry a label
`1' in their names, those with $2\times 96^3$ are labelled with `2',
and so on, up to the $2\times 324^3$ runs which are labelled with a
`5'.  The simulations have been carried out with a massively parallel
TreeSPH code. Runs at the `1'-level were done on 4 processors, and for
each higher level, we doubled the number of processors so that level-5
runs were computed on 64 processors.
\label{tabSimulations}}
\ec
\end{table*}

\subsection{Numerical simulations}

In this study, we focus on a $\Lambda$CDM cosmological model with
parameters $\Omega_0=0.3$, $\Omega_\Lambda=0.7$, Hubble constant $H_0
= 100\, h\, {\rm km\,s^{-1}\,Mpc^{-1}}$ with $h=0.7$, baryon density
$\Omega_{\rm b}=0.04$, and a scale-invariant primordial power spectrum
with index $n=1$, normalised to the abundance of rich galaxy clusters
at the present day ($\sigma_8 = 0.9$).  This ``concordance'' model
provides a good fit to a long list of current cosmological
constraints.

All of our smoothed particle hydrodynamics (SPH) simulations were
performed in cubic boxes with periodic boundary conditions, employing
an equal number of dark matter and gas particles. Besides gravity, the
gas particles interact through pressure forces and gain entropy in
hydrodynamic shocks.  We also follow radiative cooling and heating
processes for a primordial mix of hydrogen and helium, using a method
similar to \citet{Ka96}.  We adopt an external photo-ionising flux
that describes radiation from quasars as advocated by \citet{Ha96},
leading to reionisation of the Universe at redshift $z\simeq 6$
\citep[for details, see][]{Da99}.

Star formation and supernova feedback in the ISM are treated with an
effective multi-phase model, which is discussed in full detail in a
companion paper by \citet{SprHerMultiPhase}.  In brief, our model
assumes that rapidly cooling gas at high overdensity is subject to a
thermal instability which leads to the formation of cold clouds
embedded in a hot ambient medium.  Because the resulting multi-phase
structure of the ISM cannot be spatially resolved by available
cosmological simulations, we describe the dynamics of the star-forming
ISM in terms of a `sub-resolution' model, where coarse-grained
averages of fluid quantities are used to describe the medium.  In this
method, a sufficiently dense SPH fluid element will then represent a
satistical mixture of cold clouds and ambient hot gas, with a set of
equations governing the mass and energy exchange processes between the
phases.  In particular, we take the cloud material to be the reservoir
of baryons available for star formation.  We set the consumption
timescale of the gas to match the observed ``Kennicutt'' law
\citep{Ke89,Ke98} for the star formation rate in local spiral
galaxies.  In our formalism, this one free parameter simultaneously
determines a threshold density, above which the multiphase structure
in the gas and hence star formation is allowed to develop. This {\em
physical} density is $8.55\times 10^6\,h^2{\rm M}_\odot {\rm
kpc}^{-3}$ for all simulations in this study, corresponding to a {\em
comoving} baryonic overdensity of $7.7\times 10^5$ at $z=0$.

In describing star formation numerically, we spawn independent star
particles out of the multi-phase medium with mass equal to half the
original gas particle mass in a stochastic fashion, as described by
\citet{SprHerMultiPhase}. This avoids any artificial coupling of gas
and collisionless stellar material, while the total increase in
particle number over the course of a simulation is only modest.

With respect to feedback processes, we assume that massive stars
explode as supernovae on a short timescale, releasing their energy as
heat to the ambient medium of the ISM. We also assume that supernovae
evaporate cold clouds \citep{McKee77}, essentially by thermal
conduction, thereby ``cooling'' the ambient hot gas.  This process
establishes a tight self-regulation cycle for the star-forming ISM. A
model quite similar to ours in this respect has been proposed by
\citet{Ye97}. However, our technique differs significantly in the
physical parameterisation of cloud evaporation and the star formation
timescale, and also in its numerical implementation.

\begin{figure*}
\resizebox{7.5cm}{!}{\includegraphics{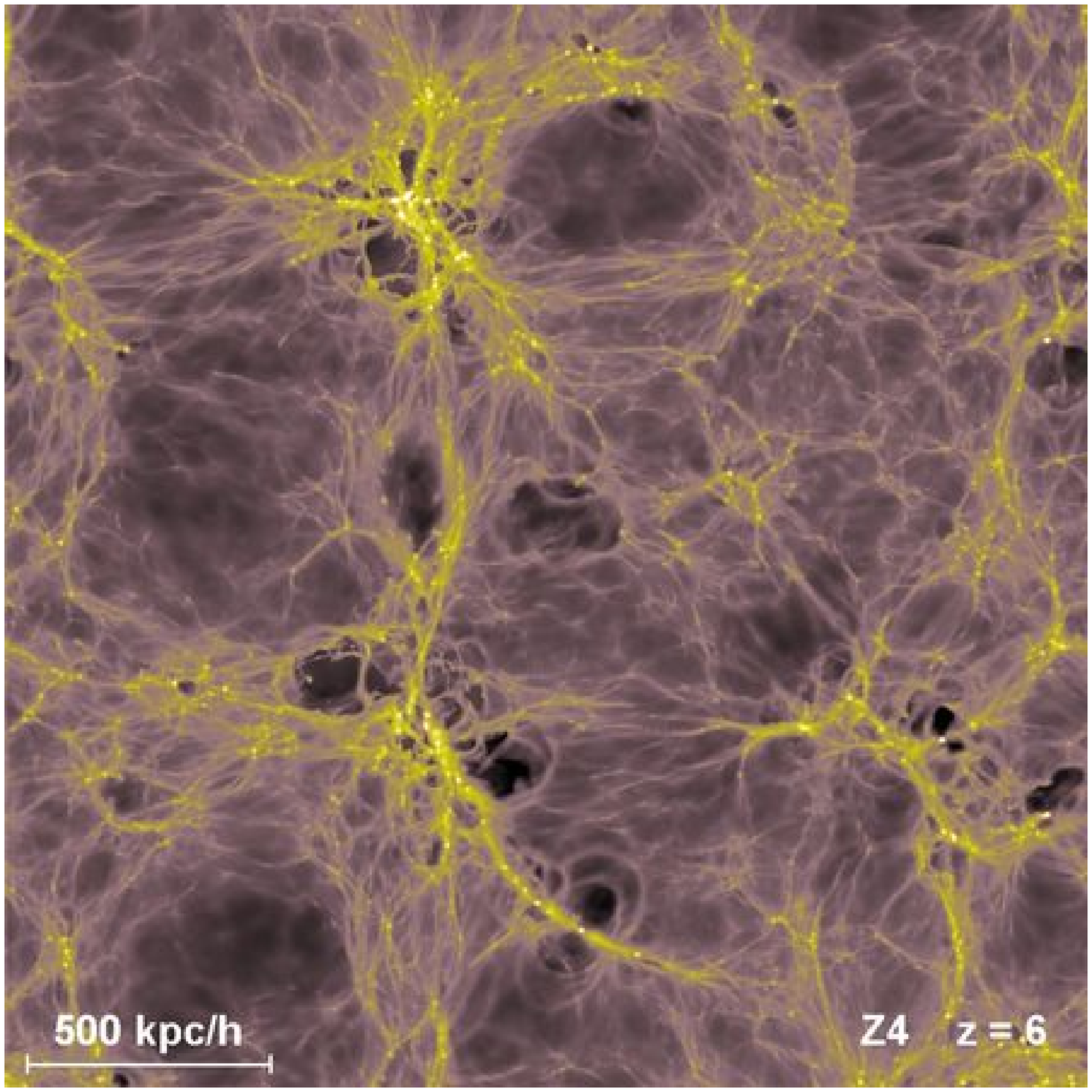}}%
\hspace{0.13cm}\resizebox{7.5cm}{!}{\includegraphics{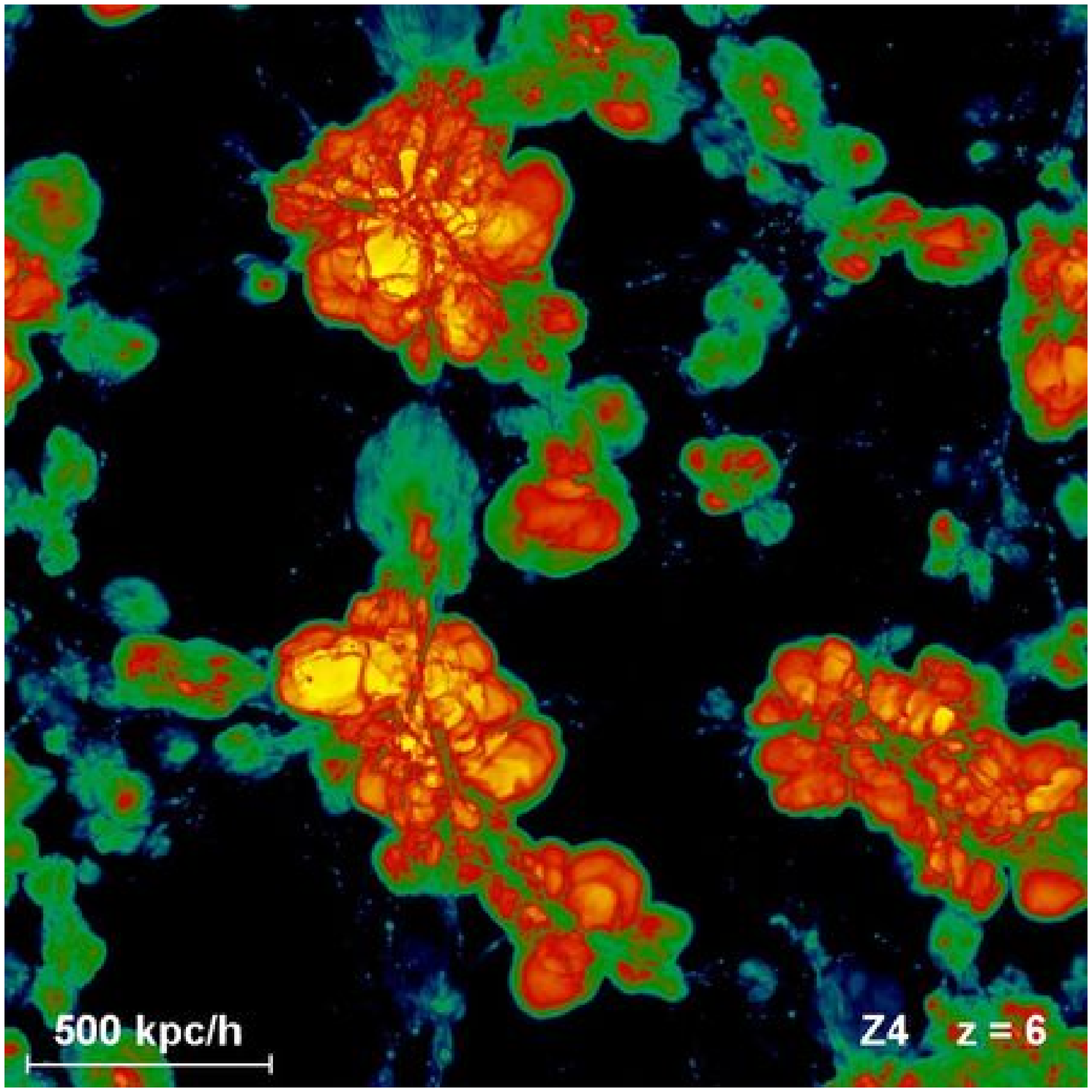}}%
\hspace{0.2cm}\resizebox{1.16cm}{!}{\includegraphics{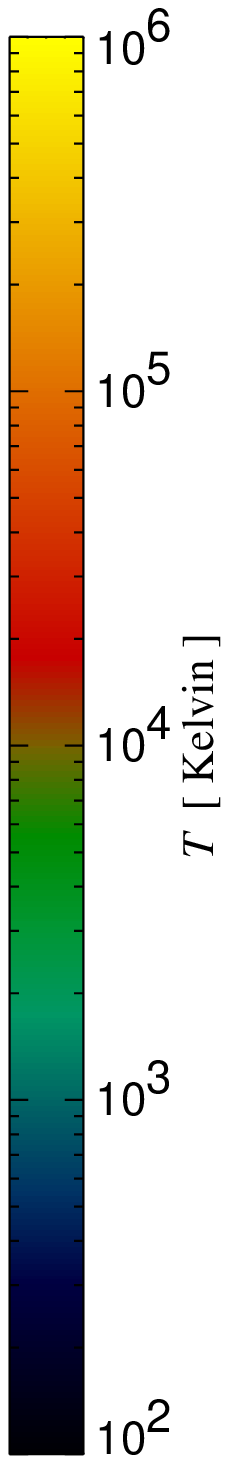}}%
\vspace{0.1cm}\\%
\resizebox{7.5cm}{!}{\includegraphics{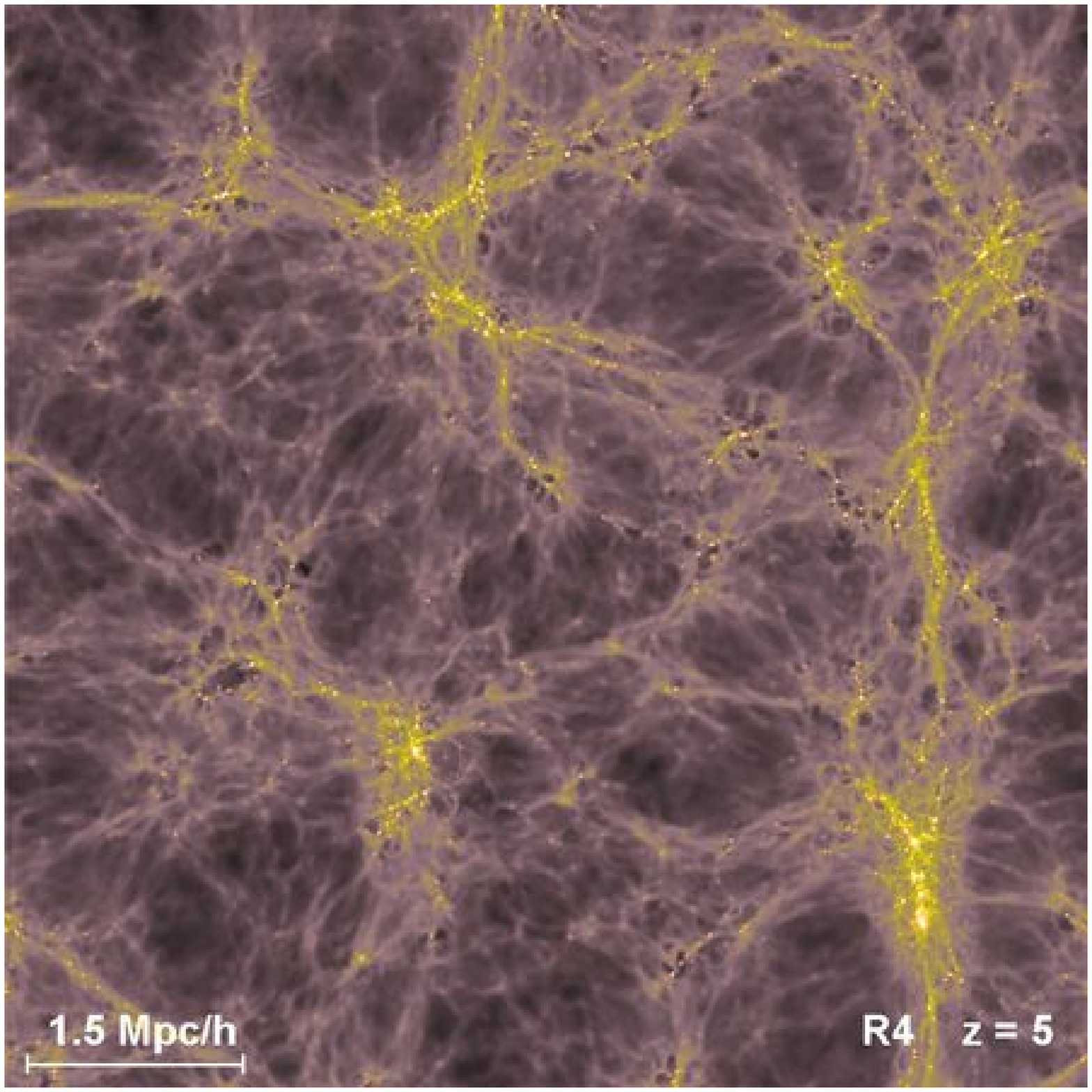}}%
\hspace{0.13cm}\resizebox{7.5cm}{!}{\includegraphics{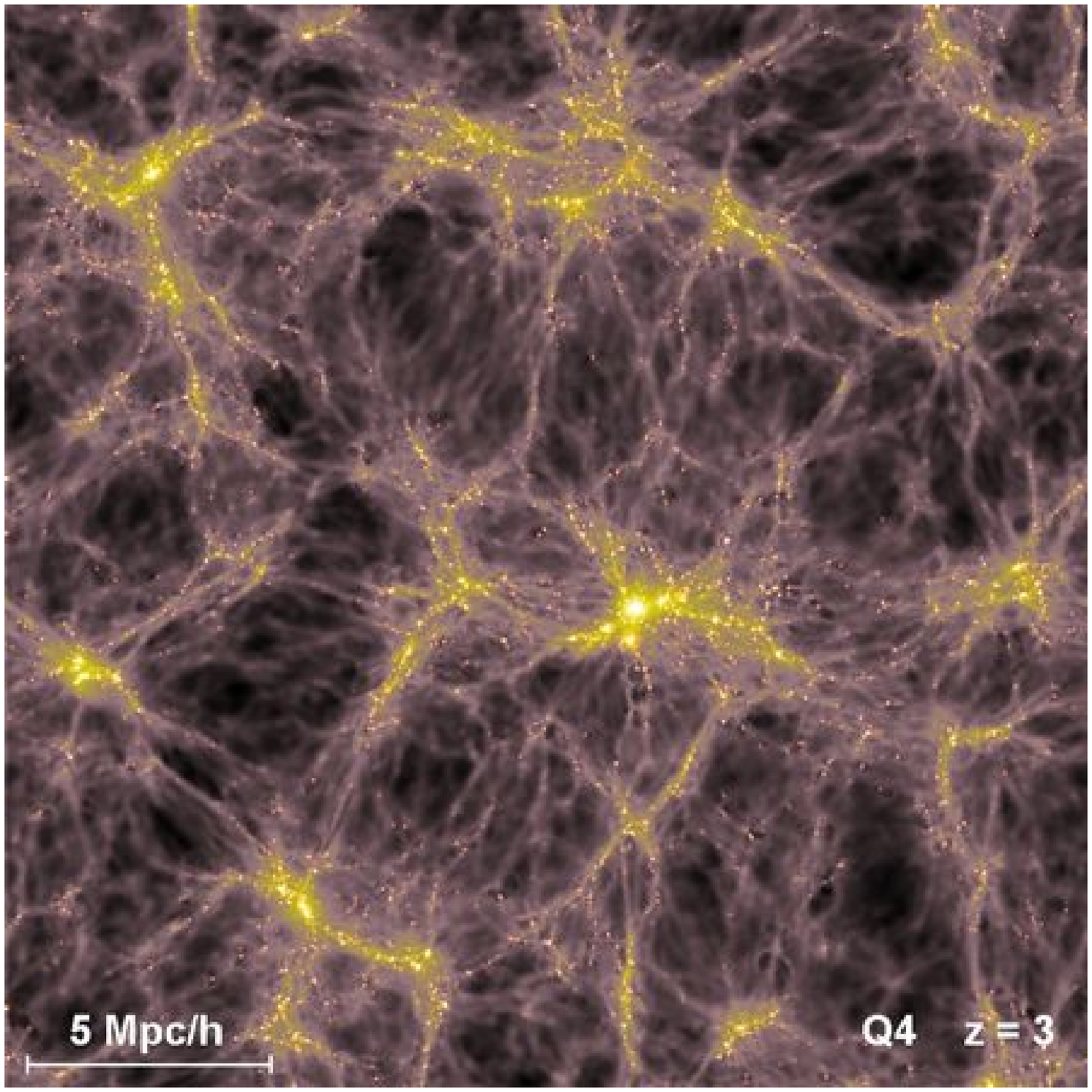}}%
\hspace*{1.36cm}\vspace{0.1cm}\\%
\resizebox{7.5cm}{!}{\includegraphics{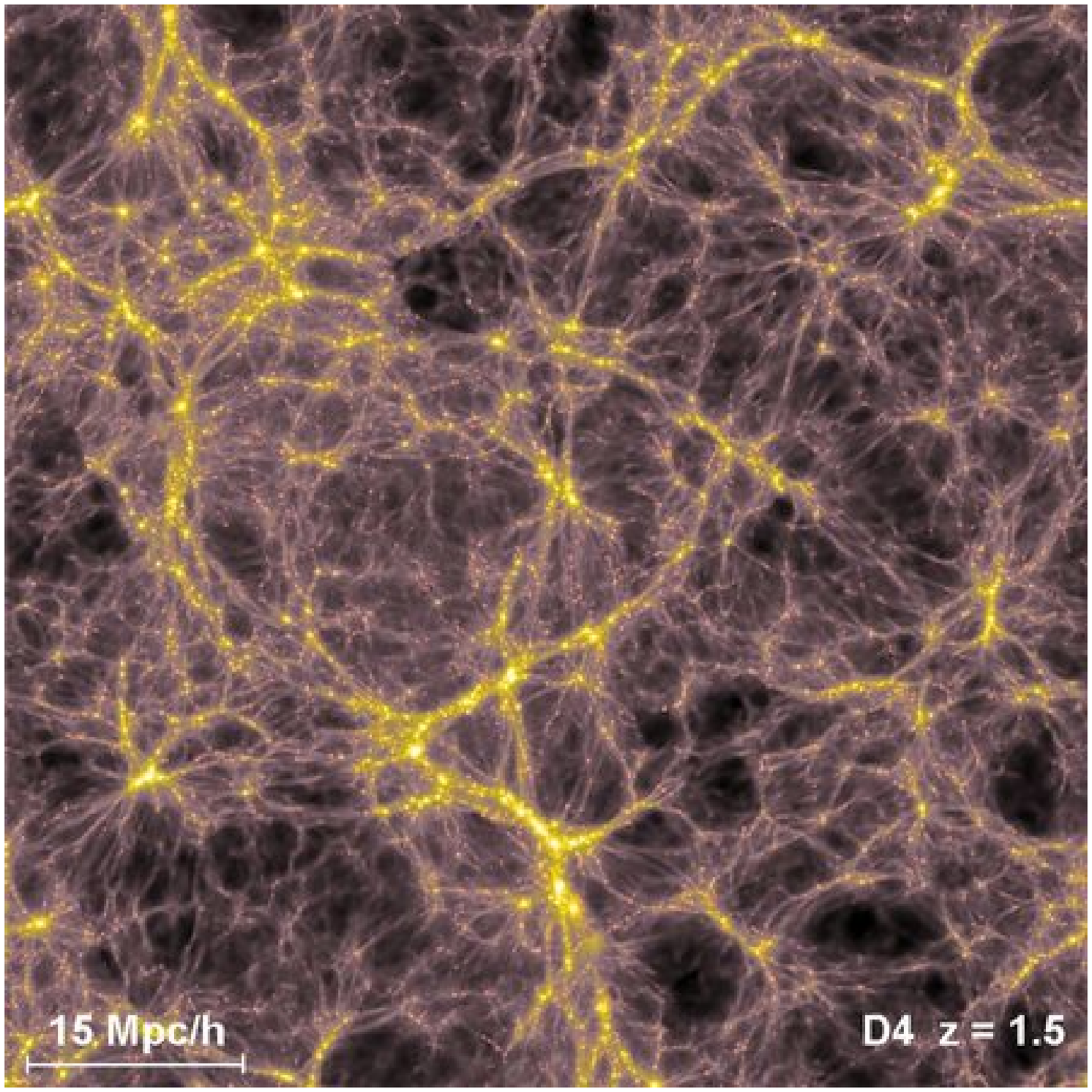}}%
\hspace{0.13cm}\resizebox{7.5cm}{!}{\includegraphics{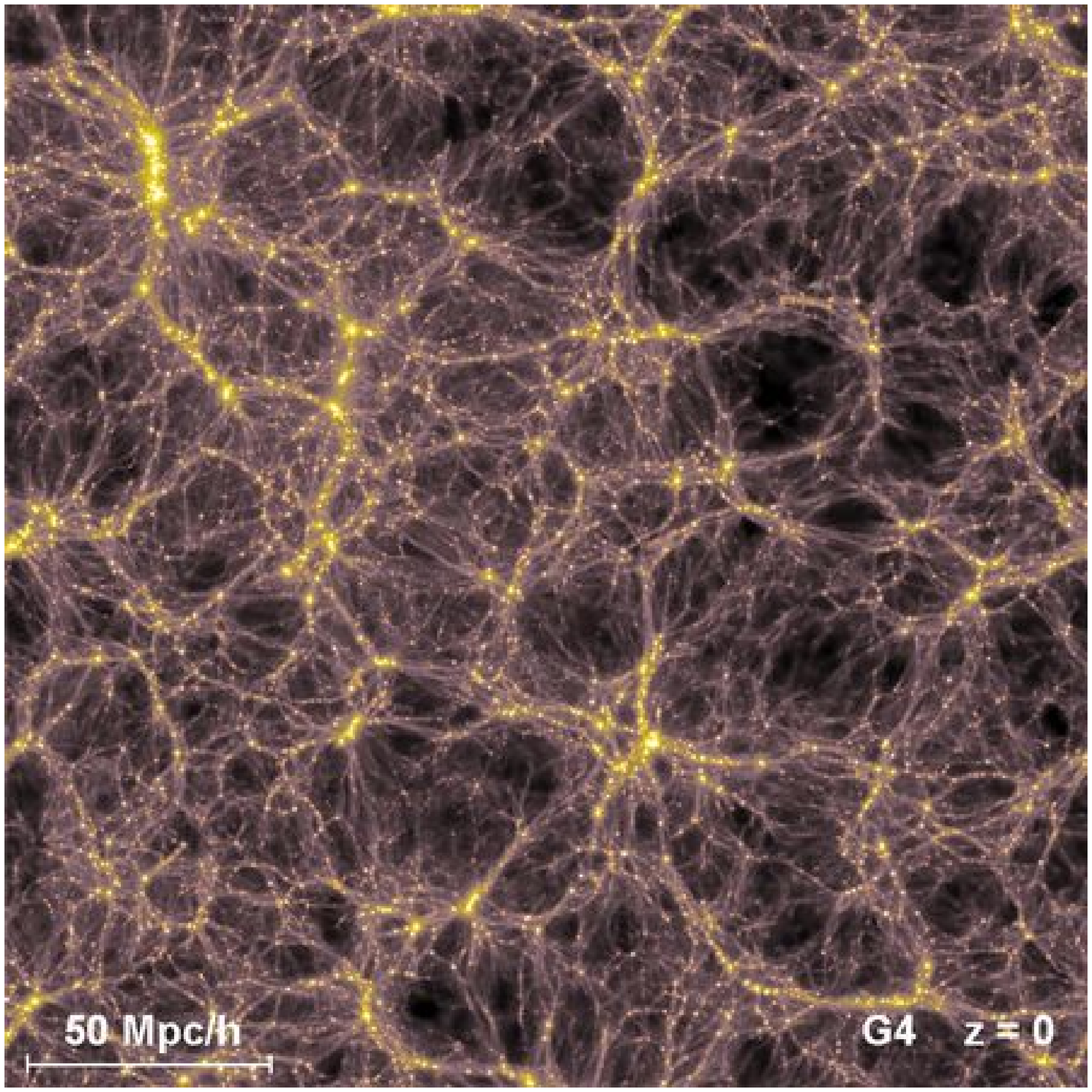}}%
\hspace*{1.36cm}\ \\%
\caption{Projected baryonic density fields in slices through a
selection of our simulations at various redshifts. In each case, the
slice has a thickness equal to $1/5$ of the box-size of the
corresponding simulation (see Table~\ref{tabSimulations}).  The Z4
simulation in the top left has the highest spatial resolution,
allowing to identify the hot ``bubbles'' in the IGM that develop as a
result of impinging galactic winds.  These bubbles are filled with gas
with temperatures up to $10^6\,{\rm K}$, as seen in the projected
mass-weighted temperature map in the top right.
\label{figSlices1}}
\end{figure*}

\begin{figure*}
\bc
\resizebox{8cm}{!}{\includegraphics{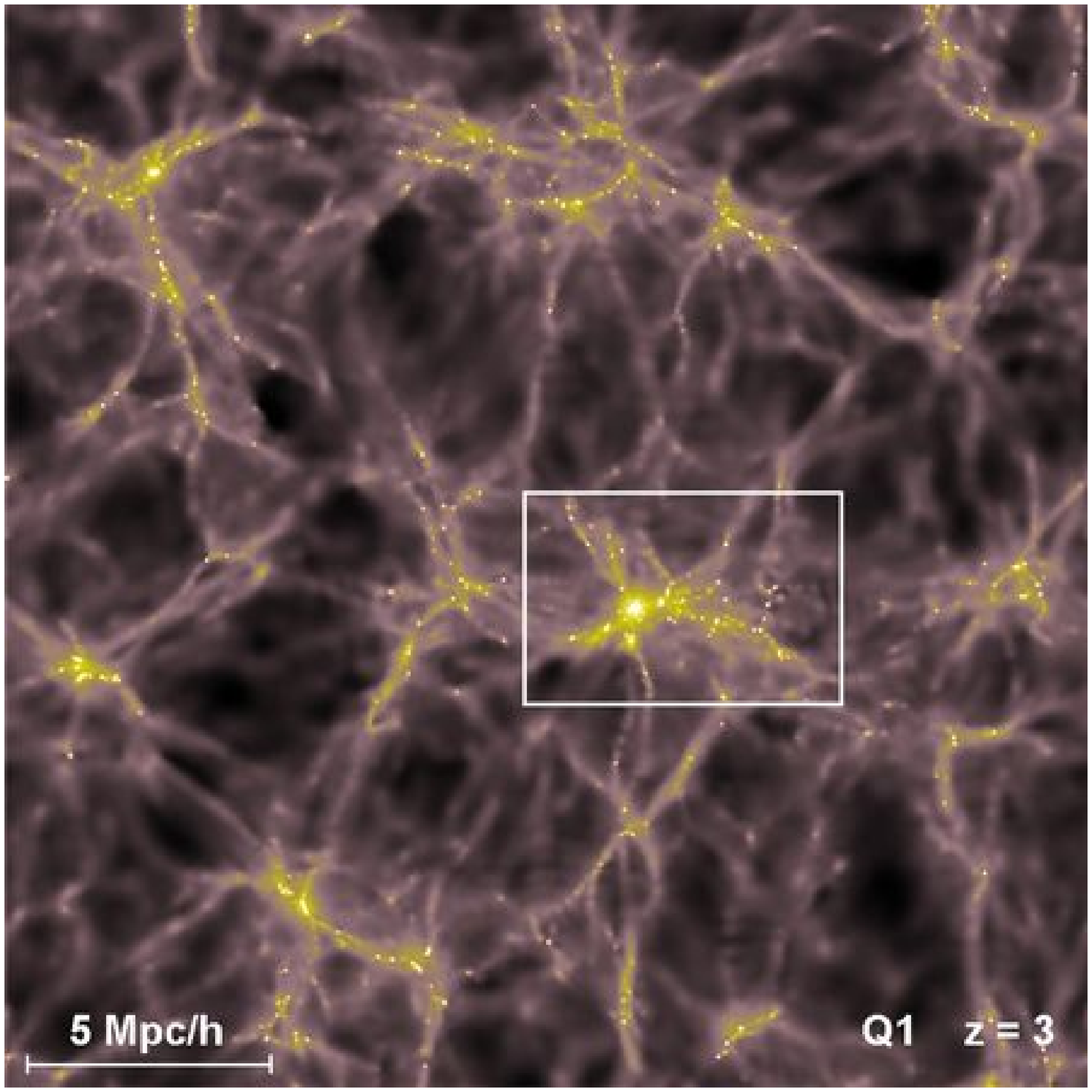}}%
\hspace{0.13cm}\resizebox{8cm}{!}{\includegraphics{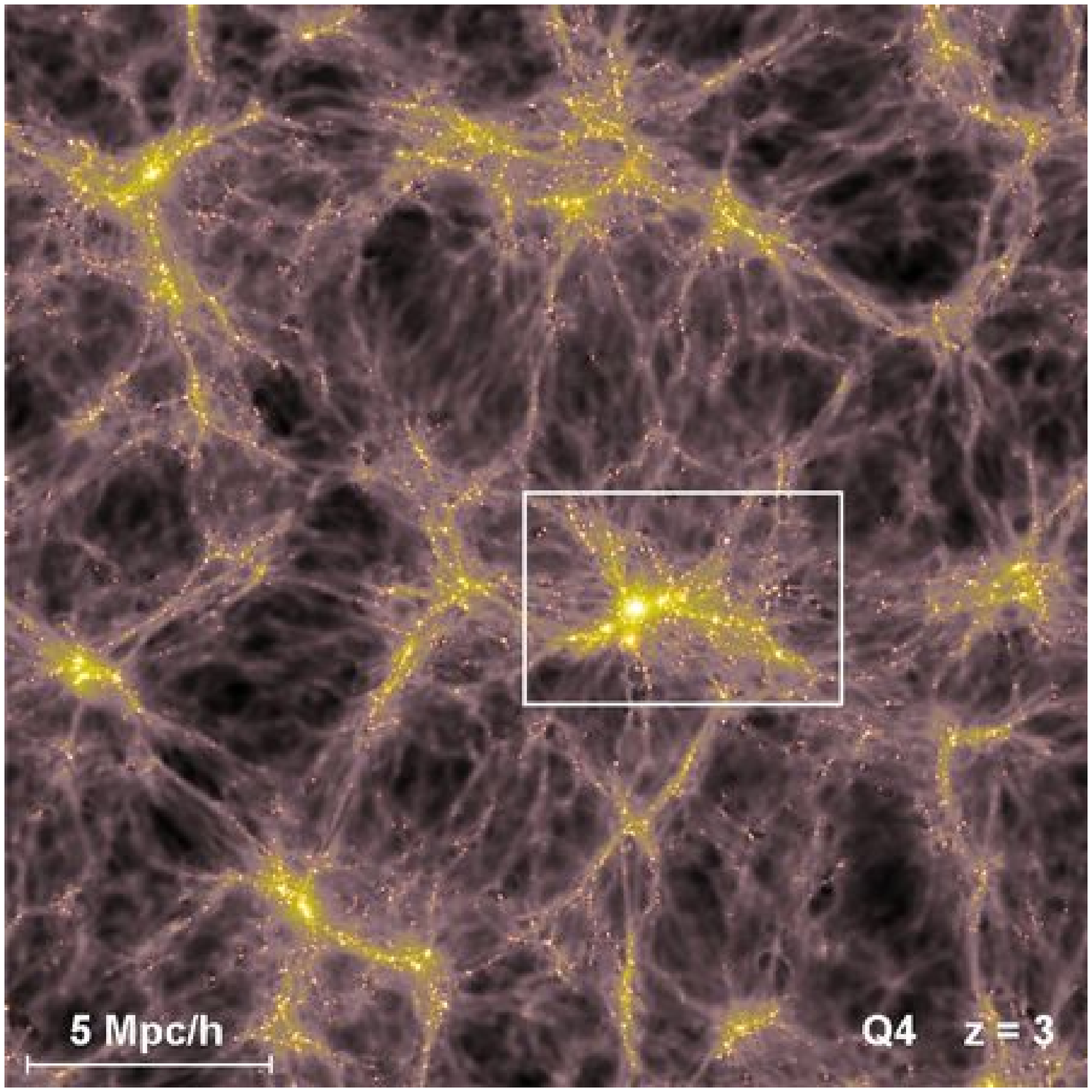}}\vspace{0.3cm}\\%
\resizebox{7.6cm}{!}{\includegraphics{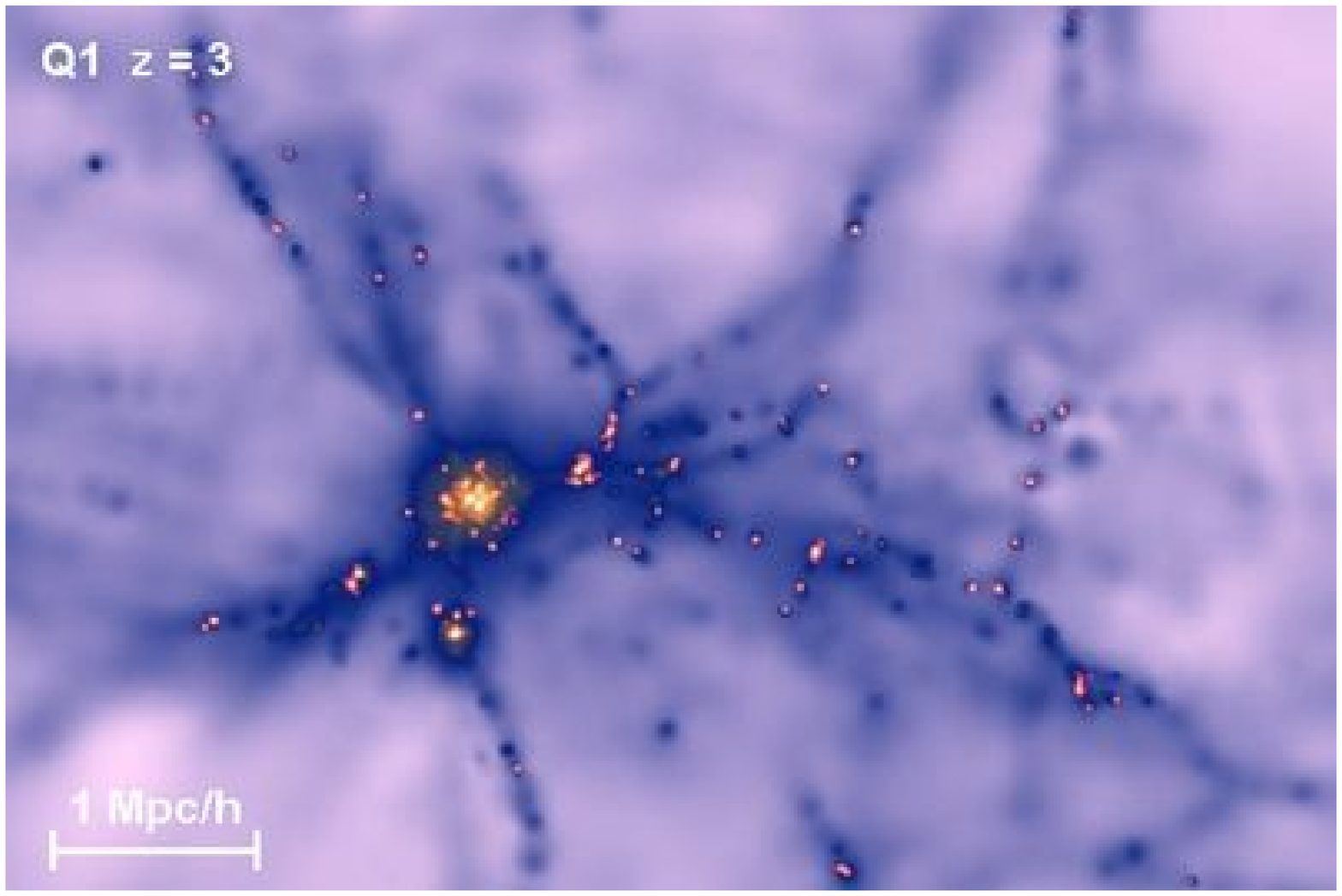}}%
\hspace{0.13cm}\resizebox{7.6cm}{!}{\includegraphics{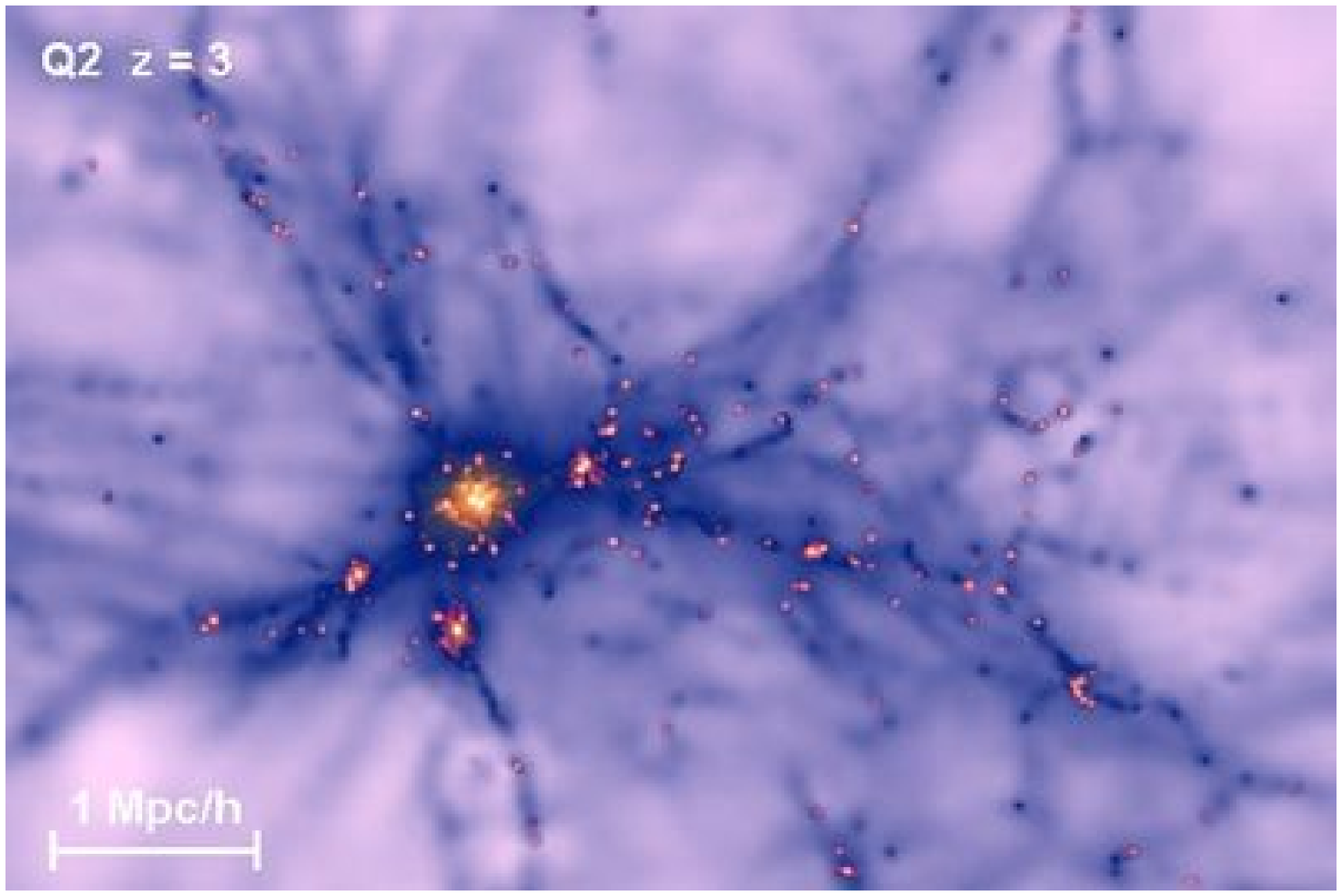}}\vspace{0.1cm}\\%
\resizebox{7.6cm}{!}{\includegraphics{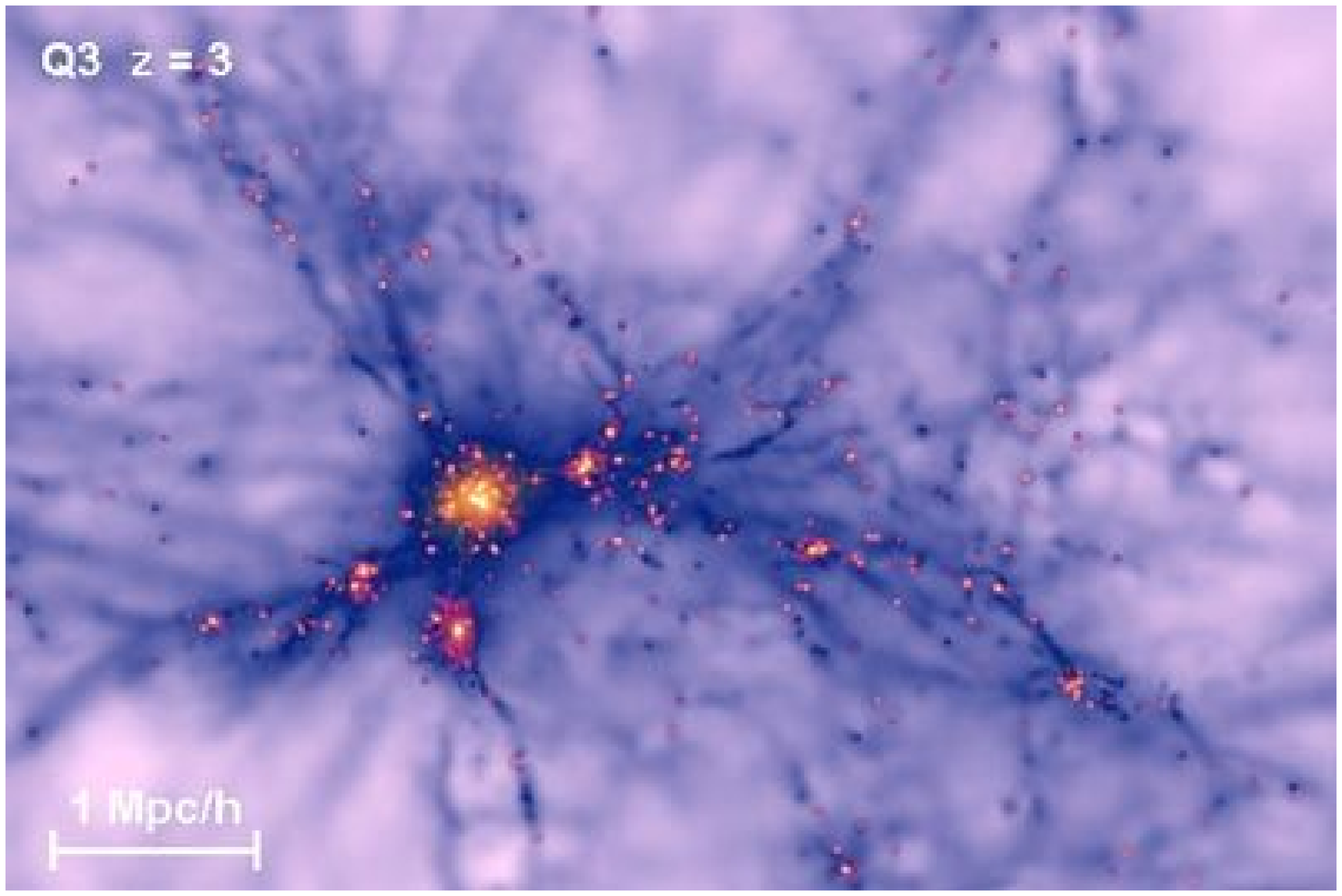}}%
\hspace{0.13cm}\resizebox{7.6cm}{!}{\includegraphics{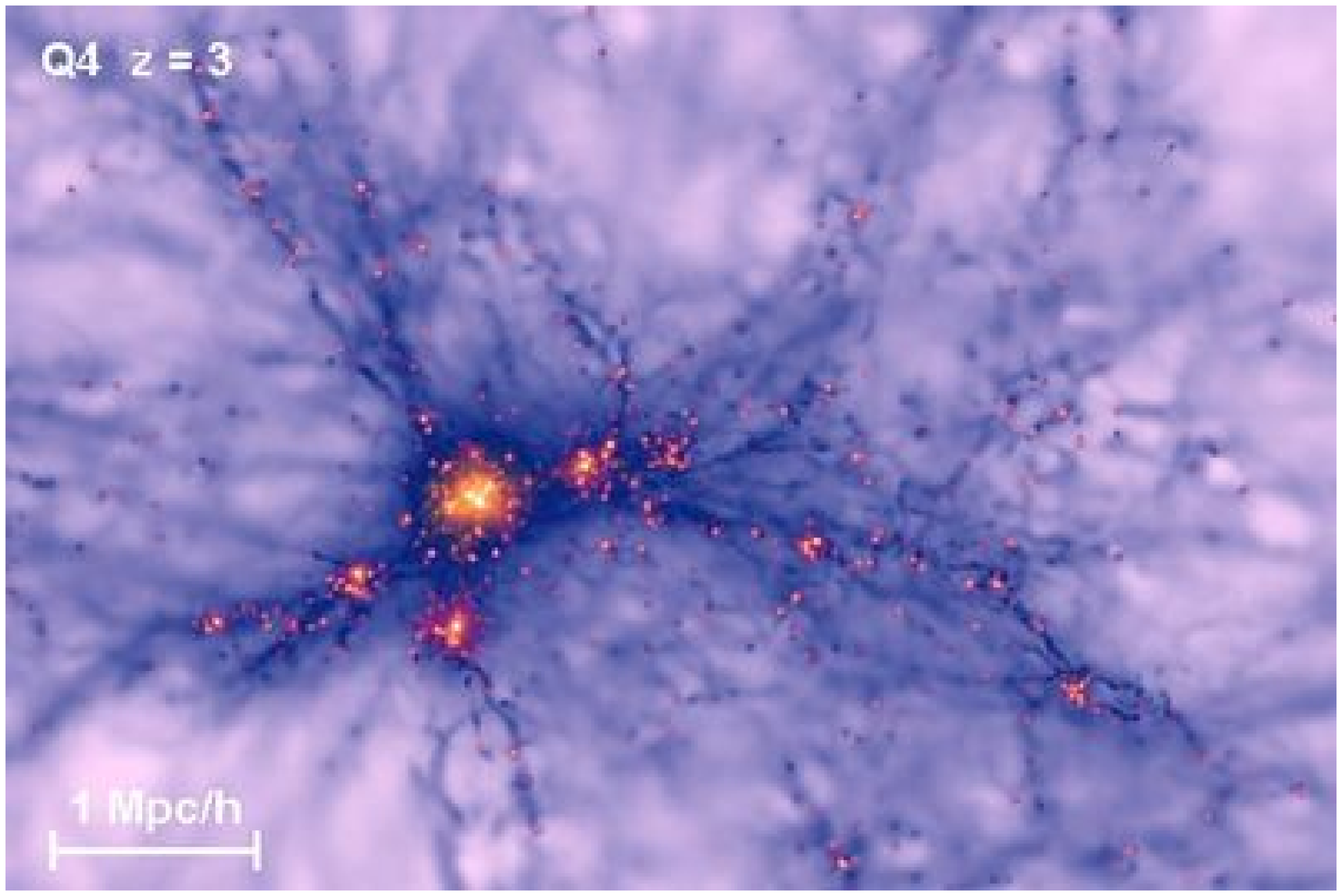}}\vspace{0.4cm}\\%
\caption{The two panels on top compare the projected baryonic density
field in slices through the Q1 and Q4 simulations at $z=3$.  It is
seen that the runs yield the same large-scale structure, as intended
by our method for generating the initial conditions.  In the bottom
four panels, we zoom in on the region marked by a white rectangle, and
show all four members of the Q-Series, where the mass resolution is
increasing by a factor of about $40$ along the sequence Q1, Q2, Q3, to
Q4. In these panels, the gas density is shown in a blue colour scale,
and the density of star particles is in red.  The higher resolution
runs in the series reproduce the objects previously seen in the low
resolution realisations, but they are able to resolve larger numbers
of low-mass halos, and they also improve the sampling of structures,
giving them a ``crisper'' appearance.  Note, however, that the
brightness of the largest objects, as measured by the density of star
particles shown in red, appears to be quite similar. This sense of
convergence will be quantified more precisely in
Section~\ref{SecConv}.
\label{figSlices2}}
\ec
\end{figure*}

In addition to this treatment of the multi-phase structure of the ISM,
we have included a phenomenological model for galactic outflows in the
simulations in our present study.  The motivation for this was
twofold.  On one hand, there is a large body of observational evidence
for the ubiquitous presence of such galactic winds
\citep[e.g.][]{Heck95,Bland95,Lehn96,Dahlem97,Mar99,Heck00}. On the
other hand, galactic-scale outflows provide one of the most attractive
mechanisms for understanding how the low-density intergalactic medium
was chemically enriched with heavy elements, and they might also play
a crucial role in the global regulation of star formation, suppressing
its overall efficiency to the ``low'' values suggested by
observational estimates of the luminosity density of the Universe.
Simulations without strong kinetic feedback invariably overpredict the
luminosity density of the Universe due to the high efficiency of gas
cooling \citep[e.g.][]{Bal01}.

In our fiducial parameterisation of this process, each star-forming
galaxy drives a wind with a mass-outflow rate equal to two times its
star formation rate, and with a wind-velocity of $v_w=484\, {\rm
km\,s^{-1}}$. These choices are deliberately ``extreme'' in the sense
that the total kinetic energy of the wind is then of order the energy
released by supernovae.  Note, however, that these wind parameters are
characteristic of the properties of outflows associated with
star-forming disks \citep[e.g.][]{Mar99}.  In addition to the
simulations listed in Table~\ref{tabSimulations}, which all employed
these fiducial values in our wind model, we performed other runs with
differing particle number and in boxes of different sizes in which we
varied only the intensity of the winds.  In this manner, we were able
to examine the sensitivity of our conclusions to the presence or
absence and strength of the winds.

On a technical level, the winds in the simulations are set up by
stochastically selecting gas particles from the star-forming ISM and
``adding them'' to a wind by modifying their velocity vectors
appropriately.  While the generation of the wind is hence purely
phenomenological, subsequent interactions with surrounding gas are
handled properly by our hydrodynamical code. The winds may thus
intercept or entrain infalling gas, shock the IGM and even produce hot
bubbles within it.  Whether or not a wind can escape from a galaxy
depends primarily on the depth of the galaxy's dark matter potential
well.  Halos with virial temperatures below $\sim 10^6\,{\rm K}$ have
central escape velocities lower than the wind speed and can lose some
of their baryons in an outflow, while more massive halos will contain
winds, to the extent that winds become progressively less important
for the regulation of star formation in more massive objects.

The fact that winds are expected to be especially important on small
mass scales points to one of the main numerical challenges encountered
in simulating hierarchical galaxy formation.  At high redshift, most
of the star formation takes place in low-mass galaxies, which form in
abundant numbers as progenitors of more massive systems.  Resolving
the earliest epochs of star formation, therefore, requires very high
mass resolution.  However, at low redshift, star formation shifts to
progressively more massive systems.  At these late times, it is
necessary to simulate a large cosmological volume in order to obtain a
fair sample of the high-end of the mass function.  Obviously, the
simultaneous requirements of high mass resolution and large simulation
volumes will quickly exhaust any computational resource.

For example, a simulation which attempts to fully resolve star
formation in all star forming halos from high-$z$ to the present would
have to follow a cosmological volume of $\sim (100\,h^{-1}\, {\rm
Mpc})^3$, at a mass resolution of about $10^6\,h^{-1}{\rm M}_\odot$,
assuming here for the moment that star formation is restricted to
halos of mass larger than $10^8 \,h^{-1}{\rm M}_\odot$, and that a
mere 100 SPH particles are sufficient to obtain a reasonably accurate
estimate of the star formation rate in a halo. This would require of
order $10^{11}$ simulation particles, which is substantially beyond
what is currently feasible.  Note that unlike in purely gravitational
bottom-up growth of structure, feedback processes associated with
winds offer the possibility that star formation on small mass scales
could influence the gas dynamics on larger scales.  Degrading the mass
resolution thus may have more problematic consequences than in purely
collisionless simulations, because it means that the effects of winds
coming from low-mass halos can be lost.

Rather than attempting to resolve the cosmic star formation history in
a single calculation, we have, therefore, selected a different
strategy.  We have simulated a series of cosmological volumes with
sizes ranging from $1\,h^{-1}{\rm Mpc}$ to $100\,h^{-1}{\rm Mpc}$ per
edge, with intermediate steps at box sizes $3.375$, $10.0$, and
$33.75\,h^{-1}{\rm Mpc}$.  The smaller boxes are run only to a
redshift where their fundamental mode starts to become non-linear.
Roughly up to this point, the boxes can at least approximately be
taken as a fair sample of a $\Lambda$CDM universe on the scales they
are intended to represent.

For each of the box sizes, we have performed a number of runs, varying
only the mass and spatial resolution. In particular, we have
repeatedly increased the mass resolution in steps of $1.5^3$, and the
spatial resolution in steps of $1.5$. For example, our lowest
resolution run of the $10.0\,h^{-1}{\rm Mpc}$ box had a resolution of
$2\times 64^3$ particles (`Q1'-run), which we then stepped up
sequentially to $2\times 96^3$ (`Q2'), $2\times 144^3$ (`Q3'),
$2\times 216^3$ (`Q4'), and finally to $2\times 324^3$ (`Q5'). In this
progression, the mass resolution in the gas is hence improved by a
factor $\sim 130$ from $4.2\times10^7\,h^{-1}{\rm Mpc}$ (Q1) to
$3.3\times10^5\,h^{-1}{\rm Mpc}$ (Q5), and the 3D spatial resolution
by a factor $\sim 5$ from $6.25\,h^{-1}{\rm kpc}$ to $1.23\,h^{-1}{\rm
kpc}$.  For all simulations which form such a {\em series}, we
generated initial conditions such that the amplitudes and phases of
large-scale waves were identical; i.e.~a higher resolution simulation
in a series had the same initial fluctuations on large scales as any
of the lower resolution fellow members of its series, with additional
waves between the old and new Nyquist frequencies sampled from the
power spectrum randomly.  In this manner, a numerical convergence
study within a given series becomes possible which is free of cosmic
variance and where an object by object comparison can be
made. Note, however, that for different box sizes we have chosen
different initial random number seeds, so that they constitute
independent realisations.

While the simulations within a given series enable us to cleanly
assess the convergence of star formation seen on a given halo mass
scale, the dynamic range of even a $2\times 324^3$ simulation is too
low to identify {\em all} the star formation occurring in the
Universe.  This is where our different box sizes come in. They provide
us with a sequence of interlocking scales and epochs, enabling us to
significantly broaden the dynamic range of our modeling. For the
$1\,h^{-1}{\rm Mpc}$ box of the ``Z-series'', we reach mass
resolutions of up to $1.1\times 10^3 h^{-1}{\rm Mpc}$, making it
possible to resolve the gas content even in ``sterile'' dark matter
halos which are never able to condense gas in their centres by atomic
line cooling.  Apart from a putative population~III
\citep[e.g.][]{Carr84,Bromm99,Abel2002}, most of them are thus never
expected to form stars.  At the other end of the spectrum, we resolve
rich galaxy clusters in our ``G-series'' runs, which have a volume
$10^6$ times larger than the ``Z-series'', but a correspondingly
poorer mass resolution.

We show a graphical illustration of the numerical simulations analysed
in our present study in Figure~\ref{figSimSet}, and give a complete
list of their properties in Table~1, together with some important
numerical parameters. As a naming convention, the simulations are
denoted by a single character that indicates their box size, and a
number that gives the particle resolution, from ``1'' for $2\times
64^3$ to ``5'' for $2\times 324^3$.  All the simulations were
performed on the Athlon-MP cluster at the Center for Parallel
Astrophysical Computing (CPAC) at the Harvard-Smithsonian Center for
Astrophysics.  We used a modified version of the parallel {\small
GADGET}-code \citep{SprGadget2000}, and integrated the entropy as the
independent thermodynamic variable \citep{SprHe01}.

\subsection{Slices through the gas density field}

In Figure~\ref{figSlices1}, we show projected baryonic density fields
for each of our five simulation volumes, illustrating the range of
scales encompassed by the simulations.  In each case, we visualise the
gas density in slices of thickness $1/5$ of the simulation box.  To
show nearly all of the simulation volume in one projection, we have
tilted the slices slightly with respect to the principal axes of the
simulation boxes (by $26.6^{\rm o}$ around the $y$-axes, and once more
by $26.6^{\rm o}$ around the $x'$-axes).  Using the periodicity of the
volumes, we can then extend the side-length of the projection to
$\sqrt 5$ times the original box size, while approximately avoiding
the repetition of structure.  In the given representation, we then
obtain a thin slice that has exactly the volume of the full simulation
box and displays about 90\% of its content uniquely.  Alternative ways
of displaying the ``full'' simulation would either have to employ
several different slices, or would have to project using a slice of
thickness equal to the box size itself, leading to a confusing
superposition of many structures.

As part of Figure~\ref{figSlices1}, we also show a mass-weighted
temperature map of a Z-series run at $z=6$.  In the corresponding
density map, it is apparent how the first star-forming galaxies drive
vigorous outflows that produce hot ``bubbles'' in the IGM.  As seen in
the temperature map, these bubbles are filled with gas at temperatures
up to $10^6\,{\rm K}$.  We plan to investigate the impact of these
outflows on the IGM in forthcoming work.  We note here, however, that
the winds are capable of stripping some gas from halos nearby, thereby
providing an indirect source of negative feedback in the form
advocated by~\citet{Scan01c}.

In Figure~\ref{figSlices2}, we show a comparison of four simulations
within the Q-Series having different mass resolution at redshift
$z=3$.  It is apparent that better mass resolution leads to more
finely resolved structure, and to a larger number of small halos that
were previously unresolved.  Note, however, that the pattern of
large-scale structure and the location of the most massive halos is
virtually unchanged, as expected, given our method for constructing
the initial conditions for the runs in this series.

\subsection{Halo definition}

Using our approach, we can directly measure the total star formation
rate in a simulation box at each timestep.  However, we are interested
not only in the cumulative star formation rate normalised to unit
comoving volume, but also in the way this rate can be broken up into
contributions from various halo mass scales.  It should be emphasised
that our requirement for the gas to be highly overdense for star
formation to occur essentially guarantees that this process is
entirely restricted to the centres of dark halos, with no stars
forming in the low-density IGM.

In order to identify virialised halos as sites of star formation, we
begin by applying a ``friends-of-friends'' (FOF) group finding
algorithm to the dark matter particles, using a fixed comoving linking
length equal to $0.2$ times the mean interparticle spacing of the dark
matter particles.  We restrict the algorithm to the dark matter
because it is unclear how one should deal with the varying particle
number and particle type used to represent the baryons in the
simulations.  Owing to the collapse of a large fraction of the gas to
very high overdensity, one could introduce biases close to the group
detection threshold if a FOF procedure were applied to all particles
on an equal footing.  It thus seems safer to restrict ourselves to
group selection based on dark matter particles alone, which to first
order will be unperturbed by the baryonic physics, permitting a
selection of the same structures as in an equivalent simulation that
followed only dark matter, to a good approximation.

\begin{figure}
\bc%
\vspace*{-0.5cm}\resizebox{8cm}{!}{\includegraphics{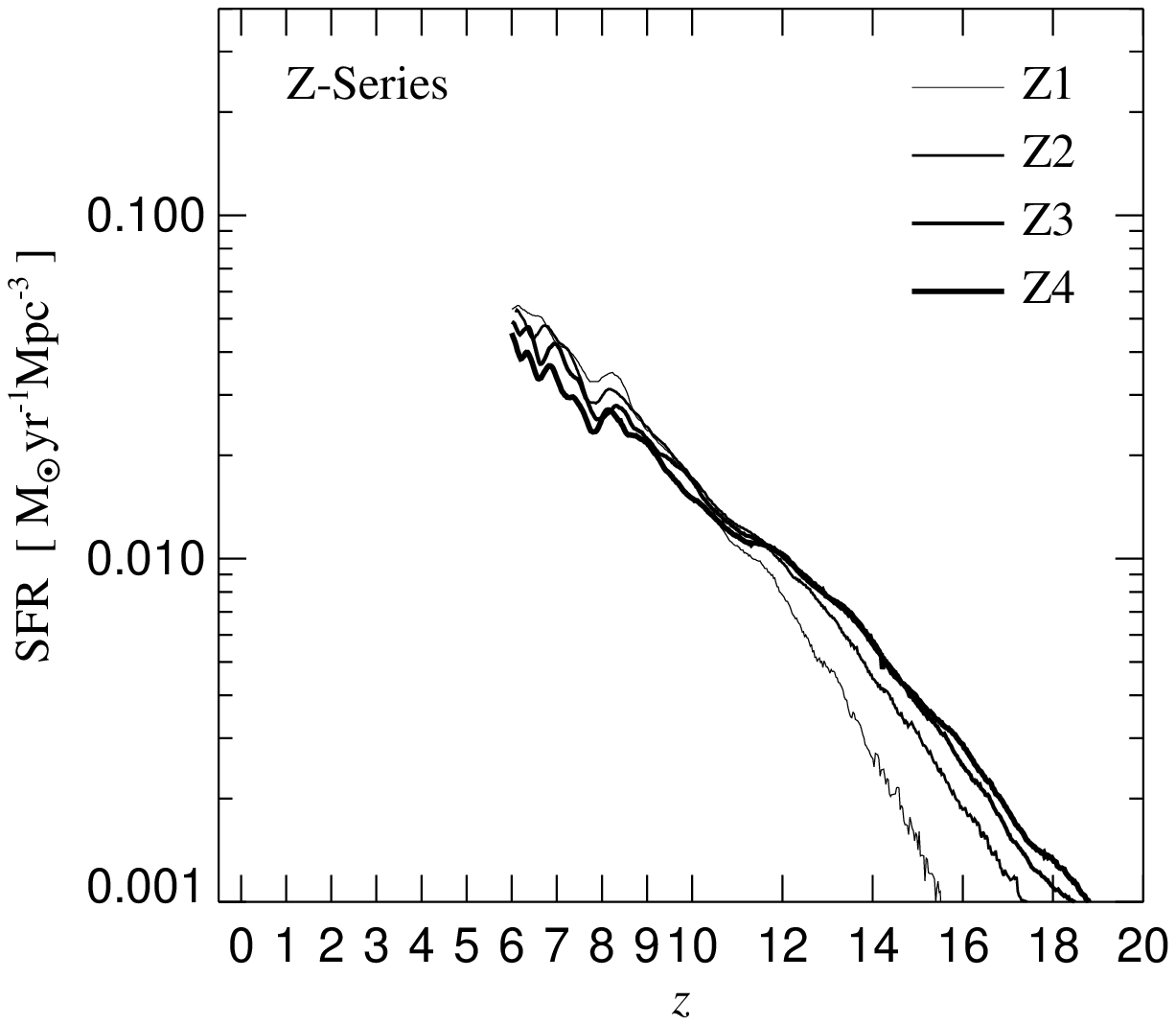}}\\%
\vspace*{-0.5cm}\resizebox{8cm}{!}{\includegraphics{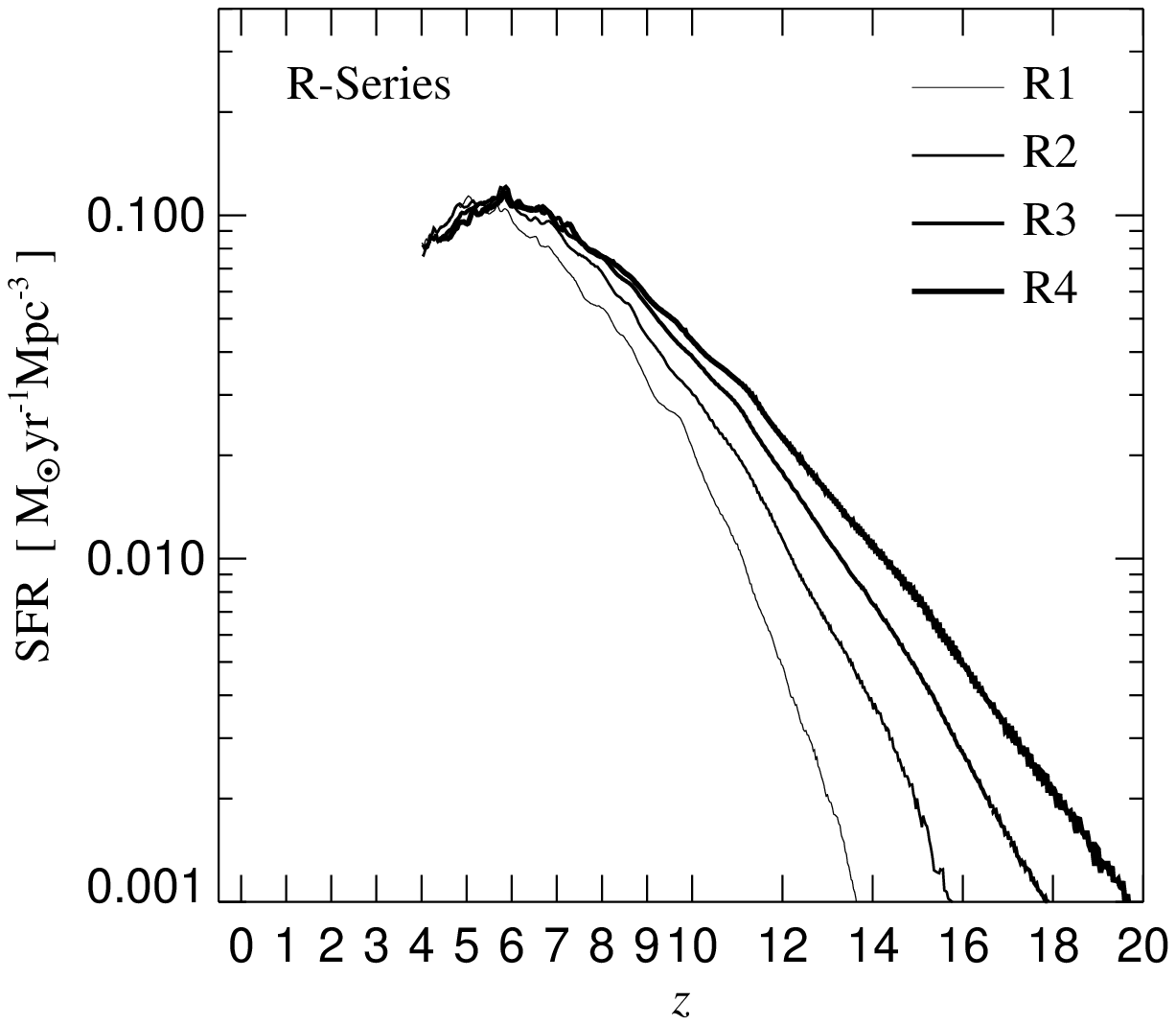}}\\%
\vspace*{-0.5cm}\resizebox{8cm}{!}{\includegraphics{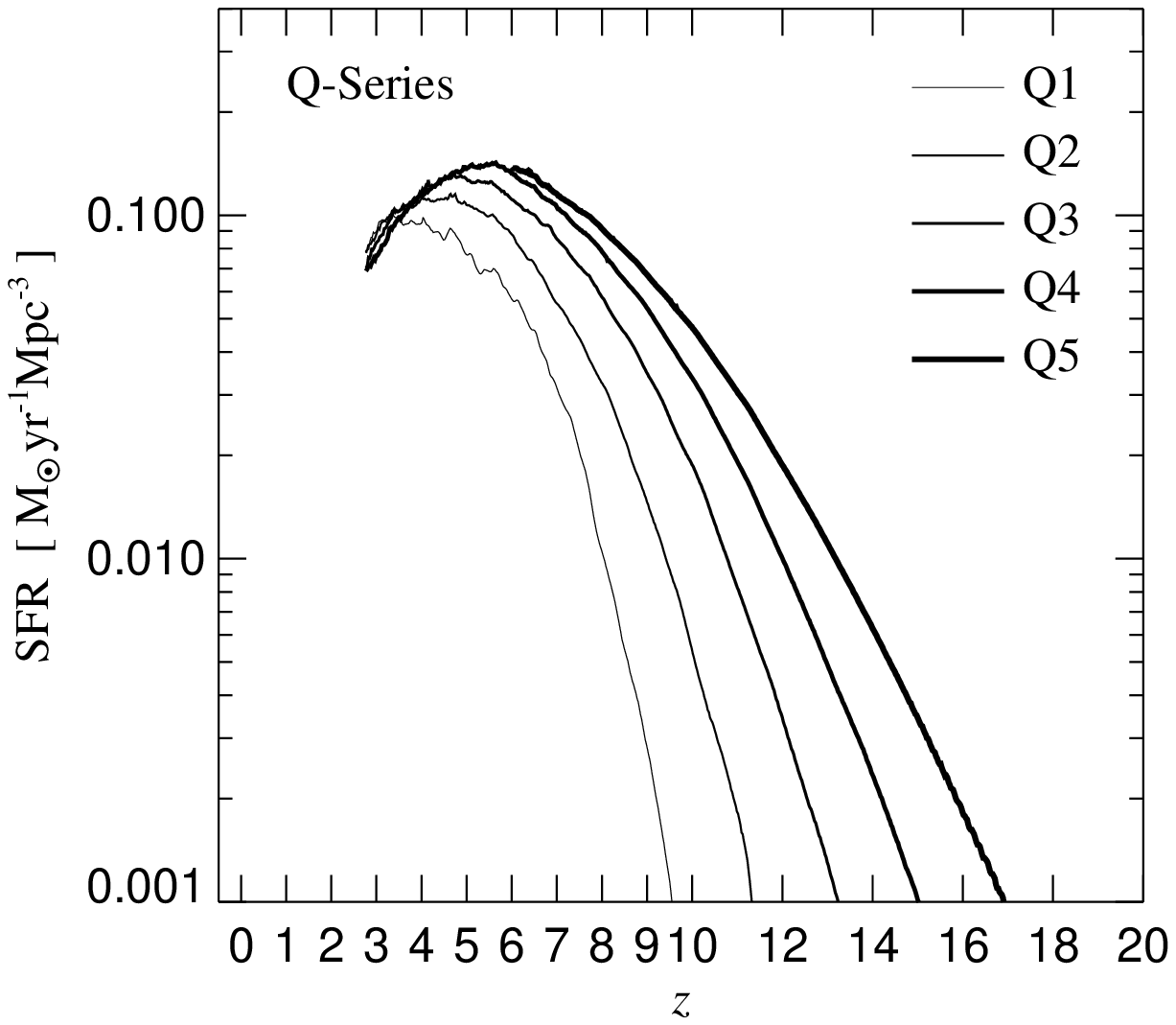}}%
\caption{Evolution of the cosmic star formation rate density in the
simulations of the Z-, R-, and Q-series. For increasing
mass-resolution, there is clearly more star formation resolved at high
redshift. However, in every case, there is a certain redshift at which
the simulations begin to agree very well. The mass resolution required
to reach this convergence point is a function of redshift and becomes
increasingly more demanding towards higher redshift.
\label{figSFR_ZRQ}}
\ec
\end{figure}

In a second step, we then associate each gas or star particle with its
nearest dark matter particle and discard all groups with fewer than 32
dark matter particles, resulting in our final ``FOF catalogue''.  Note
that \citet{Jen01} have shown that the construction of group
catalogues using the same linking length for all cosmologies and
redshifts, independent of $\Omega(z)$, leads to mass functions that
can be described by a single fitting formula.  Group catalogues
selected at a constant comoving overdensity with respect to the
background, as we do here, are thus easier to interpret than ones
obtained by trying to account for the scaling of the virial
overdensity with cosmological parameters.

For the analysis of the star formation multiplicity function discussed
in Section~\ref{SecMulti}, we will use the FOF catalogues directly,
taking the total mass of each group as a ``virial'' mass, and the sum
of the star formation rates of all a group's SPH particles as the
halo's star formation rate.  Following \citet{Mo2002}, we assign a
physical radius to a halo of mass $M$ according to \be r_{200} =
\left[ \frac{GM}{100\,\Omega_m(z) H^2(z)}\right]^{1/3}=
\frac{1}{1+z}\left[ \frac{GM}{100\,\Omega_0 H_0^2}\right]^{1/3}, \ee
and a corresponding circular velocity $V_c= (GM/r_{200})^{1/2}$.  We
also define a ``virial'' temperature as \be T=\frac{\mu V_c^2}{2k}
\simeq 36 \left(\frac{V_c}{{\rm km \, s^{-1}}}\right)^2 \, {\rm K},
\ee where $\mu \simeq 0.6 m_{\rm p}$ is the mean molecular weight of
the ionised plasma found in halos hotter than $10^4\,{\rm K}$.

However, for a comparison of halos on an object-by-object basis
between the different runs of a given series, it is advantageous to
reduce the noise in the virial mass assignment by using the spherical
overdensity algorithm to define the positions and virial radii of
halos.  To this end, we find the minimum of the gravitational
potential of each FOF group, and define the location of the
corresponding particle as the halo center. We then grow spheres until
they enclose a comoving overdensity of 200 with respect to the mean
background density, with the star formation rate of the halo being
simply defined as the sum of all the star formation rates of the
enclosed SPH particles. In order to match objects between different
simulations of a series, we consider the halos of one of the
simulations in declining order of mass, and find for each of them the
nearest halo in the other simulation.  Each halo is only allowed to
become a member of one matching pair in this process.

\begin{figure*}
\bc
\resizebox{!}{6.6cm}{\includegraphics{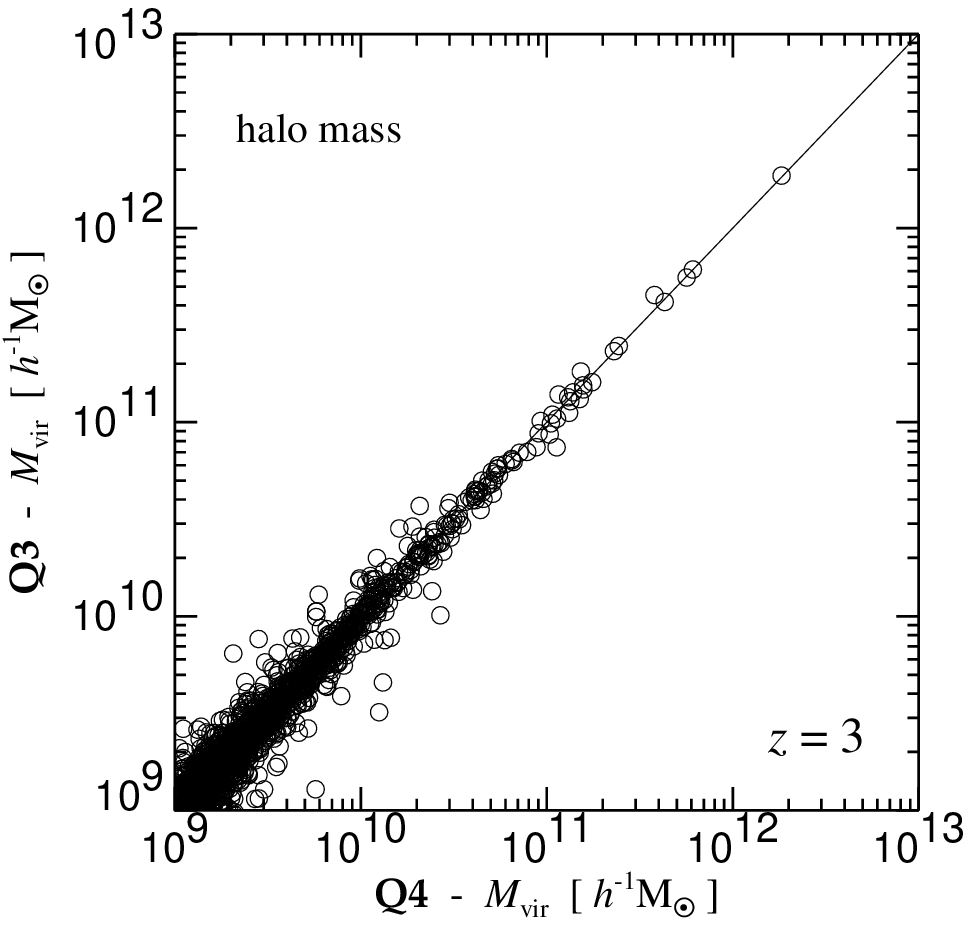}}%
\hspace*{-0.3cm}\resizebox{!}{6.6cm}{\includegraphics{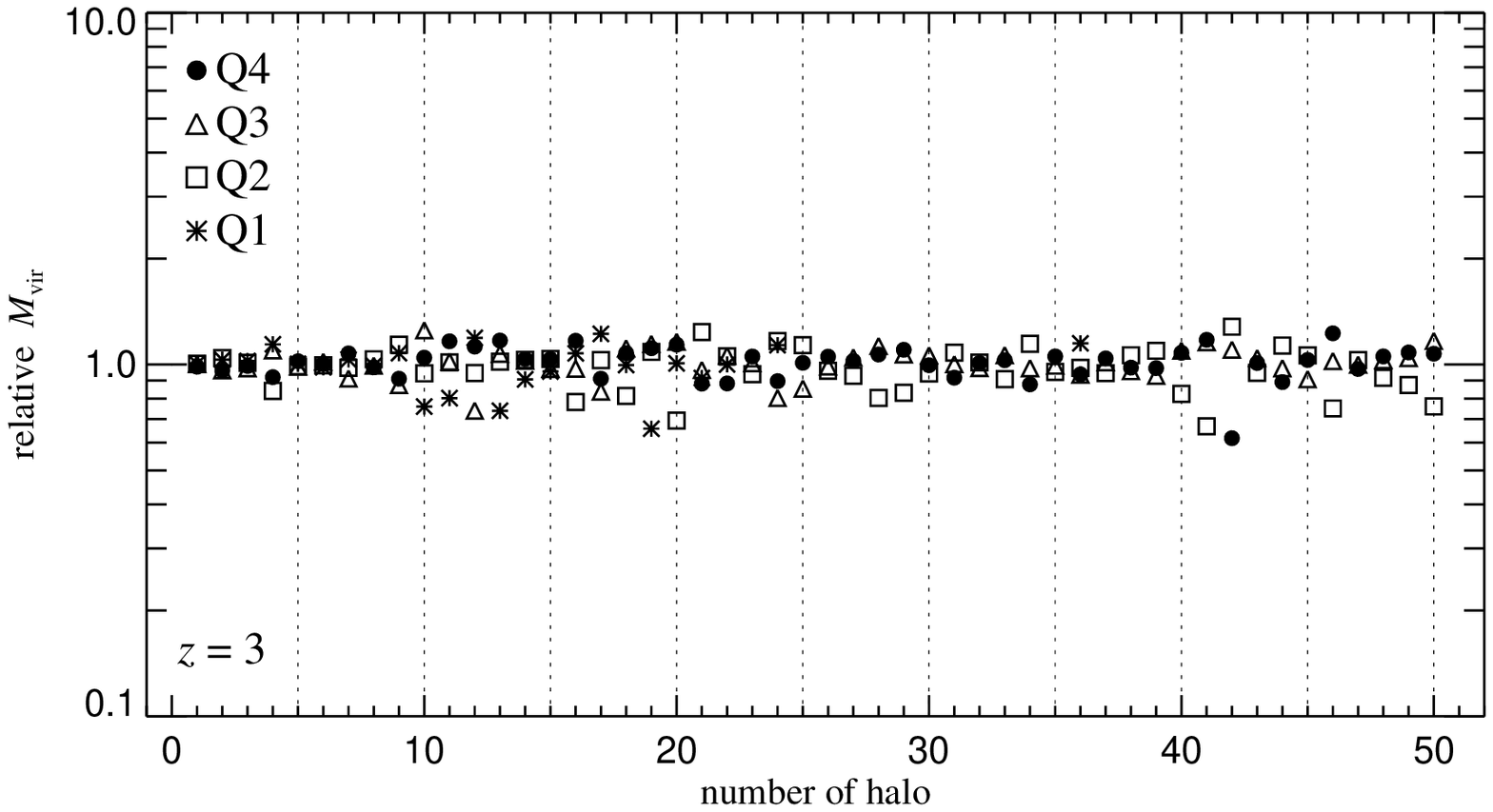}}\\%
\resizebox{!}{6.6cm}{\includegraphics{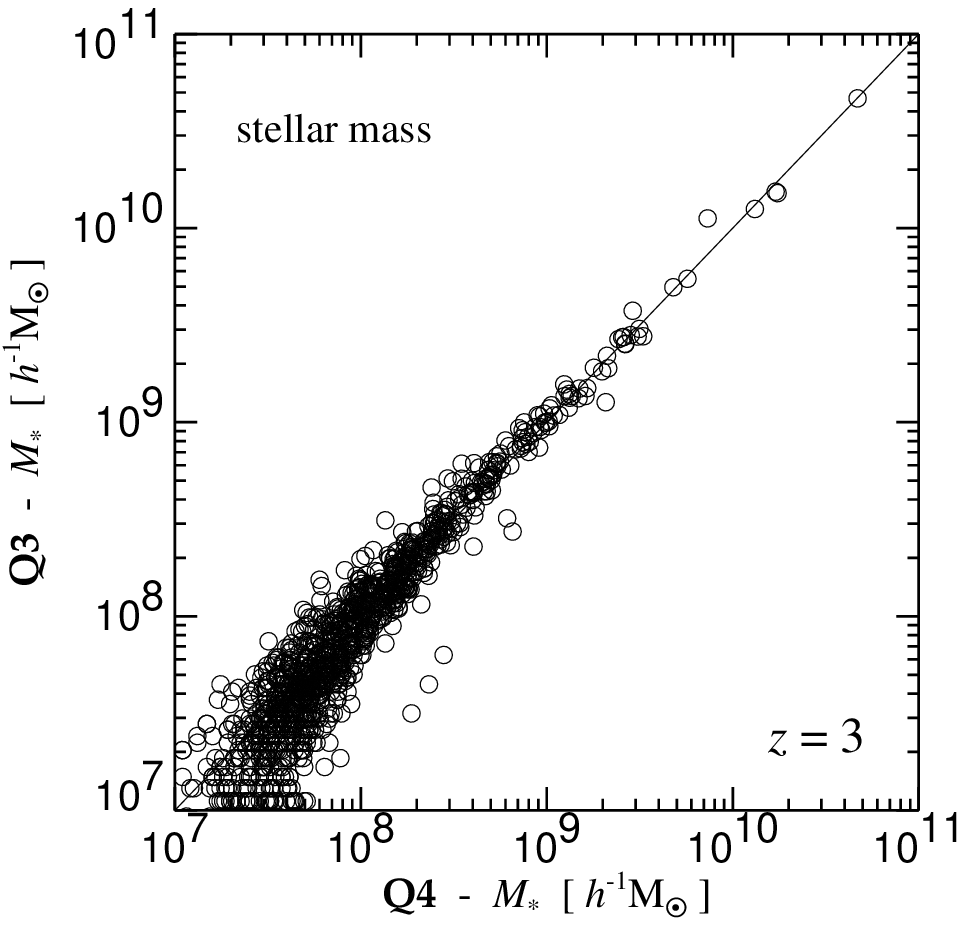}}%
\hspace*{-0.3cm}\resizebox{!}{6.6cm}{\includegraphics{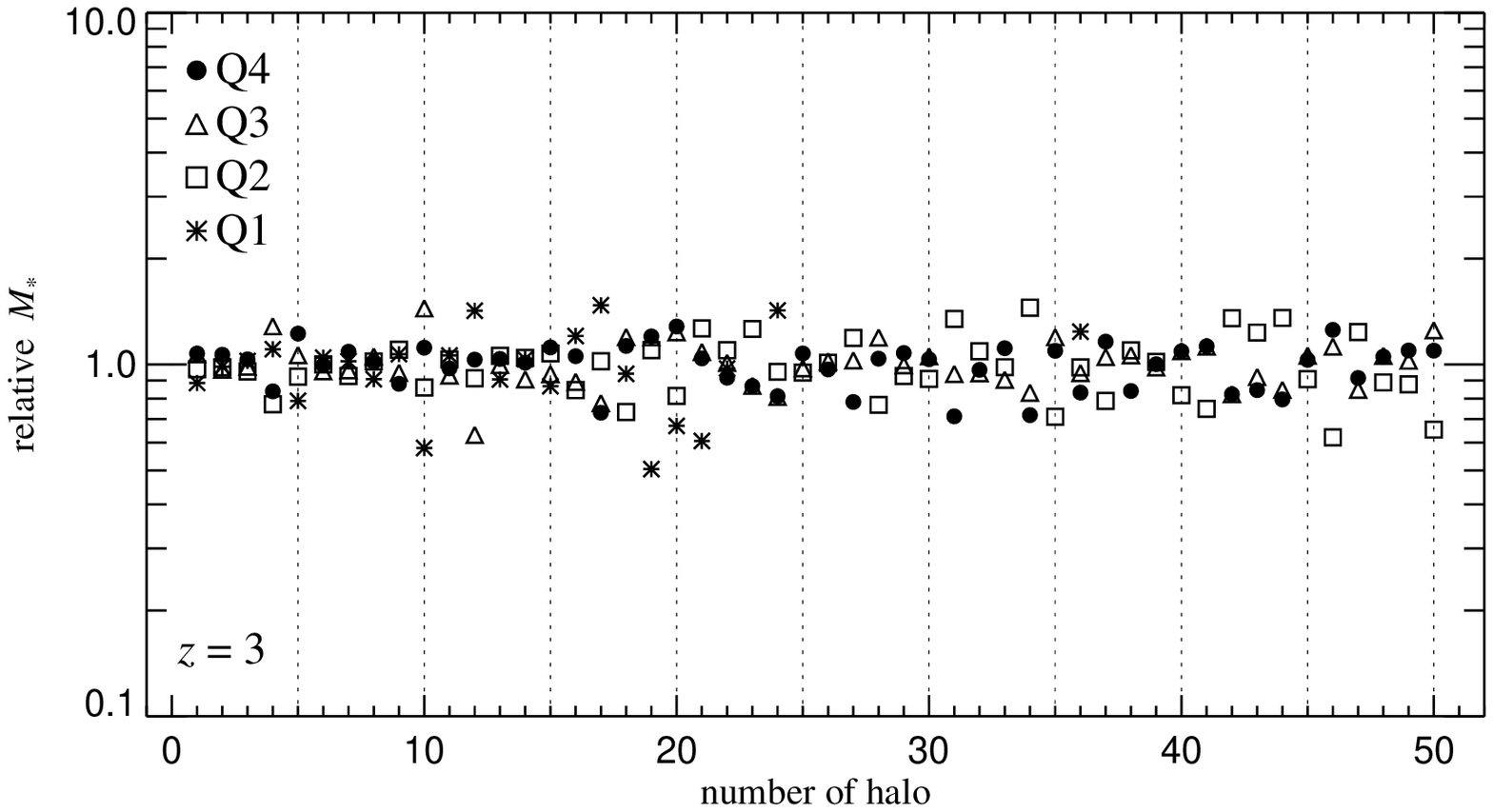}}\\%
\resizebox{!}{6.6cm}{\includegraphics{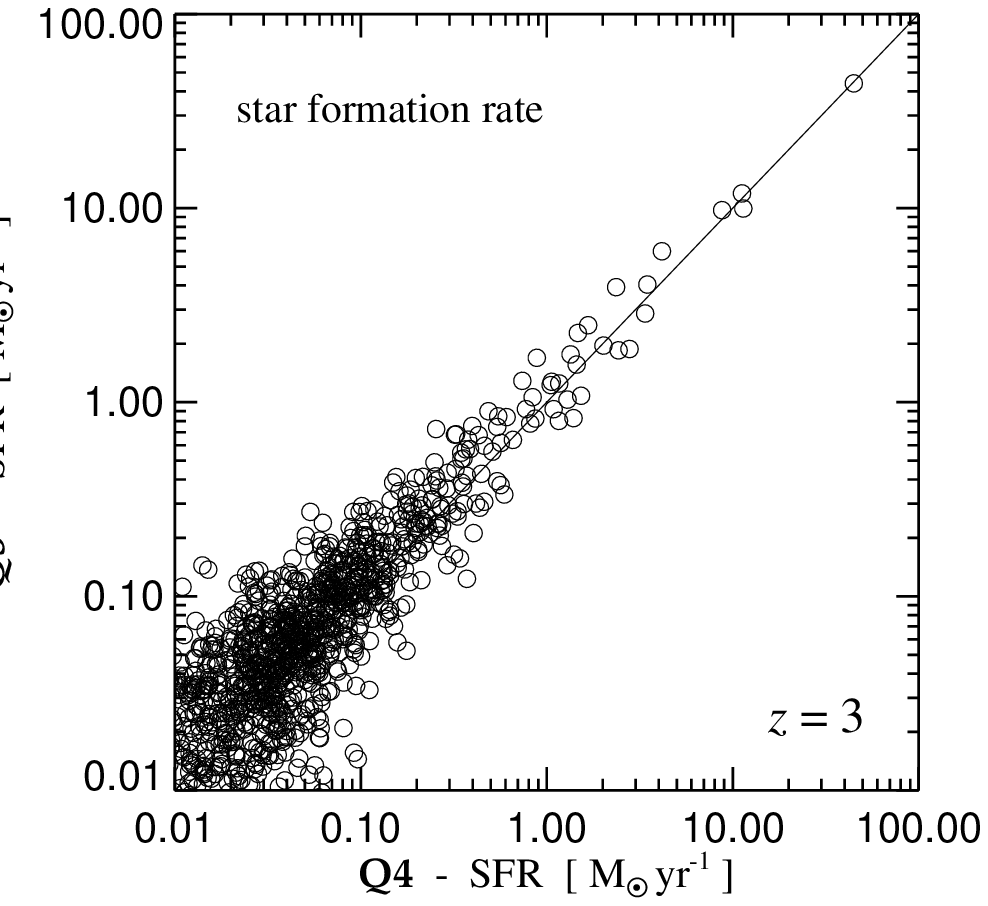}}%
\hspace*{-0.3cm}\resizebox{!}{6.6cm}{\includegraphics{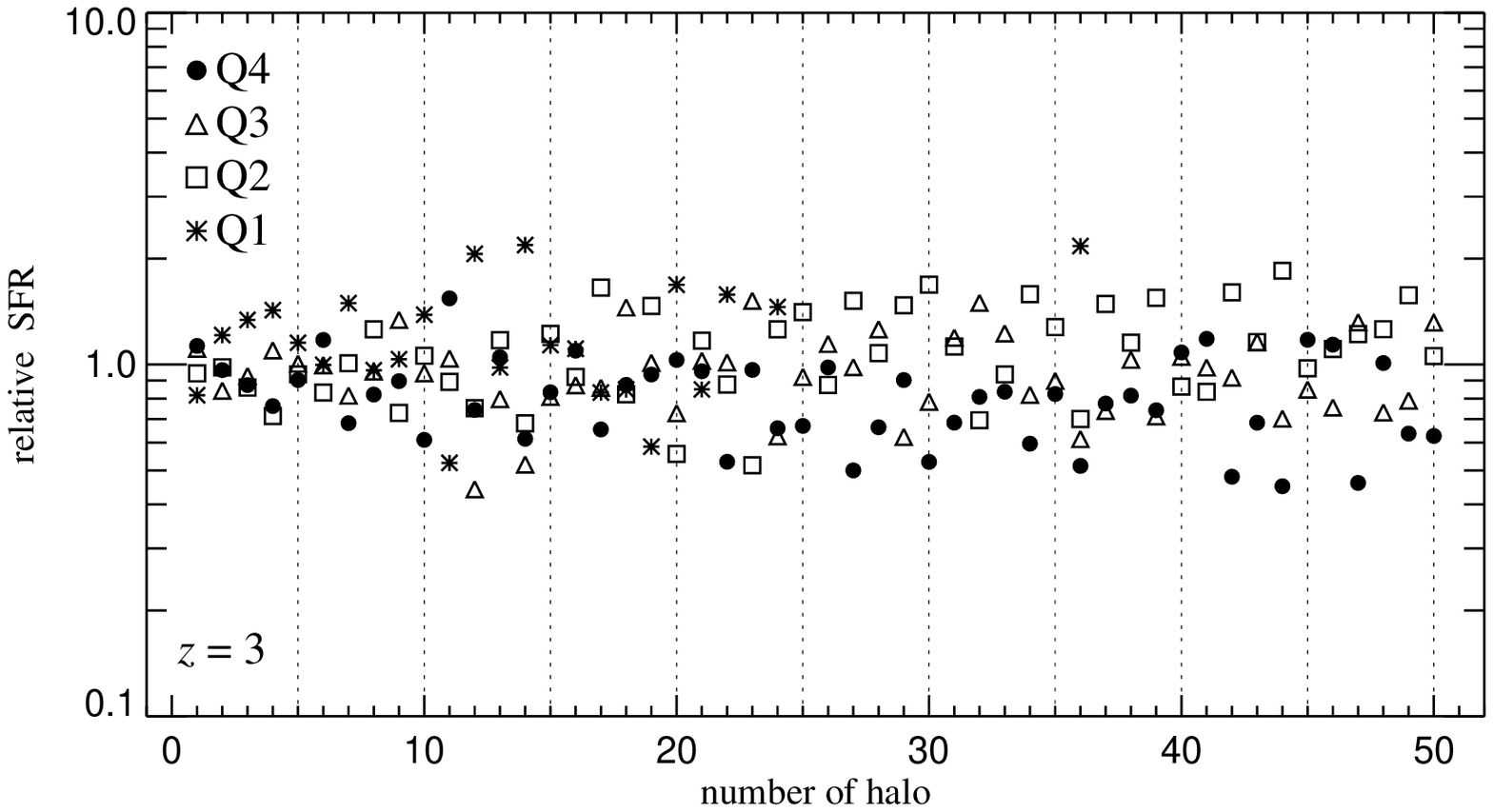}}%
\caption{Object-by-object comparison of halos in the Q-Series at
redshift $z=3$.  We analyse the convergence of the virial mass (top
row), the total mass of stars inside the virial radius (middle row),
and the total star formation rate within the virial radius (bottom
row).  In the panels on the left, we plot a direct comparison of
measurements for the Q3 and Q4 runs for each of these three
quantities. In the panels on the right, we compare results for the 50
most massive halos of all four simulations of the series.  Here, the
horizontal axes give the number of each halo, in decreasing order of
mass. On the vertical axes, we plot for each run the relative
deviation of the measured quantity for each halo from the mean of all
four runs.
\label{figObjCompQ}}
\ec
\end{figure*}

\begin{figure*}
\bc
\resizebox{!}{6.6cm}{\includegraphics{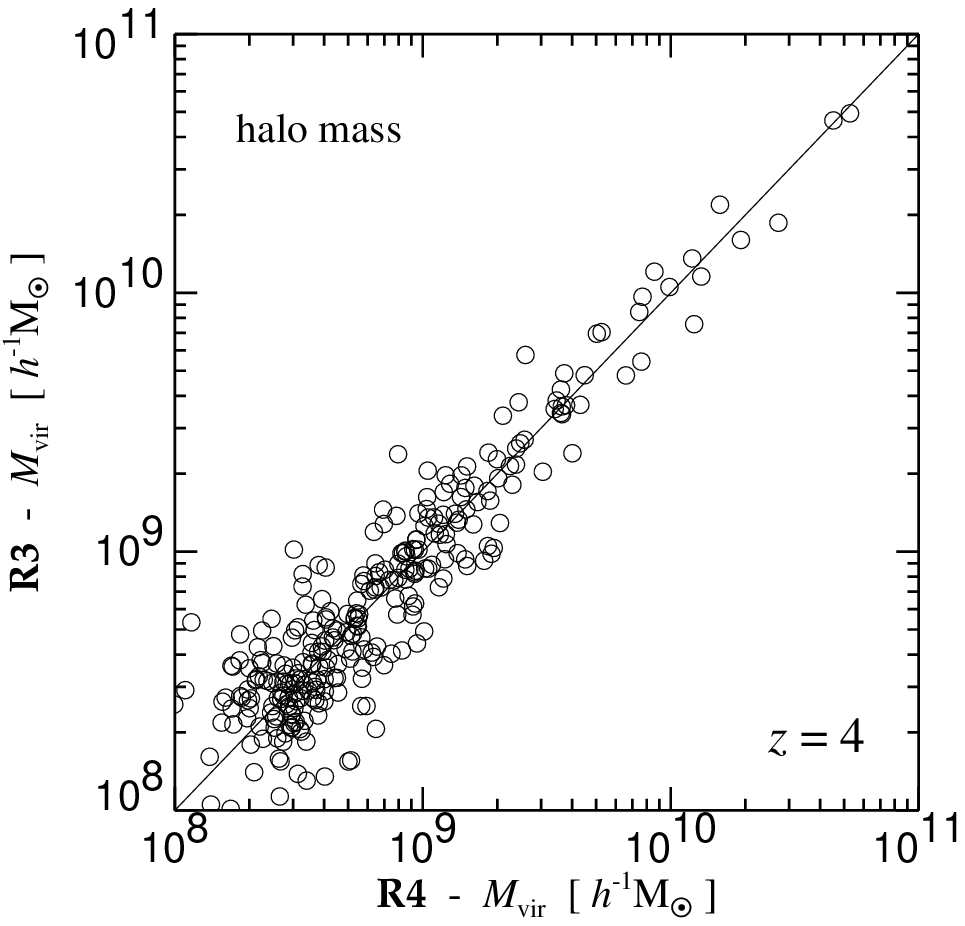}}%
\hspace*{-0.3cm}\resizebox{!}{6.6cm}{\includegraphics{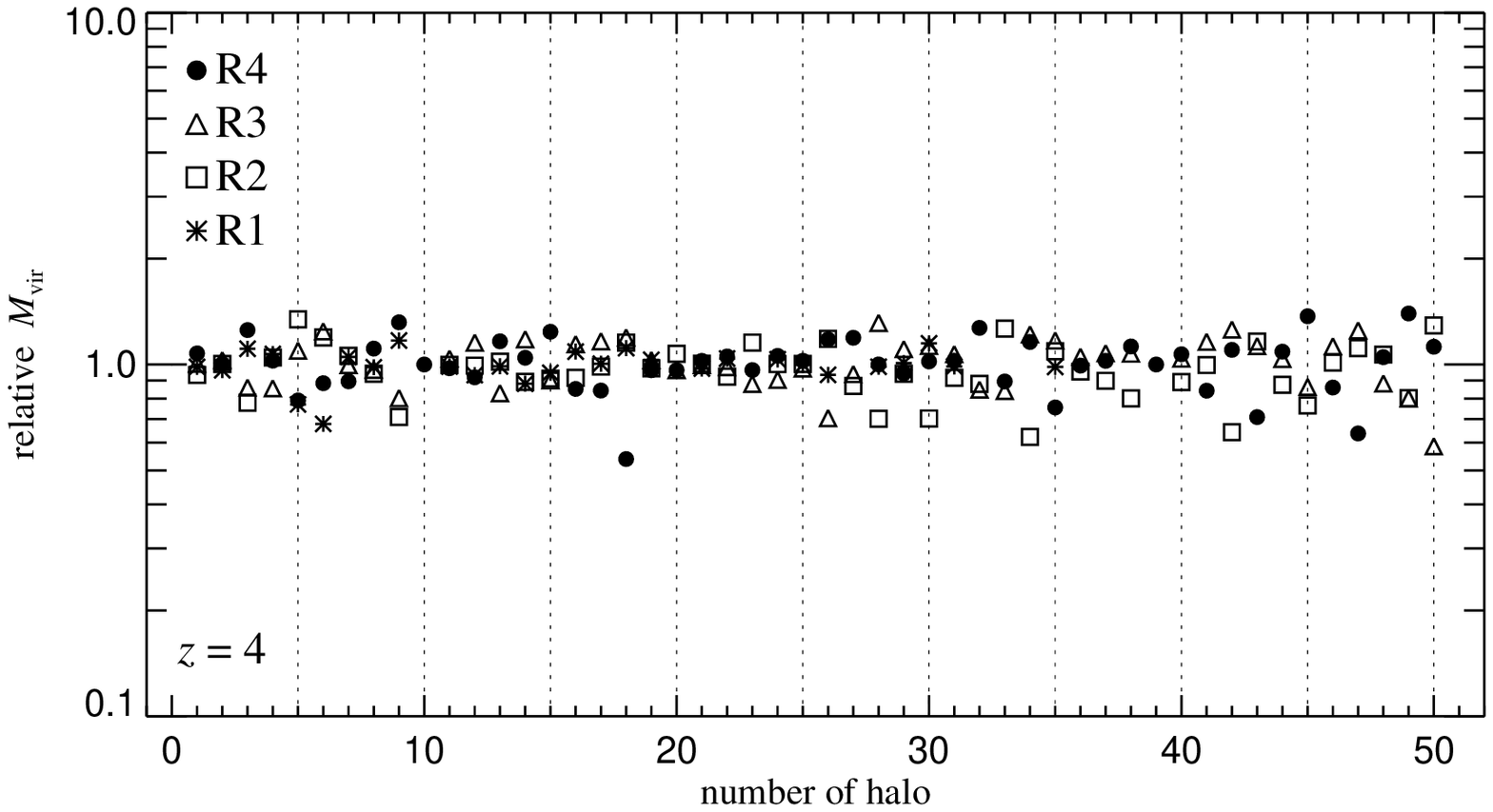}}\\%
\resizebox{!}{6.6cm}{\includegraphics{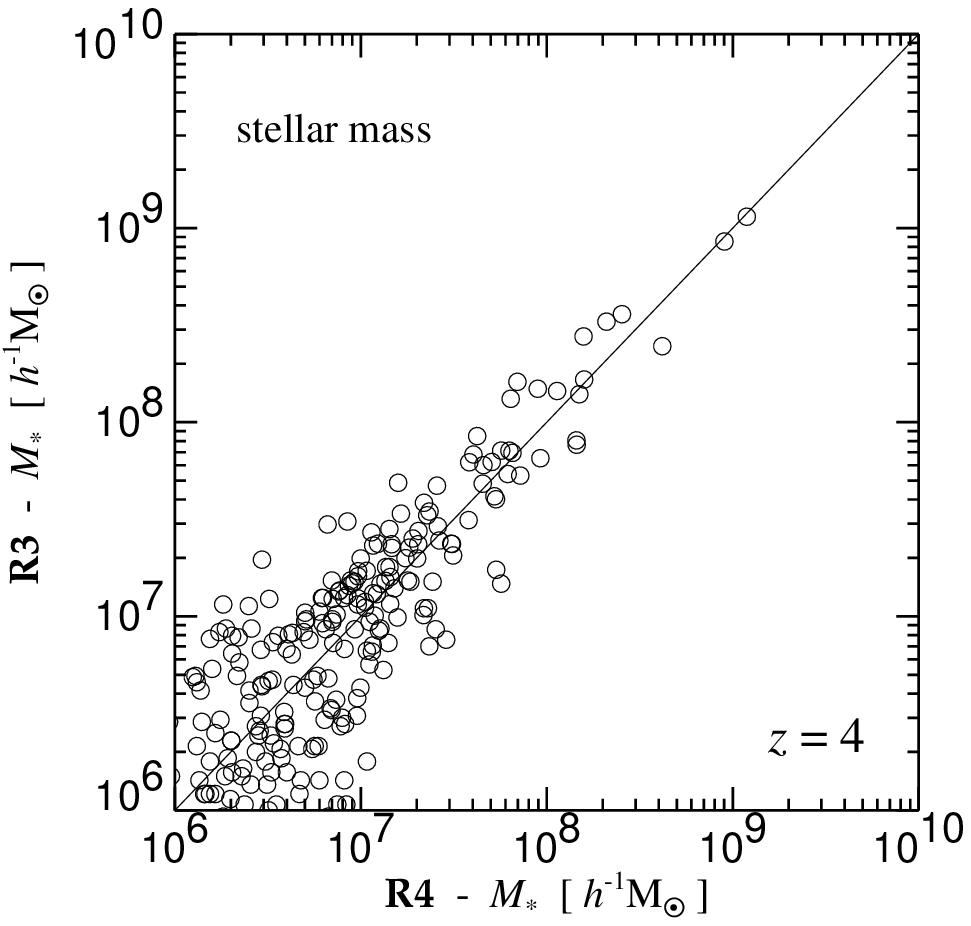}}%
\hspace*{-0.3cm}\resizebox{!}{6.6cm}{\includegraphics{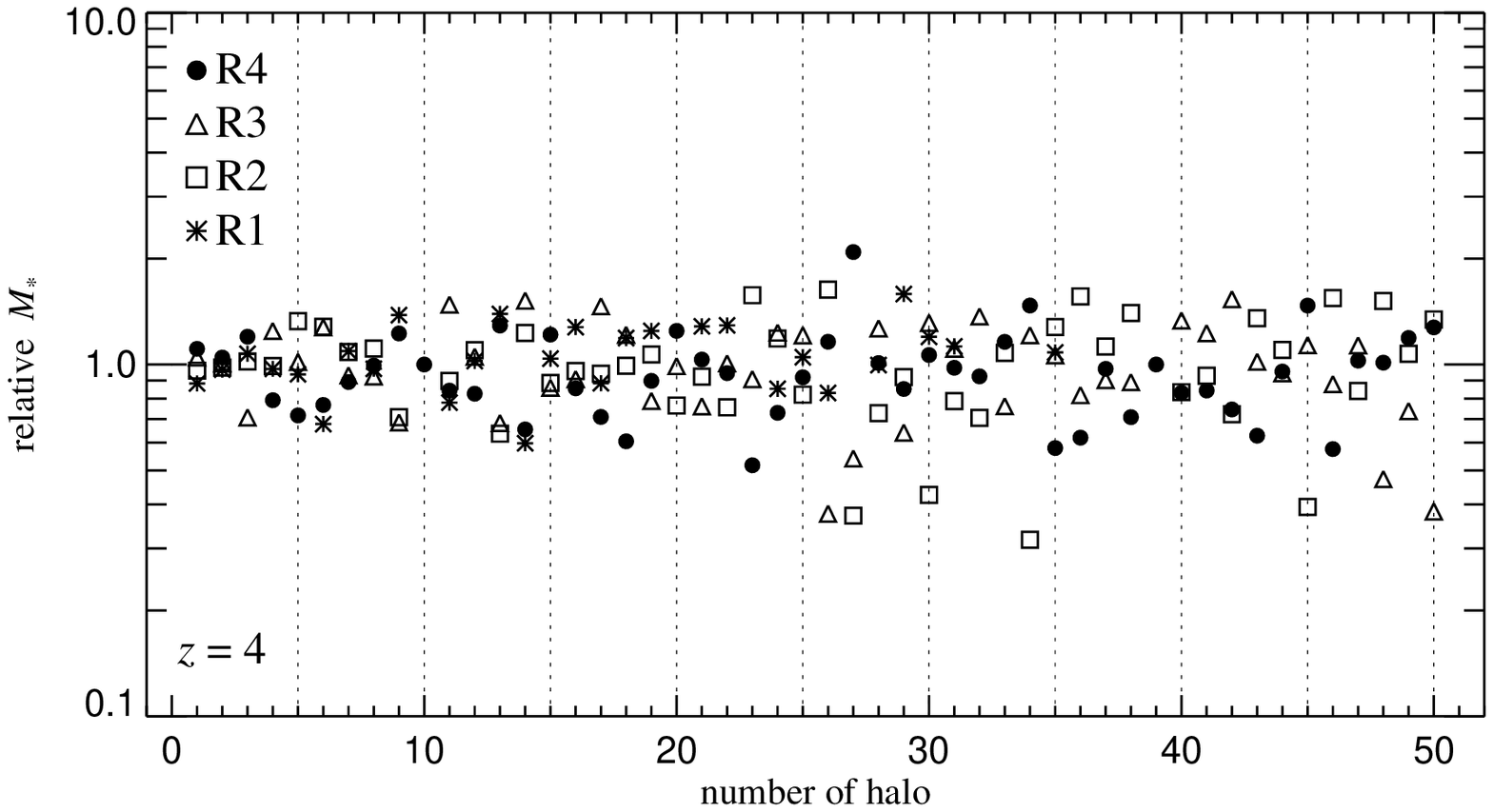}}\\%
\resizebox{!}{6.6cm}{\includegraphics{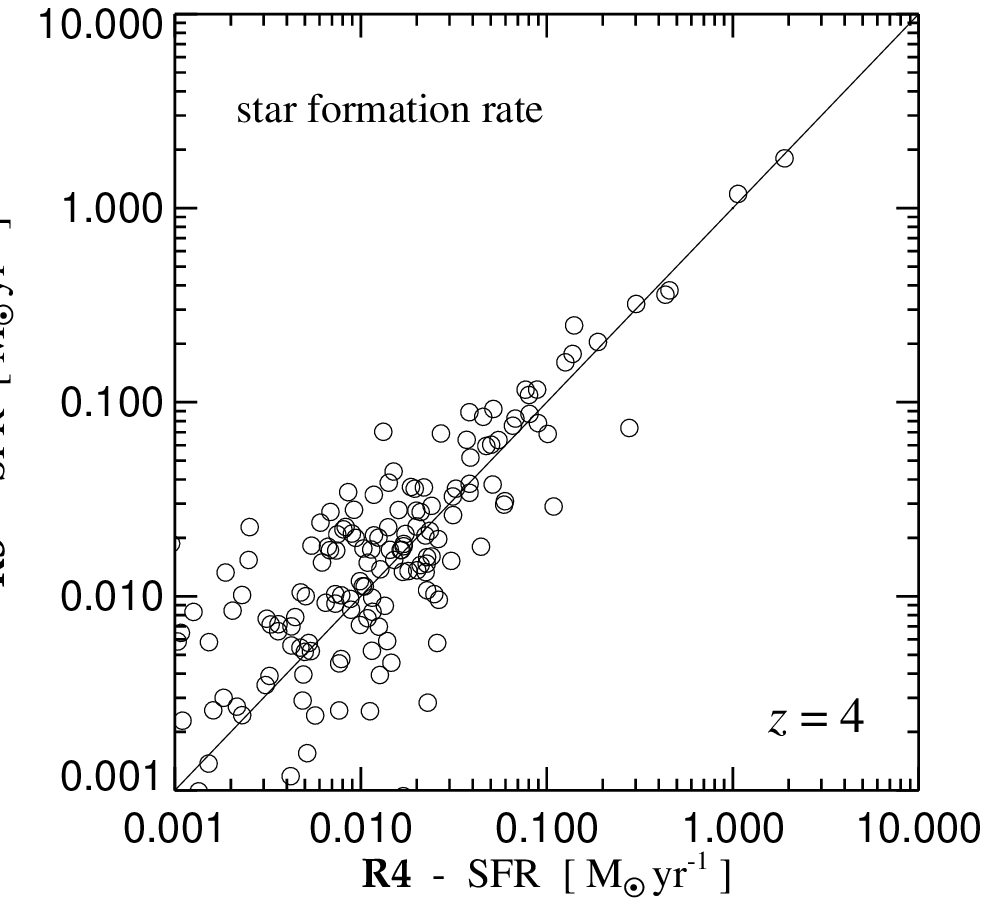}}%
\hspace*{-0.3cm}\resizebox{!}{6.6cm}{\includegraphics{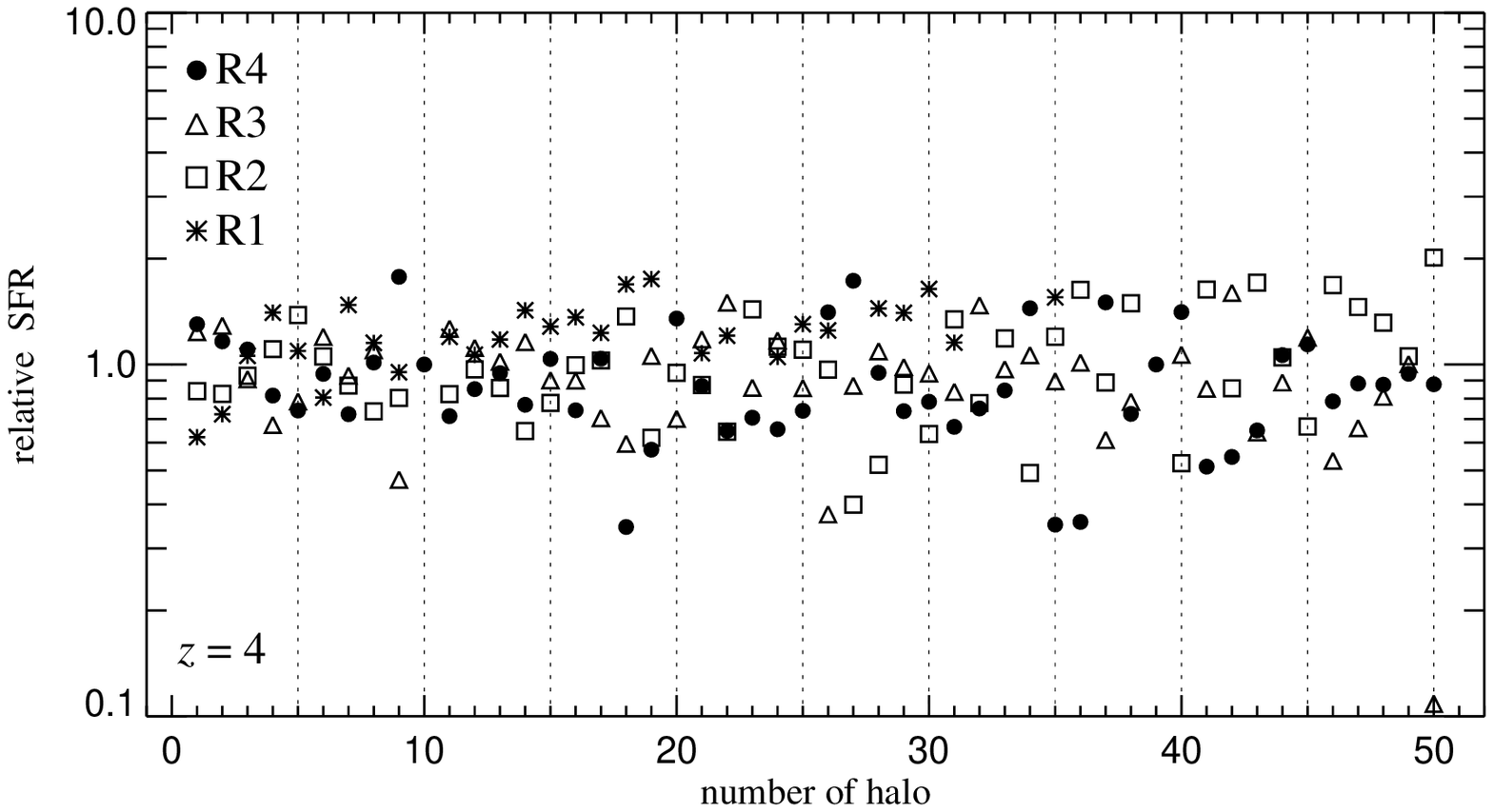}}%
\caption{Object-by-object comparison of halos in the R-series at
redshift $z=4$.  As in Figure~\ref{figObjCompQ}, we analyse the
convergence of the virial mass (top row), the total mass of stars
inside the virial radius (middle row), and the total star formation
rate within the virial radius (bottom row).  In the panels on the
left, we plot for each of these three quantities a direct comparison
of measurements for the R3 and R4 runs. In the panels on the right, we
compare results for the 50 most massive halos of all four simulations
of the series, where the horizontal axes simply give the number of
each halo, in decreasing order of mass. On the vertical axes, we plot
for each run the relative deviation of the measured quantity for each
halo from the mean of all four runs.
\label{figObjCompZ}}
\ec
\end{figure*}

\section{Numerical convergence tests} \label{SecConv}

In this section, we examine the numerical convergence of our
simulation results for star formation rates.  Figure~\ref{figSFR_ZRQ}
shows the evolution of the cosmic star formation rate density in the
simulations making up the Z-, R-, and Q-Series.  In all three cases,
the higher resolution simulations predict substantially more star
formation at high redshift, but at sufficiently low redshift, the
results of the different runs begin to agree quite well.

This behaviour is not surprising.  It is clear that simulations with
higher mass resolution are able to measure star formation rates in
halos of mass smaller than previously resolved.  Furthermore, towards
higher redshift, the bulk of the star formation shifts to
progressively lower mass scales, as expected for hierarchical growth
of structure, making it ever more important to resolve low mass
objects.

While the convergence between the simulations within a given series at
low redshift is encouraging, the disagreement at high redshift implies
that none of our simulations {\em alone} can be used to reconstruct
the star formation history from low to very high redshift.
Fortunately, however, the mass resolution needed to achieve
convergence at a given redshift is a function of epoch.  It becomes
increasingly less demanding to do this at lower redshift because at
that point increasingly more massive halos dominate the star formation
rate.  This opens up the possibility of combining simulations on
different scales to obtain a composite result for the full history of
star formation.  This is the approach we use in the present analysis.

However, this strategy can succeed only if we can demonstrate that
reliably converged star formation rates can be obtained for individual
objects. In Figure~\ref{figObjCompQ}, we compare halos in the
simulations of the Q-series directly with one another, and in
Figure~\ref{figObjCompZ} we perform the same check for the R-series.
In particular, we compare the virial mass, the total mass of stars
inside the virial radius, and the total star formation rate within the
virial radius, where the properties of halos have been defined using
the spherical overdensity algorithm.  For each of these three
quantities, we show scatter plots that directly compare the $2\times
144^3$ and $2\times 216^3$ resolutions of the series.  In addition, we
consider the 50 most massive halos explicitly, and compare all four
simulations of the series with one other in the relevant panels.

From this comparison it is clear that for the most massive systems the
agreement between the simulations is quite good, without any obvious
bias towards lower or higher values of the measured quantities as a
consequence of resolution effects. Overall, we consider the comparison
to be an important success.  It shows that we have indeed formulated a
numerically well controlled model for cooling, star formation and
strong feedback by galactic outflows.  We are not aware that other
simulations of cosmological galaxy formation have been demonstrated to
pass similar tests with equally good results.

Only close to the resolution limit do we notice a tendency for lower
resolution simulations to overpredict the star formation rate.  This
effect will show up again in the analysis of Section~\ref{SecMulti}.
It can be understood as a consequence of the lack of star formation
and feedback in unresolved progenitor halos.  As a result, a newly
formed halo in a low-resolution run will initially have a gas fraction
that is essentially equal to the universal baryon fraction, while in a
corresponding higher-resolution simulation the gas fraction for the
same halo can be lower than this because its progenitors were already
able to form stars and lose some of their baryons in galactic
outflows.

In summary, the results presented in this section give us confidence
that we can obtain meaningful, numerically well-converged results for
the star formation rate in a given object at a given epoch. The reason
we cannot easily obtain a similarly well-converged result for the
cosmic star formation density with a single simulation lies in our
failure to obtain a complete sampling of the cosmological mass
function together with sufficient mass resolution in one simulation
box.  A way to circumvent this difficulty will be discussed in the
next section.

\section{The multiplicity function of cosmic star formation}  
\label{SecMulti}

It is clear that the evolution of the cosmic star formation rate
density, $\dot \rho_\star(z)$, is of fundamental importance to
cosmology.  At a given epoch, the function $\dot \rho_\star(z)$ will
take on a single value, and, hence, does not describe which objects
are responsible for the bulk of star formation occurring at that
redshift.  To address this issue, we develop the notion of a
distribution function of star formation with respect to halo mass; a
quantity we will refer to as the ``multiplicity function of cosmic star
formation''.  In defining this property of star forming objects, we
implicitly assume that the mass in the Universe can be meaningfully
partitioned among gravitationally bound halos, and that star formation
is restricted to these halos.  The problem thus naturally decomposes
into determining the number distribution of halos as a function of
mass, the so-called mass function, and then measuring the average star
formation rate in halos of a given mass.

We start by briefly reviewing the concept of a cosmological mass
function.  Let $F(M,z)$ denote the fraction of mass that is bound at
epoch $z$ in halos of mass smaller than $M$.  It follows that the
comoving differential mass function of halos is given by \be n(M,z) =
\frac{\overline{\rho}}{M}\,\frac{{\rm d}F}{{\rm d}{M}}, \ee where
$\overline{\rho}$ is the comoving background density, and
$n(M,z)\,{\rm d}M$ gives the number density of halos in a mass
interval ${\rm d}M$ around $M$.

\begin{figure*}
\bc
\resizebox{8.5cm}{!}{\includegraphics{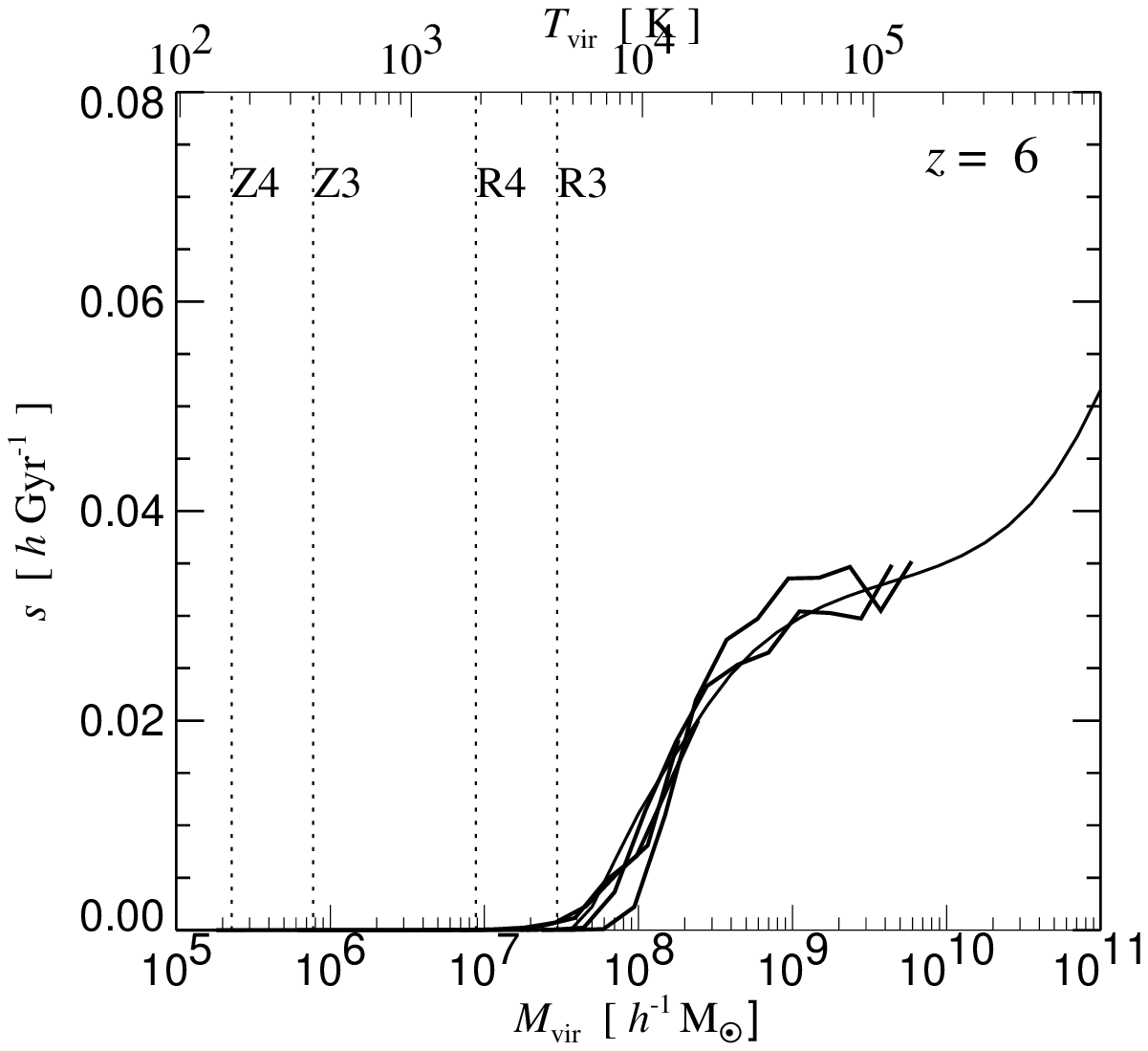}}%
\resizebox{8.5cm}{!}{\includegraphics{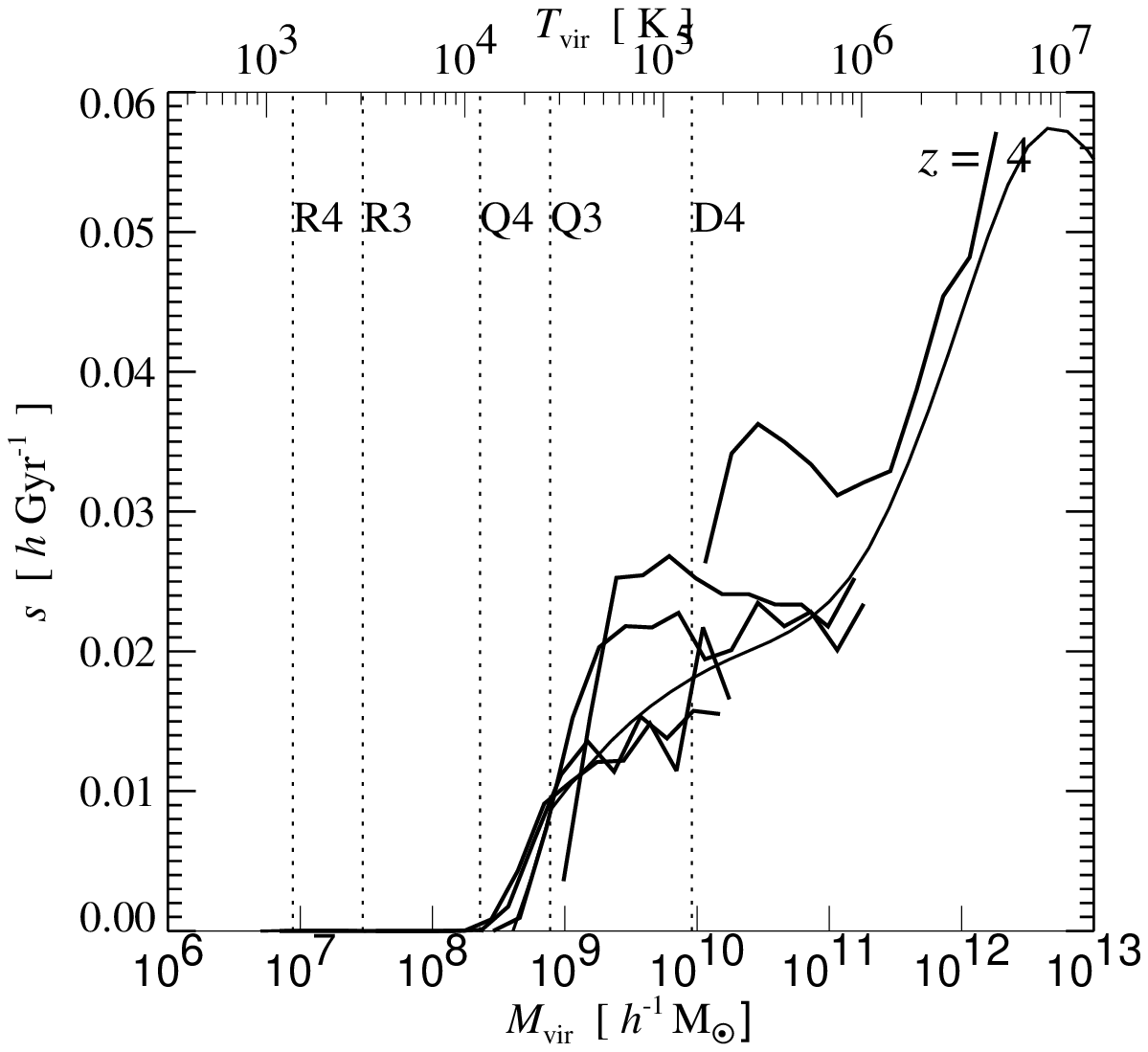}}\\%
\resizebox{8.5cm}{!}{\includegraphics{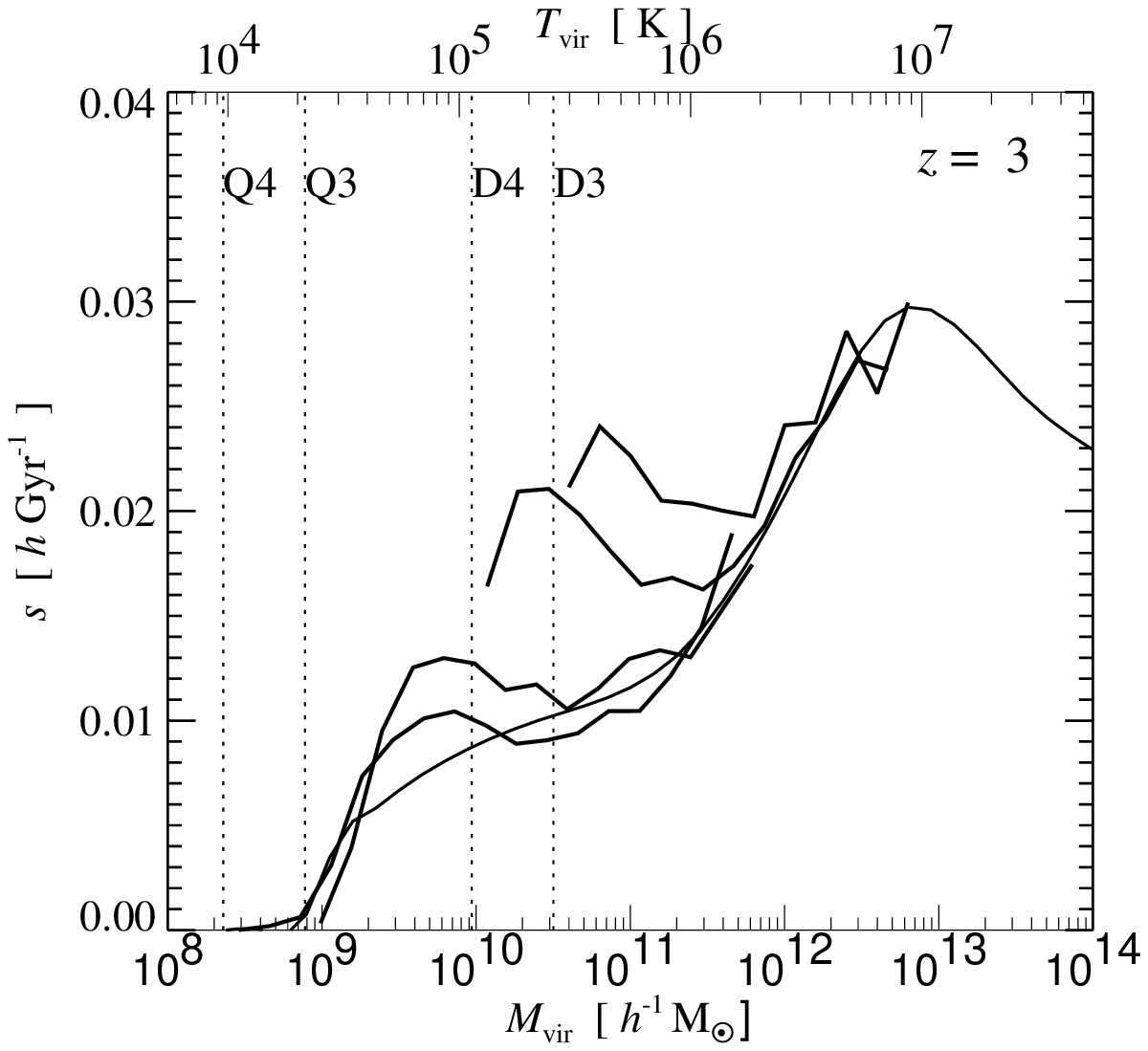}}%
\resizebox{8.5cm}{!}{\includegraphics{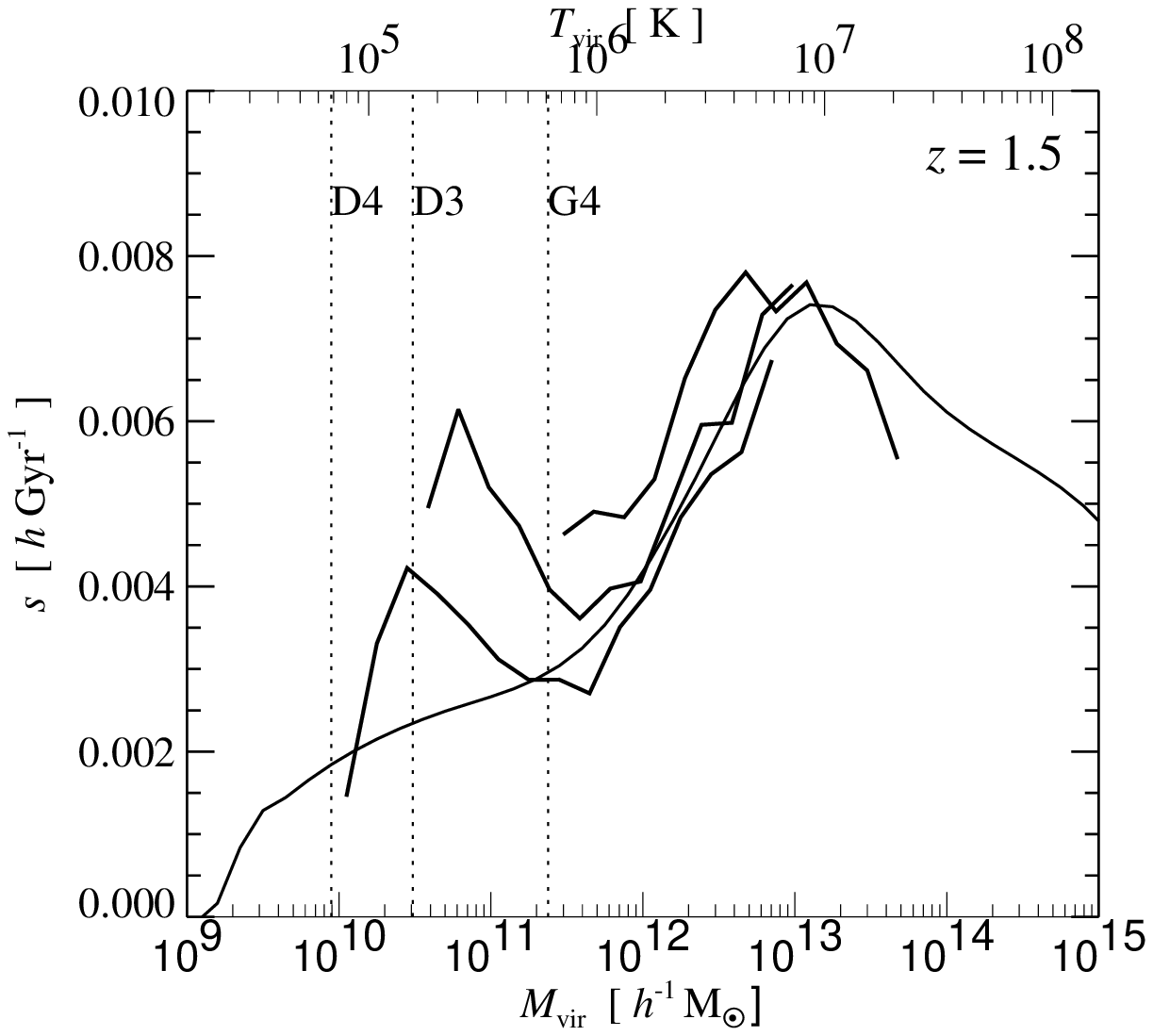}}\\%
\caption{Mean star formation rate $s(M,z)=\left<\dot M_\star\right>/M$ in
halos, normalised to the halo mass. Each panel shows results for a
different redshift.  Halos have been binned into logarithmic intervals
of width ${\rm dlog} M=0.2$, with the thick lines showing the mean in
each bin for several different simulations. The thin line is the
spline fit for our best estimate of $s(M,z)$.  Vertical lines in each
panel mark the mass resolution limits (32 particles) of the FOF
catalogues of the simulations shown.
\label{figSfrVsMass}}
\ec
\end{figure*}

\citet{Pre74} proposed that $F(M,z)$ can be approximated by \be F(M,z)
= {\rm erf}\left[\frac{\delta_c}{\sqrt{2}\,\sigma (M,z)}\right]
,\ee where $\delta_c= 1.686$ is the (linearly extrapolated) critical
overdensity for top-hat collapse, and $\sigma (M,z)$ describes the rms
fluctuations of the linearly evolved density field at redshift $z$,
smoothed by a top-hat sphere that just encloses a background mass
equal to $M$.  The variance of the filtered density field is given by
\be \sigma^2(M,z) =\int P_z(k)\, |W_{R}(k)|^2 \,{\rm d}^3k, \ee where
$W_R(k)=3j_1(kR)/(kR)$, $R= [3M/(4\pi\overline{\rho})]^{1/3}$, and
$P_z(k)$ denotes the power spectrum linearly evolved to redshift $z$.

Overall, the Press-Schechter (PS) mass function provides a
surprisingly good description of numerical N-body simulations of
non-linear gravitational growth of structure.  However, recent high
precision work has also clearly established that there are
non-negligible discrepancies between PS theory and simulations.  The
PS formula overpredicts the abundance of halos at intermediate
mass-scales, and underpredicts the abundance of both low and high mass
halos.  This deficiency has led to the development of more accurate
fitting formulae.

In particular, \citet{ShTo99} have found a significantly improved
parameterisation, based on an excursion set formalism.  The
\citet{ShTo99} mass function can be expressed as \be n(M,z)=
\frac{\overline{\rho}}{M^2}\,\frac{{\rm d}\ln\sigma^{-1}}{{\rm d}\ln
M\;\,\,}\, f(\sigma), \ee where \be f(\sigma) = A\sqrt{\frac{2a}{\pi}}
\left[ 1+ \left(\frac{\sigma^2}{a\delta_c^2}\right)^p\right]
\frac{\delta_c}{\sigma}
\exp\left(-\frac{a\delta_c^2}{2\sigma^2}\right), \ee with $A=0.3222$,
$a=0.707$, and $p=0.3$. \citet{SMT01} and \citet{ShTo02} also showed
that this modified mass function can be understood in terms of an extension
of the excursion set formalism that allows for ellipsoidal collapse.

In a comprehensive analysis of a number of N-body simulations,
\citet{Jen01} showed that the Sheth \& Tormen mass function works
remarkably well over at least four decades in mass.  In addition, the
quality of the fit is nearly independent of cosmology, power spectrum,
and epoch, provided that $\delta_c=1.686$ is taken in all cosmologies
and the mass function is measured at fixed comoving overdensity,
independent of $\Omega(z)$. If instead the expected cosmological
scaling of the virial overdensity for top-hat collapse is used in the
construction of the halo catalogue itself, the universality of the
mass function is slightly broken. It is clearly preferable to have a
universal mass function for all cosmologies; hence, we have identified
our halos at fixed comoving overdensity.

The mass function may also be expressed in terms of the multiplicity
function $g(M)$ of halos, which we define as \be g(M)=\frac{{\rm
d}F}{{\rm d}\log M}.  \ee This quantity gives the fraction of mass
that is bound in halos per unit logarithmic interval in mass. In a
plot of $g(M)$ versus $\log M$, the area under the curve then directly
corresponds to the fraction of mass bound in the respective mass
range.  For the Sheth \& Tormen mass function, the multiplicity
function is given by \be g(M)= f(\sigma)\, \frac{{\rm
d}\ln\sigma^{-1}}{{\rm d}\log M\,}. \ee

We now introduce a similar quantity for the cosmic star formation rate
density. Let us define \be s(M,z)=\frac{\left< \dot M_\star
\right>}{M} \ee to be the mean star formation rate of halos of mass
$M$ at epoch $z$, normalised to the masses of the halos.  With this
normalisation, $s(M,z)$ makes possible a direct comparison of the mean
efficiency of star formation between halos of different mass.  Much of
the effort of the numerical work of this paper is focused on trying to
measure this function from our set of simulations over a broad range
of mass scales and redshifts.

Once $s(M,z)$ is known, the cosmic star formation rate density can be
computed from \be \dot \rho_\star(z) = \overline{\rho} \int s(M,z)\,
g(M,z)\, {\rm d}\log M . \label{eqm1} \ee We call the function \be
S(M,z) = s(M,z)\, g(M,z)\label{eqm2} \ee the multiplicity function of
star formation, in analogy to the multiplicity function of the halo
mass distribution.

There are a number of advantages to factoring the star formation rate
in the manner suggested by equations (\ref{eqm1}) and (\ref{eqm2}).
First, a plot of $S(M)$ versus $\log M$ cleanly illustrates which halo
mass scales are responsible for the bulk of star formation occurring
at a given epoch.  Second, even if a numerical simulation can provide
an accurate measurement of $s(M,z)$ over an extended range of mass
scales, it is difficult if not impossible for any simulation to
accurately determine the mass function $g(M,z)$ at the same time. This
is because accurate measurements of $s(M,z)$ primarily require high
mass resolution, which typically necessitates the use of rather small
simulation volumes, resulting in a poor representation of the
cosmological mass function.  In particular, at the high-mass end, one
will typically suffer from completeness problems.  However, the
analytic form of the mass function can then be used to compute an
estimate of the expected star formation density in a simulation in
which the mass function was sampled precisely. Provided that $s(M,z)$
varies sufficiently slowly with $M$, it may be also be possible to
extrapolate $s(M,z)$ into regions not well probed by the simulation,
although this procedure clearly requires some caution.

\begin{figure}
\bc
\resizebox{8.0cm}{!}{\includegraphics{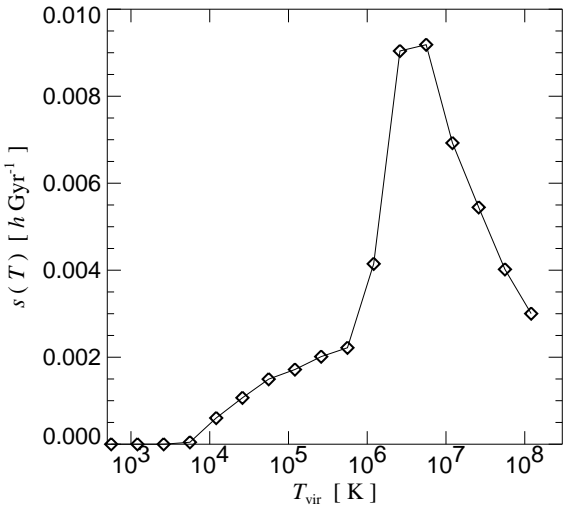}}%
\caption{Normalised star formation rate $s=\left<\dot
M_\star\right>/M$ in a series of small test simulations of isolated
NFW halos at $z=0$.  The halos have been set-up in virial equilibrium
with masses between $10^7$ and $10^{15}\,h^{-1}{\rm M}_\odot$, all
with a self-similar structure initially.  We have then evolved them
with the same model for cooling, star formation, and feedback as in
the cosmological simulations, and measured their star formation rates
at a time $2.5\,{\rm Gyr}$ after the simulations were started.
Despite the simplicity of these simulations, a qualitatively similar
behaviour as for the function $\tilde s_z(T)$ measured from the full
cosmological simulations is observed.  Note in particular the
suppression of star formation in halos with virial temperatures below
$10^6\,{\rm K}$ (where galactic winds can escape), the complete
absence of star formation for virial temperatures below $10^4\,{\rm
K}$ (due to the absence of radiative cooling mechanisms), and the peak
at a temperature $\sim 10^7\,{\rm K}$ (due to the decline of the
efficiency of cooling flows at still higher temperatures).
\label{figEffIsolatedHalo}}
\ec
\end{figure}

\begin{figure*}
\bc
\resizebox{5.8cm}{!}{\includegraphics{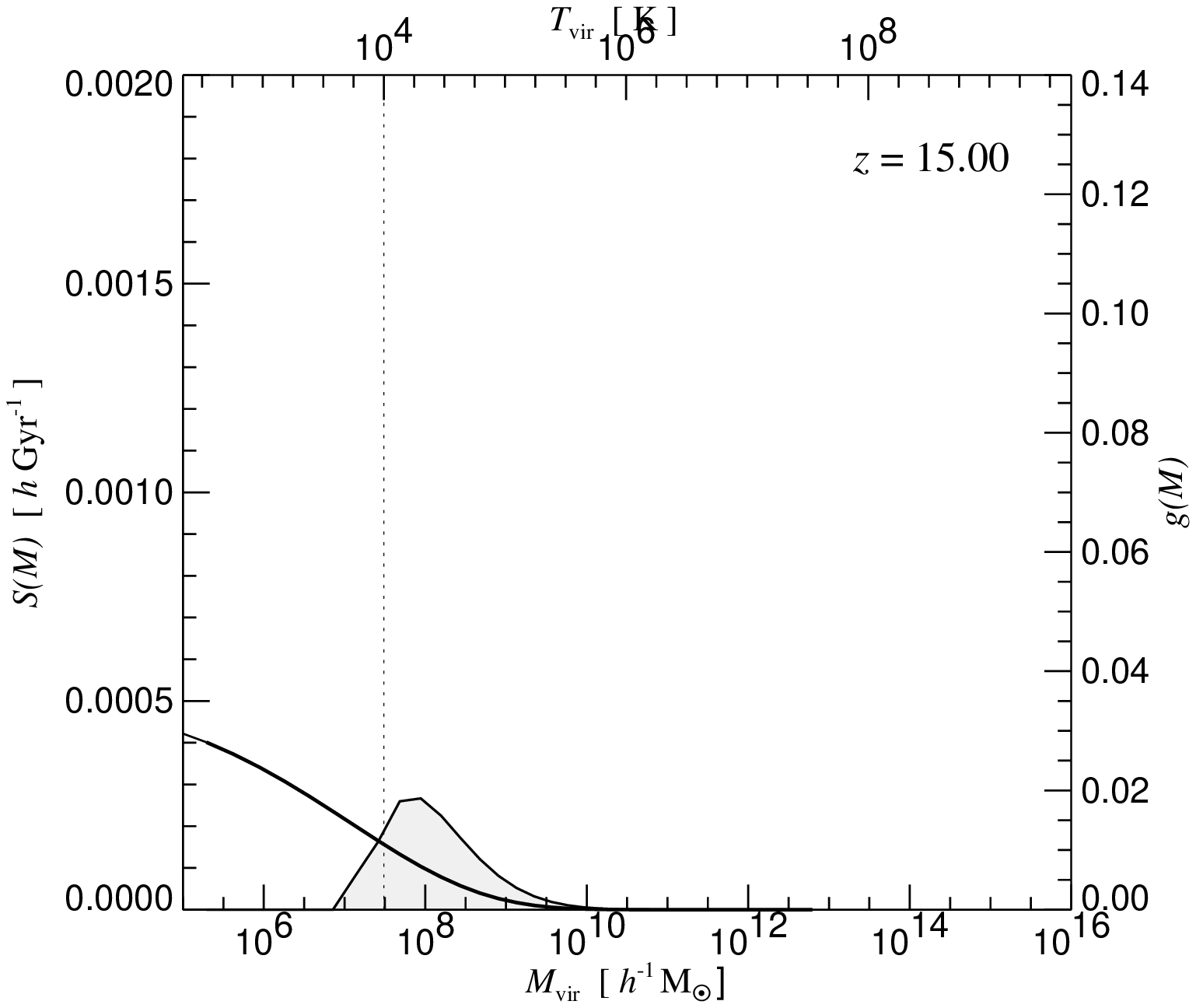}}%
\resizebox{5.8cm}{!}{\includegraphics{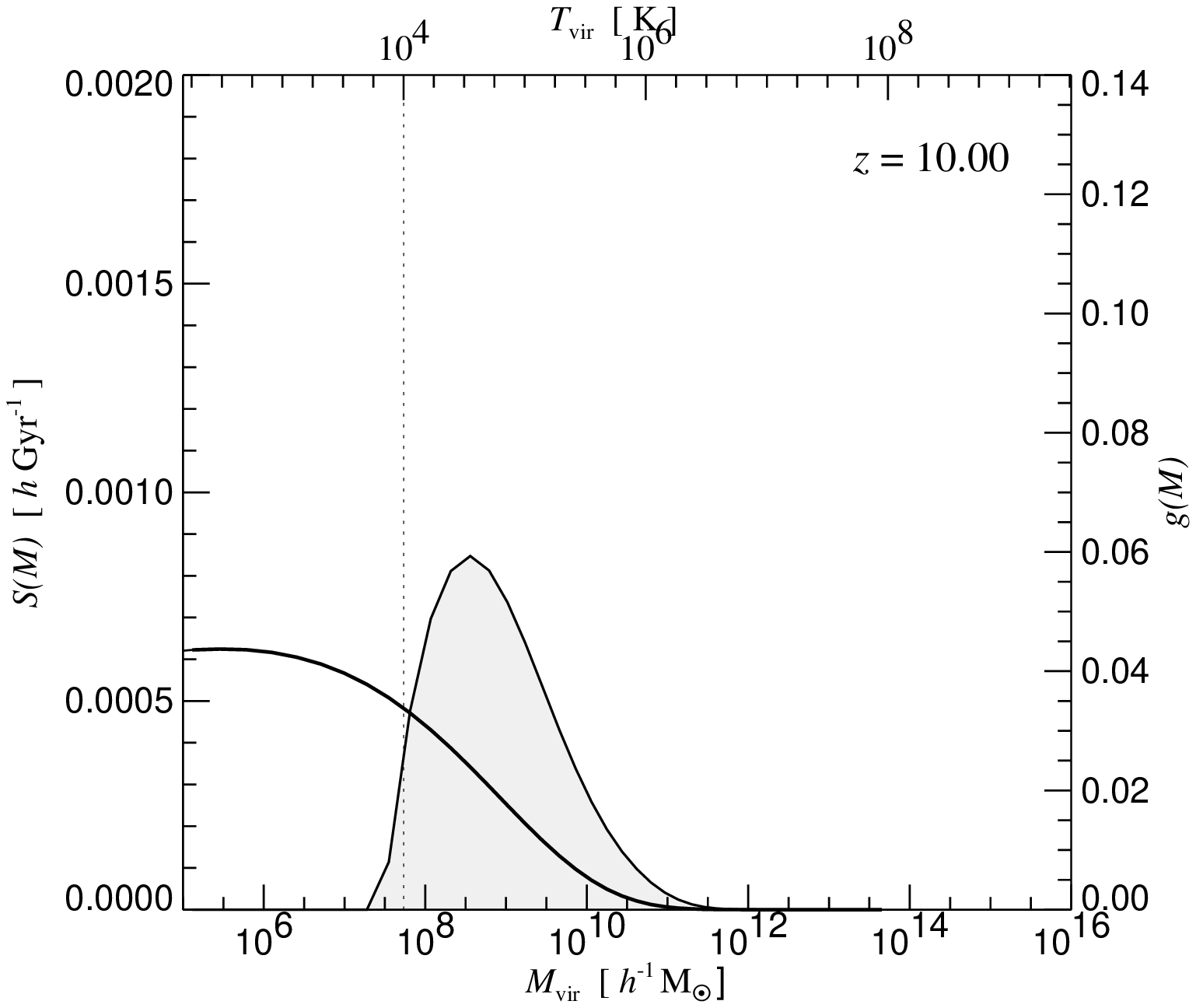}}%
\resizebox{5.8cm}{!}{\includegraphics{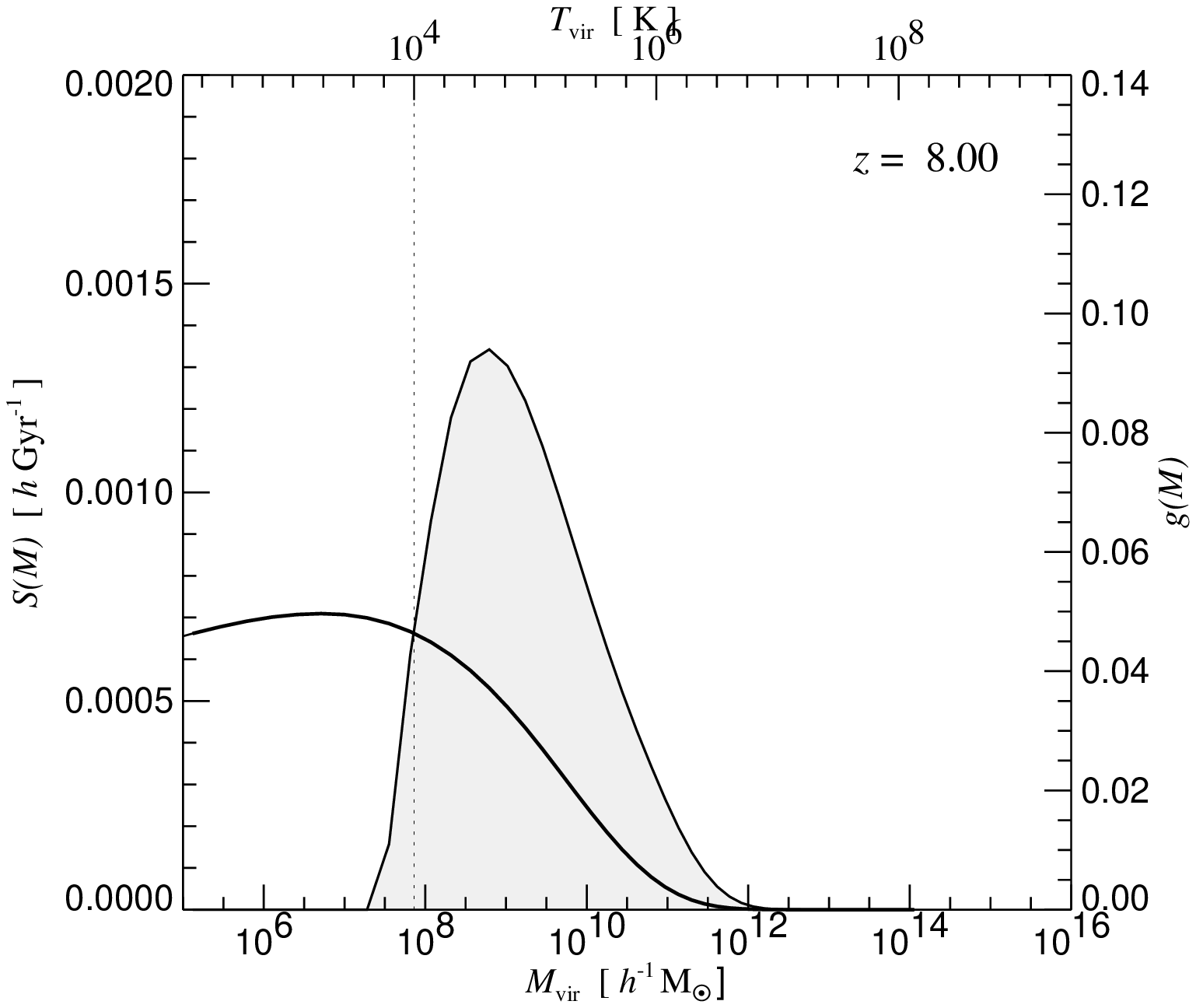}}\\%
\resizebox{5.8cm}{!}{\includegraphics{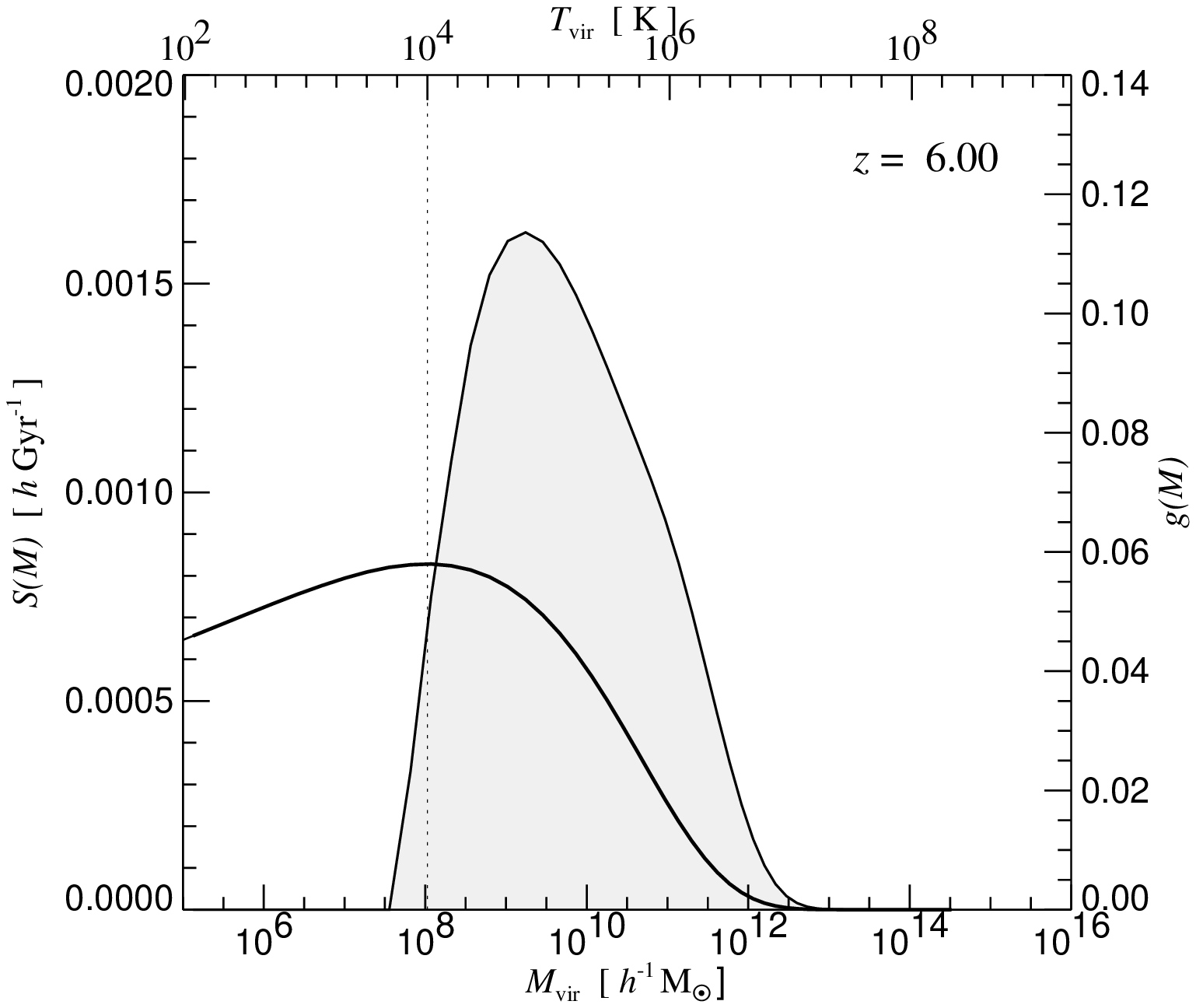}}%
\resizebox{5.8cm}{!}{\includegraphics{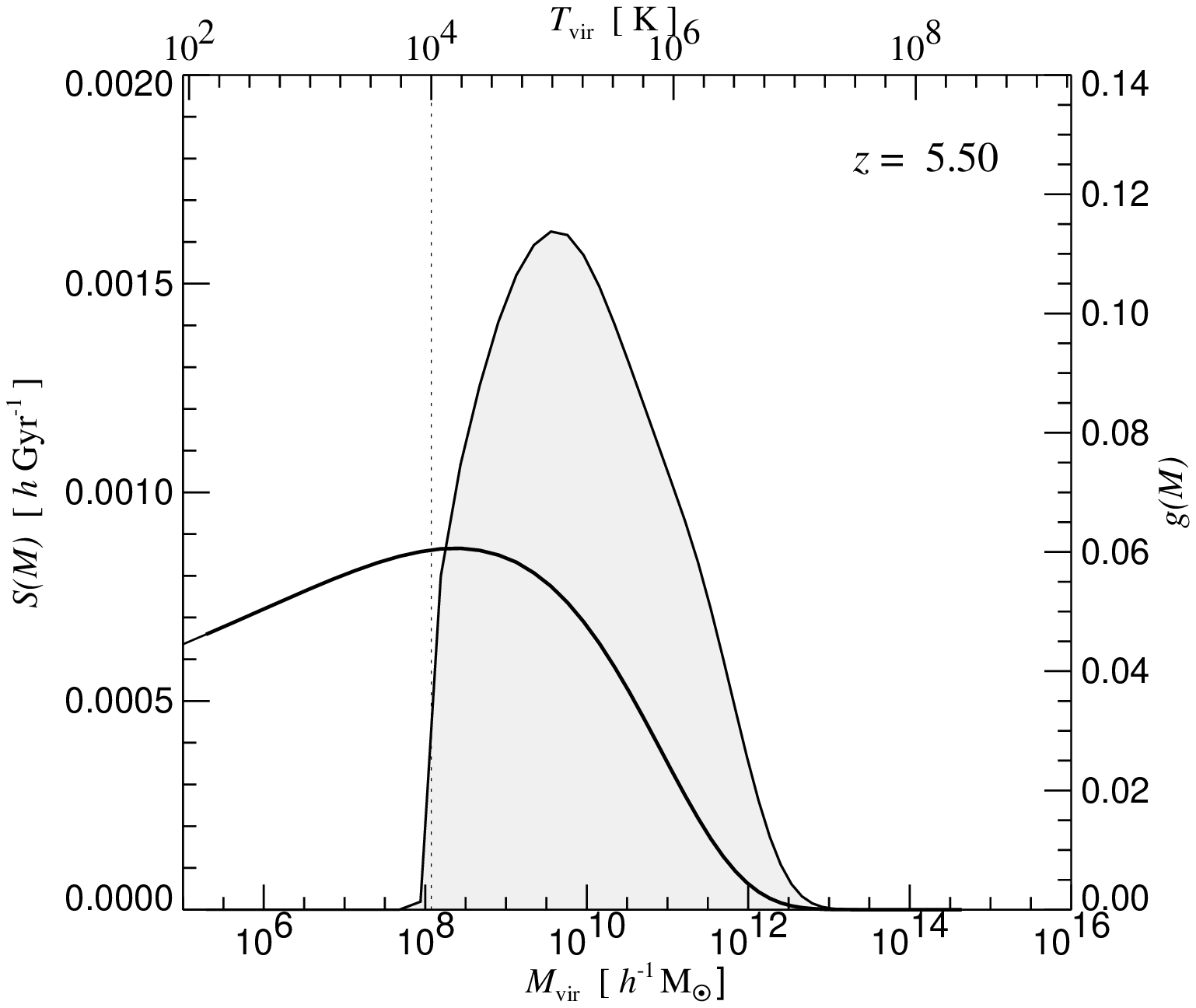}}%
\resizebox{5.8cm}{!}{\includegraphics{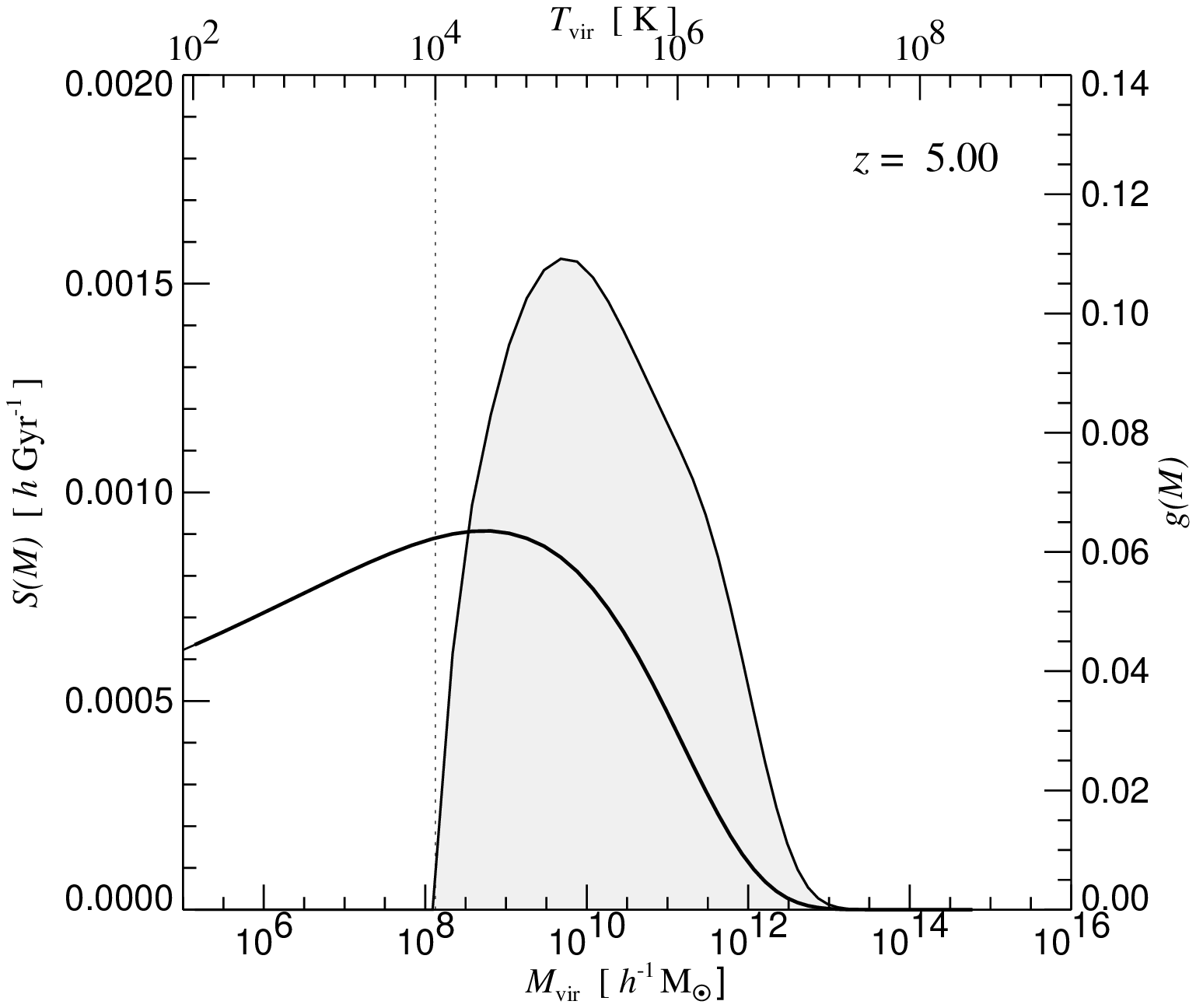}}\\%
\resizebox{5.8cm}{!}{\includegraphics{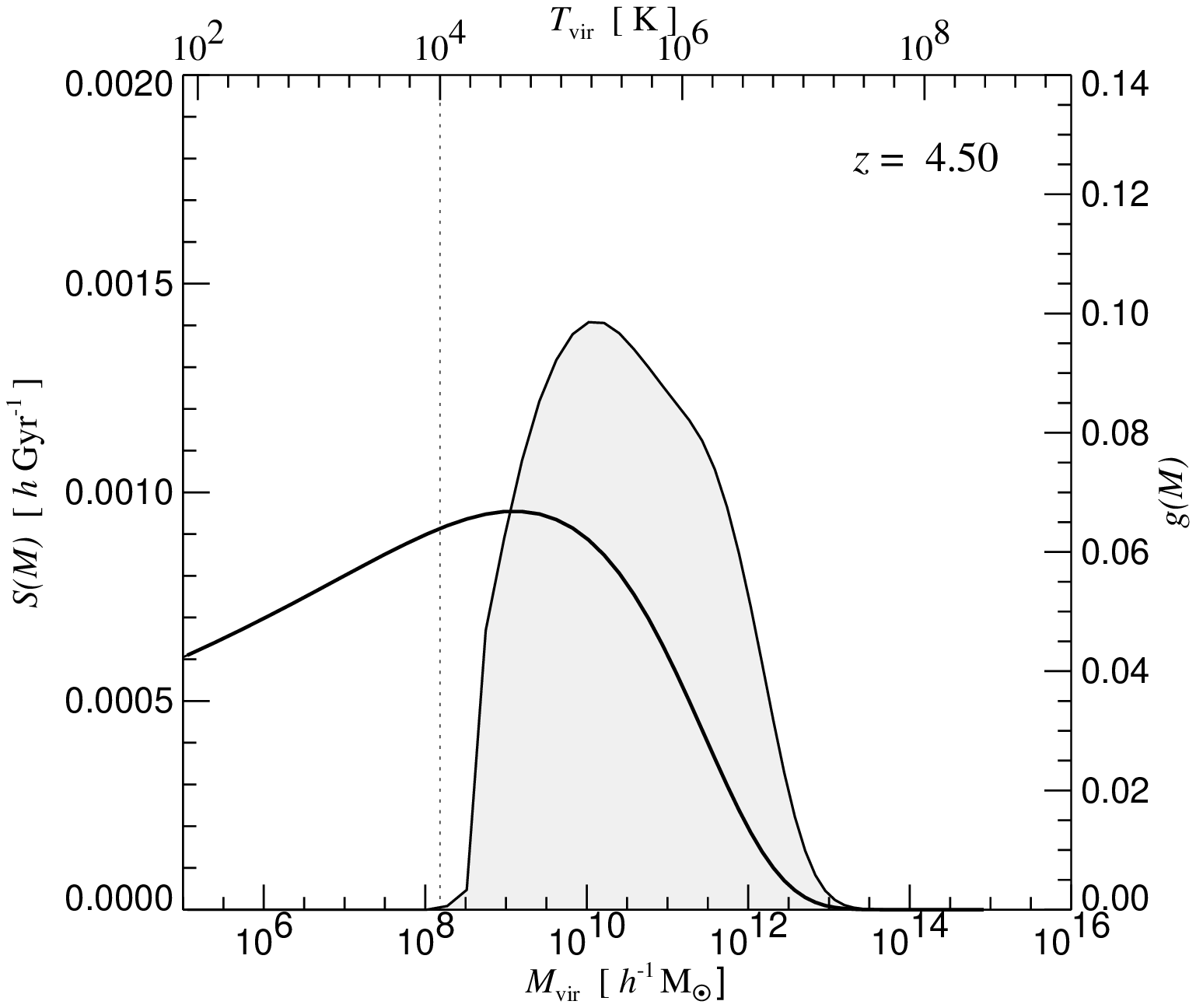}}%
\resizebox{5.8cm}{!}{\includegraphics{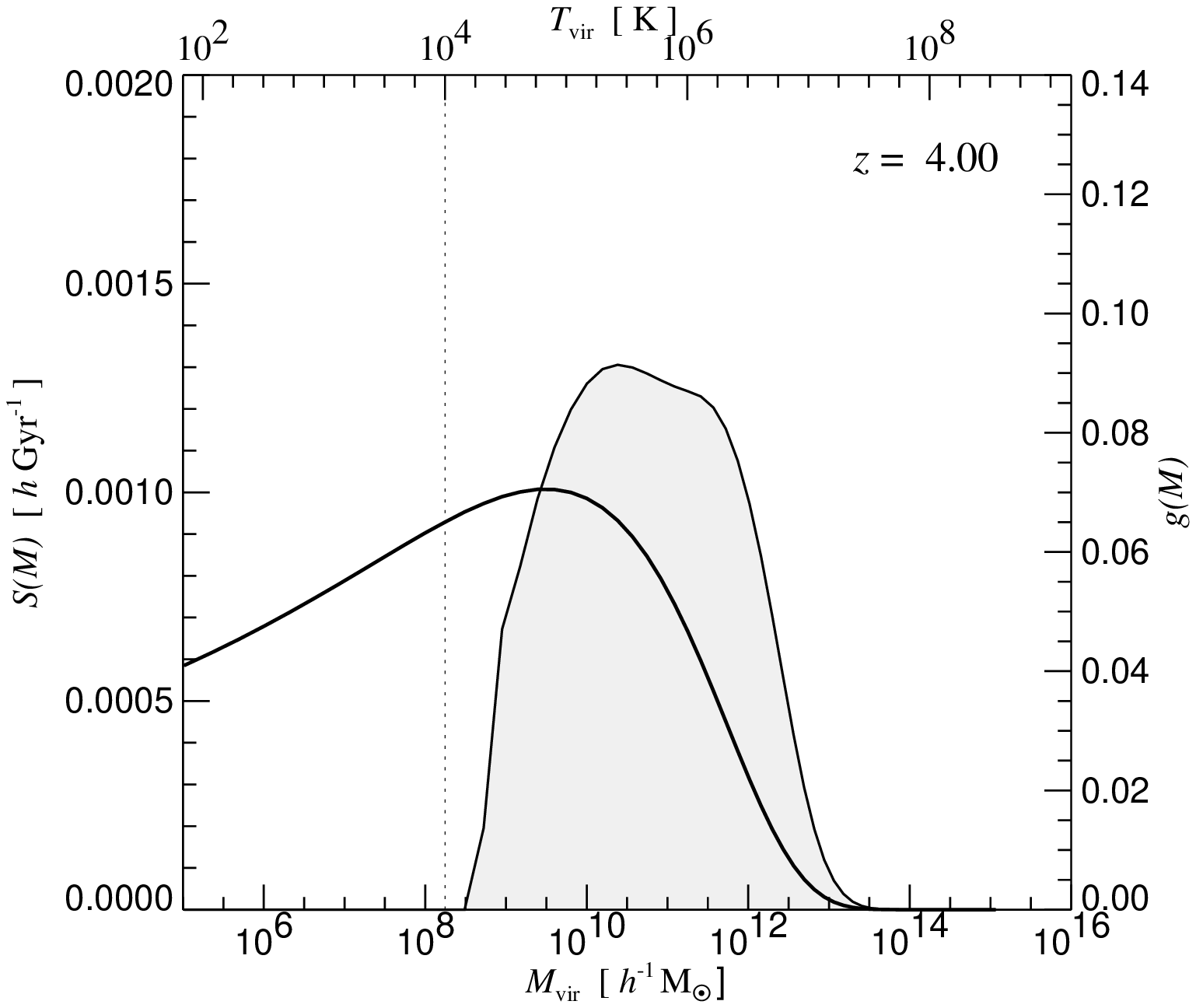}}%
\resizebox{5.8cm}{!}{\includegraphics{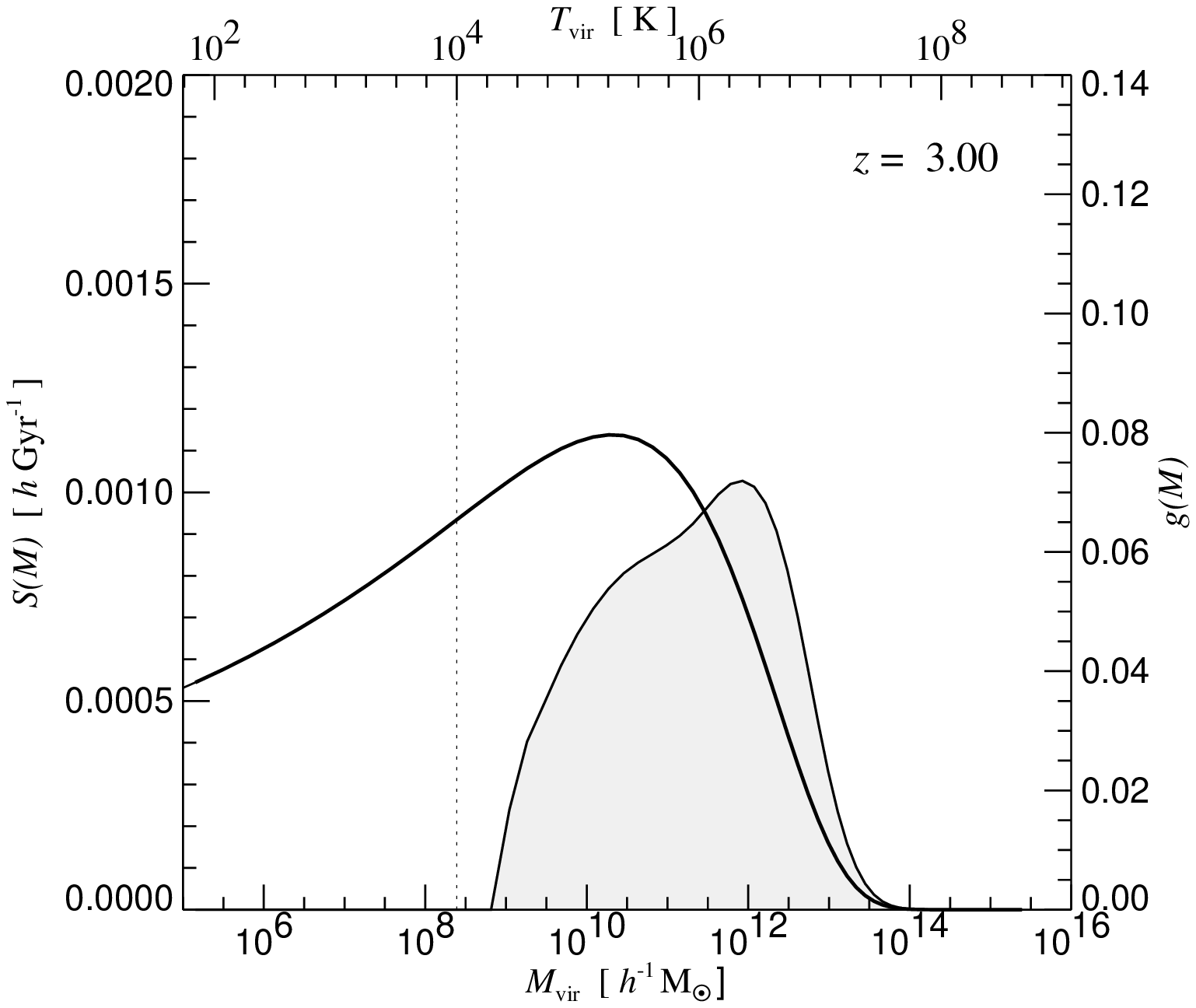}}\\%
\resizebox{5.8cm}{!}{\includegraphics{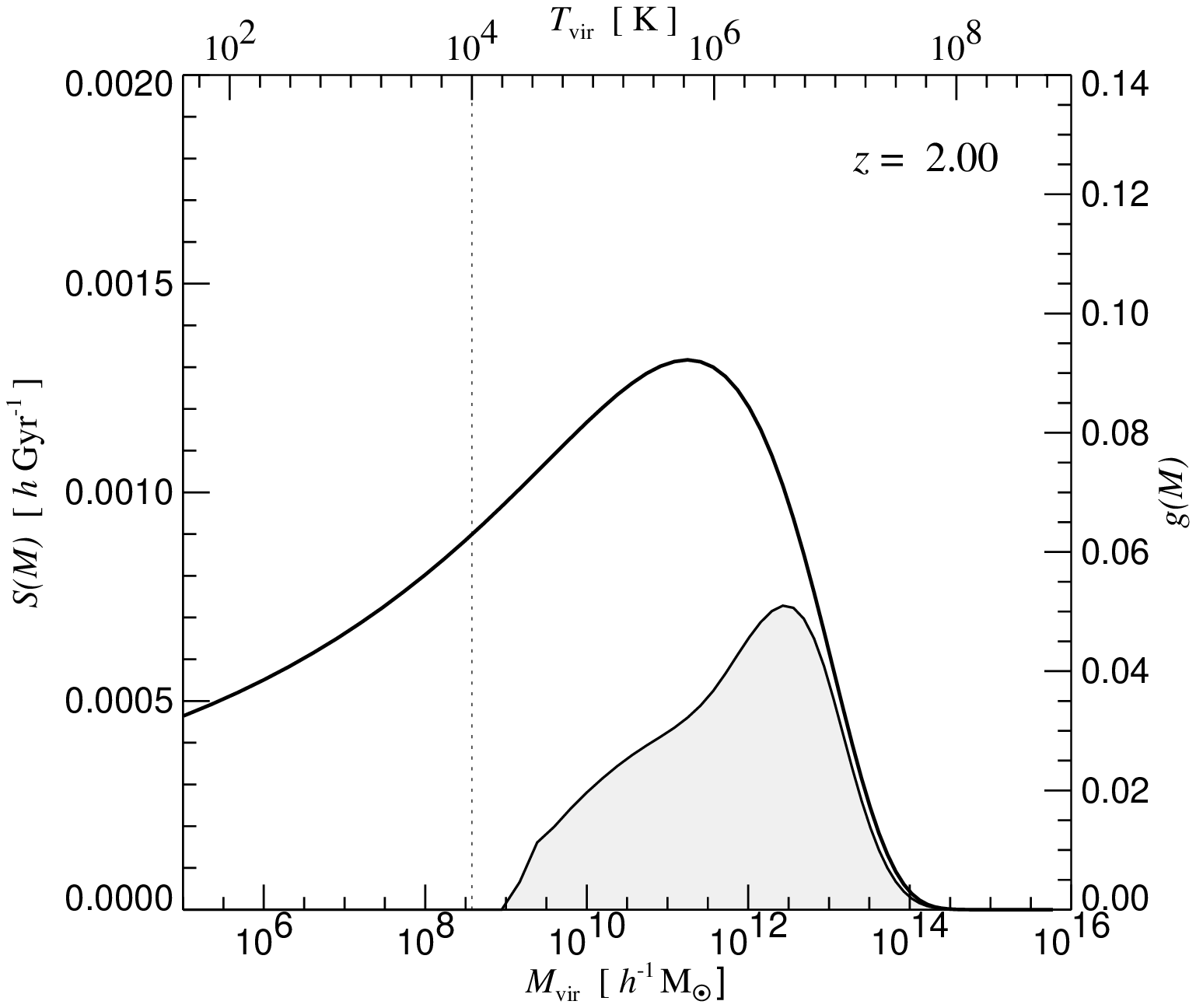}}%
\resizebox{5.8cm}{!}{\includegraphics{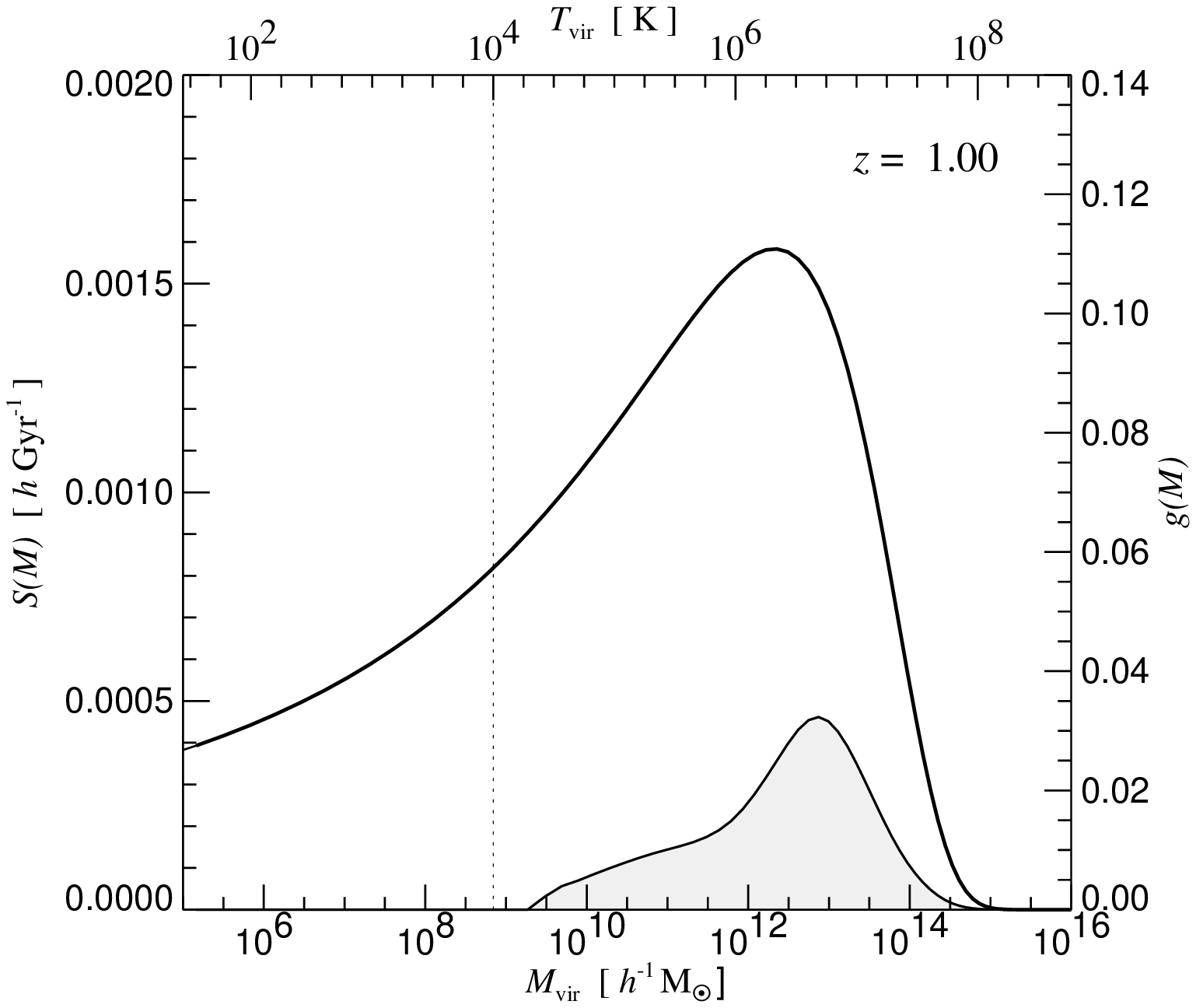}}%
\resizebox{5.8cm}{!}{\includegraphics{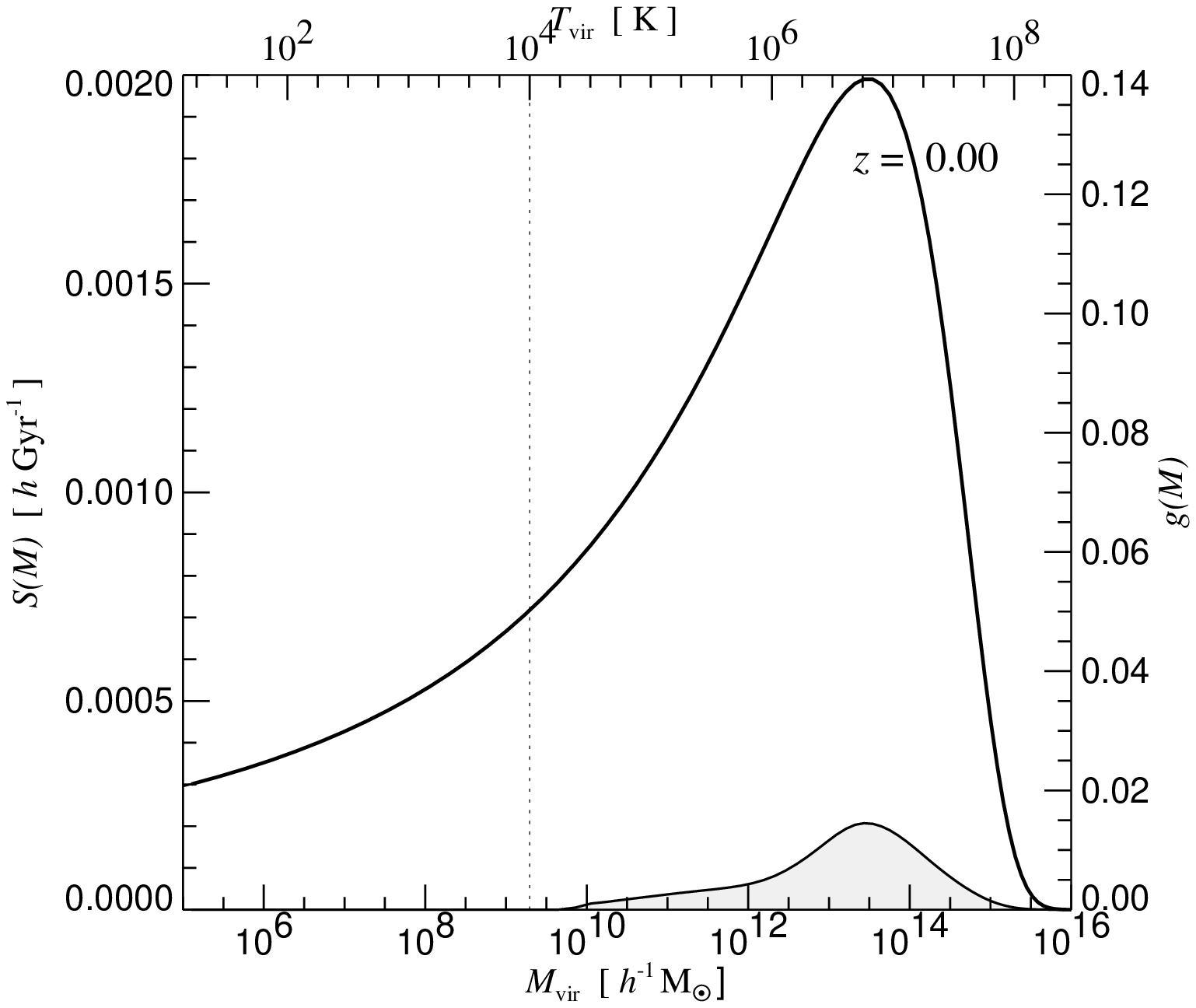}}\\%
\caption{Multiplicity function of star formation (shaded) as a
function of redshift.  This quantity is defined as the product of our
measurements for $s(M,z)=\left< \dot M_\star \right>/M$ with the Sheth
\& Tormen multiplicity function (solid line) for the bound mass in
halos.  On the top axes of each panel, we give the corresponding
virial temperature scale at each epoch, and we indicate the
temperature $10^4\,{\rm K}$ by a dotted vertical line.
\label{figMultiplicity}}
\ec
\end{figure*}

In Figure~\ref{figSfrVsMass}, we show examples of our measurements of
$s(M,z)$ from the simulations.  At each redshift, we have binned the
halos by mass in logarithmic intervals of width ${\rm dlog} M=0.2$,
and computed the arithmetic mean of $\dot M_\star/M$ for each
bin. Using the different simulation results available at each
redshift, we then fitted a smooth spline function to our results,
which represents our best estimate for $s(M,z)=\left< \dot M_\star
\right> / M$.  Note that this is not an entirely straightforward
process, as a cursory look at the example measurements in
Figure~\ref{figSfrVsMass} shows.

In particular, simulations that cannot resolve {\em all} star forming
halos above $T_{\rm vir}\simeq 10^4\,{\rm K}$ exhibit a biased result
at their resolution limit, in the form of an overestimate of $\left<
\dot M_\star \right> / M$.  This is because star formation and
feedback in halos of still lower mass were missed in these
simulations, so the gas fraction has not been reduced yet by galactic
outflows in progenitor halos, unlike in simulations at higher mass
resolution.  There is hence a ``compensation effect'' in simulations
with poor mass resolution.  Some of the star formation missed in
unresolved halos is shifted towards the first generation of halos that
are resolved, partly making up for what was missed.

Obviously, these biases complicate the measurement of $s(M,z)$
considerably, particularly if only a single redshift is considered.
However, we are aided by our large set of simulations of different
mass resolution, and also by the fact that $s(M,z)$ appears to
maintain its shape surprisingly well when expressed as a function of
virial temperature, i.e.~when $s(M,z)$ is factored as \be s(M,z)=
q(z)\, \tilde s_z (T_{\rm vir}), \ee where $T_{\rm vir}$ is the virial
temperature of a halo of mass $M$ at redshift $z$.  Here $q(z)$
describes an ``amplitude'' function, and $\tilde s_z(T)$ is the
``shape'' of the normalised star formation rate as a function of
virial temperature.  To a very good approximation, this shape appears
to be independent of epoch, i.e.~all the redshift dependence can be
absorbed into the global scaling parameter $q(z)$. Only at virial
temperatures between $\simeq 10^4\,{\rm K}$ and $\simeq
5\times10^4\,{\rm K}$ is this not fully true.  Here, the photoionising
background reduces the efficiency of star formation during
reionisation, which we take into account as a residual dependence of
$\tilde s_z(T)$ on redshift.

Using our large set of simulations at many different output redshifts,
we find that there is a rapid rise of $\tilde s_z(T)$ at the onset of star
formation, followed by a shallow increase towards more massive halos.
At a temperature of about $10^6\,{\rm K}$, $\tilde s_z(T)$ begins to increase
more rapidly.  This scale is very likely related to the velocity of
the galactic winds that star forming regions can generate in our
simulations.  For halos with virial temperatures above $10^6\,{\rm
K}$, winds are expected to be unable to flow out into the IGM.  This
renders the feedback induced by outflows much less efficient.
Finally, at a temperature of about $10^7\,{\rm K}$, the rate of star
formation begins to fall again, presumably because halos start to
experience less efficient cooling flows.  Note that the peak can only
be explored at low redshift, when the corresponding halos have
actually formed.  At high redshift, we formally assume that it still
exists in our parameterisation of $\tilde s_z(T)$, but whether or not this is
actually true is unimportant, because the halo multiplicity function
vanishes at the corresponding mass scales at high redshift, so that no
contribution to $\dot\rho_\star(z)$ arises.

We have also carried out a number of small test simulations of
isolated halos to see whether the behaviour of $\tilde s_z(T)$ measured in
the cosmological simulations can be qualitatively recovered by
studying isolated systems.  To this end, we set up a series of
isolated NFW-halos \citep{NFW,NFW2} consisting of dark matter and gas,
with the gas initially being distributed like the dark matter.
Initial particle velocities and gas temperatures were assigned so that
the halos begin in virial equilibrium to very good approximation, with
concentration $c=10$, baryon mass fraction $f_b=0.13$, and spin
parameter $\lambda=0.1$.  We set up 17 halos with total masses spaced
logarithmically between $10^7\,h^{-1}{\rm M}_\odot$ and
$10^{15}\,h^{-1}{\rm M}_\odot$, using 40000 gas and 80000 dark matter
particles, and with physical sizes corresponding to their expected
extent at $z=0$.  We then evolved these halos in isolation using the
same model for cooling, star formation, and feedback as used in the
cosmological runs, except that there was no ionising UV background.
Note that the initial conditions for all the halos were identical
except for an overall scaling, i.e.~without the physics of cooling and
star formation they would evolve in a self-similar manner.

In Figure~\ref{figEffIsolatedHalo}, we show the normalised star
formation rate in these models at a time $t=2.5\,{\rm Gyr}$ after the
simulations have been started.  At this point, the halos form stars at
an approximately steady rate, with the initial rapid phase of disk
formation already being largely completed.  We expect that a
measurement at this time should be qualitatively similar to $\tilde
s_z(T)$ inferred from the cosmological simulations, and this is indeed
the case. Note, in particular, the interesting maximum at a
temperature of around $10^7\,{\rm K}$.  More massive halos exhibit a
lower star formation rate because their cooling flows are less
efficient when normalised to the total gas content of the halo.  These
results demonstrate that we are able to faithfully match the
properties of individual forming galaxies in our cosmological volumes.

\begin{figure*}
\bc
\resizebox{16.0cm}{!}{\includegraphics{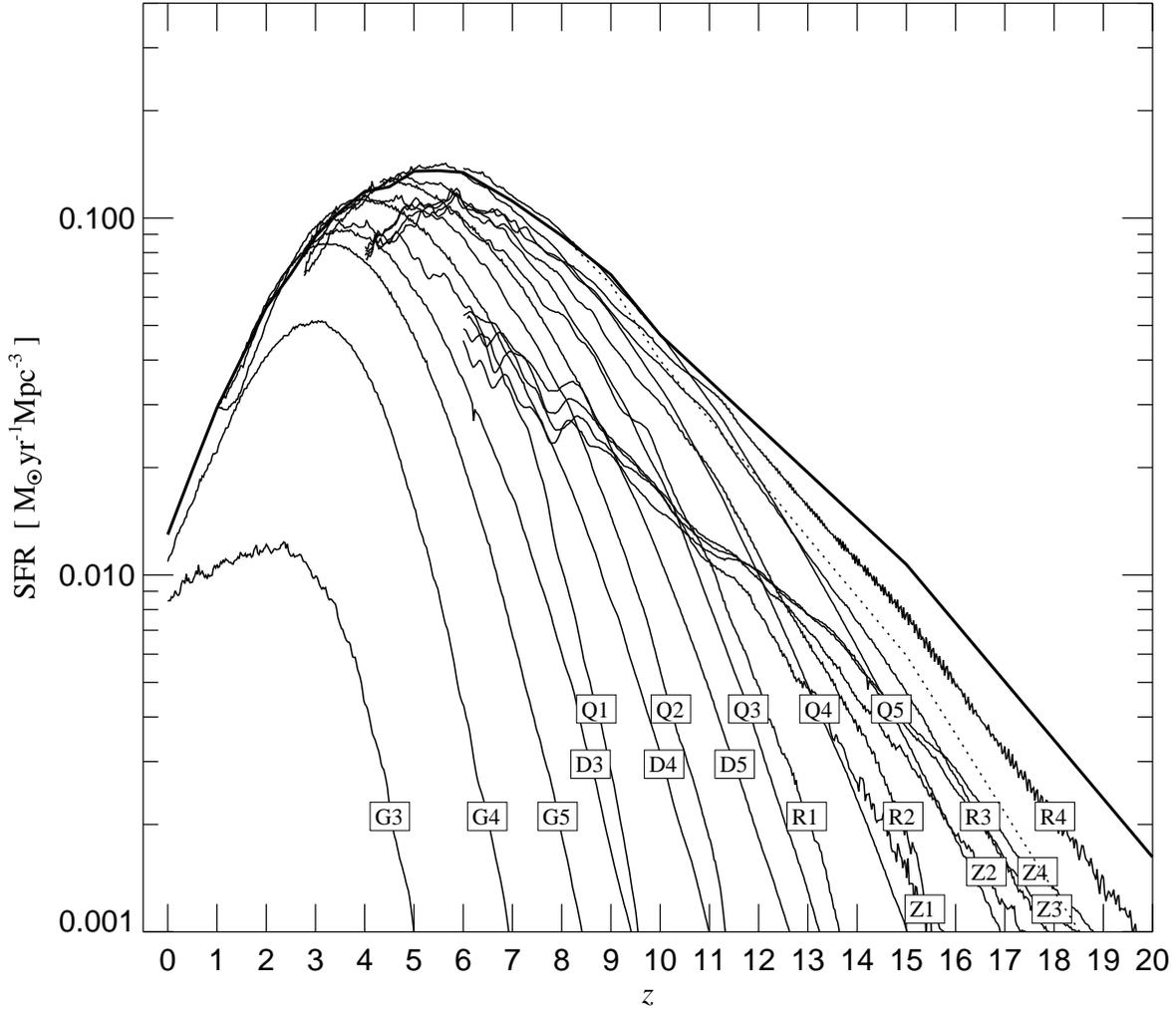}}%
\caption{Evolution of the comoving cosmic star formation rate density
in all of our simulations on a common plot. Individual runs are
labelled at the bottom. It is seen that the collection of runs forms a
common ``envelope'', with each simulation of low mass-resolution
eventually breaking away from the envelope at sufficiently high
redshift. A simulation may also underpredict the cosmic star formation
density when its volume is too small to resolve a fair sample of the
high end of the mass spectrum of halos expected at the given
epoch. This is the case for the Z-series at ``low'' redshifts of
$z<15$.  At these epochs, the star formation is dominated by the
rarest massive objects, of which the Z-series does not contain a
proper number.  Interestingly, the common envelope formed by the
simulations is very well reproduced if we multiply our measurement for
the average star formation rate in halos of a given mass, $s(M,z)=
\left<\dot M_\star\right>/M$, with the Sheth \& Tormen mass function,
and integrate the resulting multiplicity function to obtain $\dot
\rho_\star(z)$.  This result is shown as the bold line, and may be
viewed as our prediction for the star formation history, corrected for
incomplete sampling of the halo mass function. The dotted line shows
the result of this computation if the Press \& Schechter mass function
is used instead at high redshift.
\label{figSFRallSims}}
\ec
\end{figure*}

It is also interesting to note that the ``amplitude'' $q(z)$ measured
in the cosmological simulations scales approximately as
$q(z)\propto(1+z)^{3}$ over an extended redshift interval between
$z\sim 1-8$, with the scaling being somewhat slower at very low and
very high redshift. This result can be understood if at intermediate
redshifts the halos are modelled as self-similar, with the star
formation rate following the cosmological scaling of the cooling time.

In Figure~\ref{figMultiplicity}, we show our final result for the
multiplicity function of the cosmic star formation rate density.
Following equation~(\ref{eqm2}), we have defined it as the product of
our spline fits for $s(M,z)$ obtained from our full set of simulations
with the Sheth \& Tormen multiplicity function for the $\Lambda$CDM
model.  Note that in the representation of
Figure~\ref{figMultiplicity}, the area under the curve of the
multiplicity function (shaded) is proportional to the star formation
rate density. In each redshift panel, we also show the multiplicity
function of the mass distribution in bound halos at the corresponding
epoch.  It is seen how the history of $\dot\rho_\star(z)$ arises by
the interplay of two opposing trends.  The gradual shift of the mass
multiplicity function towards higher mass scales leads to an ever
increasing fraction of mass bound in star-forming halos with virial
temperatures well above $10^4\,{\rm K}$. On the other hand, star
formation in halos of a given mass becomes less efficient with time
because of the decline of the mean density within halos as a result of
cosmic expansion.  Together, these effects then produce a maximum of
the star formation rate at an intermediate epoch, with a fall-off from
there towards both low and high redshifts.

\section{Evolution of the cosmic star formation rate density} \label{SecSFR}

\subsection{Numerical results}

In Figure~\ref{figSFRallSims}, we show the evolution of the cosmic
star formation rate per unit comoving volume, as directly measured
from all of our simulations.  Interestingly, the collection of runs
appears to describe an ``envelope'', which we tentatively interpret as
the result for a fully converged numerical simulation with good
sampling of the entire cosmological mass function.  Simulations of low
mass-resolution eventually break away from this envelope towards high
redshift.  The poorer the resolution, the lower the redshift at which
this occurs. This trend is consistent with our expectations developed
in Section~\ref{SecMulti}, based on the joint analysis of the
multiplicity functions of star formation and mass.

Note, however, that a simulation may also underpredict the cosmic star
formation rate density at a given epoch when its volume is too small
to contain a fair sample of the high end of the mass function.  This
is the case for the Z-series at ``low'' redshifts of $z<15$.  At these
epochs, the star formation rate density is dominated by the rarest
objects in the exponential tail of the mass function, which the
Z-series does not properly sample, owing to insufficient volume.

It is now interesting to see whether the common envelope traced out by
the simulations is reproduced if we multiply our measurements for the
average star formation in halos of a given mass, $s(M,z)= \left<\dot
M_\star\right>/M$, with the Sheth \& Tormen multiplicity function, and
integrate to obtain $\dot \rho_\star$ using equation~(\ref{eqm1}).
The result we obtain from this procedure is the bold line drawn in
Figure~\ref{figSFRallSims}.  The good agreement between this curve and
the ``envelope'' from the individual simulations indirectly shows that
the Sheth \& Tormen mass function provides an adequate description of
the halo abundance in our simulations.  Only at very high redshift do
the simulations fall slightly short of the expectation from the
analytic mass function.  In this regime, the cosmic mean of the star
formation rate density is dominated by very rare objects in the
exponential tail of the mass function, making it difficult to obtain a
representative sample in a small simulation volume. It is also possible
that the Sheth \& Tormen multiplicity function becomes less accurate
at these high redshifts, where it is largely untested. In fact, if we
use the Press \& Schechter mass function to integrate the expected
star formation density, we obtain the result indicated by the dotted
line in Figure~\ref{figSFRallSims}, which lies substantially below
the Sheth \& Tormen result at very high redshift. At intermediate
redshifts of around $z \sim 10$, the difference essentially vanishes,
while at lower redshift, the Press \& Schechter result lies higher.
This is consistent with \citet{Jang2001}, who found that the halo
abundance measured in collisionless simulations at redshift $z=10$
matches the Press \& Schechter prediction quite well.  Overall, the
results in Figure~\ref{figSFRallSims} give us confidence that we can
identify our measurements of $s(M,z)$ from our simulations convolved
with the Sheth \& Tormen mass function as the best estimate for the
expected star formation rate density in fully resolved simulations of
essentially arbitrarily large volume.  In the remainder of our
analysis we will therefore adopt this composite result as the
prediction of our numerical simulations.

\begin{figure}
\bc
\resizebox{8.0cm}{!}{\includegraphics{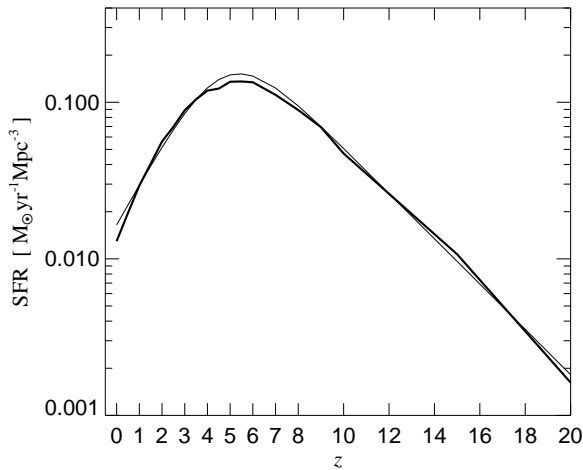}}%
\caption{Cosmic star formation rate density of our composite
simulation result (thick line) compared to the analytic fit (thin
line) given in equation~(\ref{eqnfit}). The accuracy of the
double-exponential fit is better than about $15\%$ over the full
redshift range $0\le z\le 20$.
\label{figSFRanalyticFit}}
\ec
\end{figure}

Interestingly, we have found that the composite simulation result can
be remarkably well-fitted by a double exponential of the form \be
\dot\rho_\star(z)= \dot\rho_m\,
\frac{\beta\exp\left[\alpha(z-z_m)\right]}
{\beta-\alpha+\alpha\exp\left[\beta(z-z_m)\right]}, \label{eqnfit}\ee
where $\alpha= 3/5$, $\beta=14/15$, $z_m=5.4$ marks a break redshift,
and $\dot\rho_m= 0.15\,{\rm M}_\odot{\rm yr}^{-1}{\rm Mpc}^{-3}$ fixes
the overall normalisation. In Figure~\ref{figSFRanalyticFit}, we
compare this fit to our measurement, and find it to be accurate to
within $\simeq 15\%$ in the range $0\le z \le 20$.  We suspect that
the variation of the parameter values of the fitting function with
cosmological parameters can be understood in terms of simple scaling
arguments, an idea that we will pursue further elsewhere.

\subsection{Comparison with observational measurements}

\begin{figure*}
\bc
\resizebox{13.0cm}{!}{\includegraphics{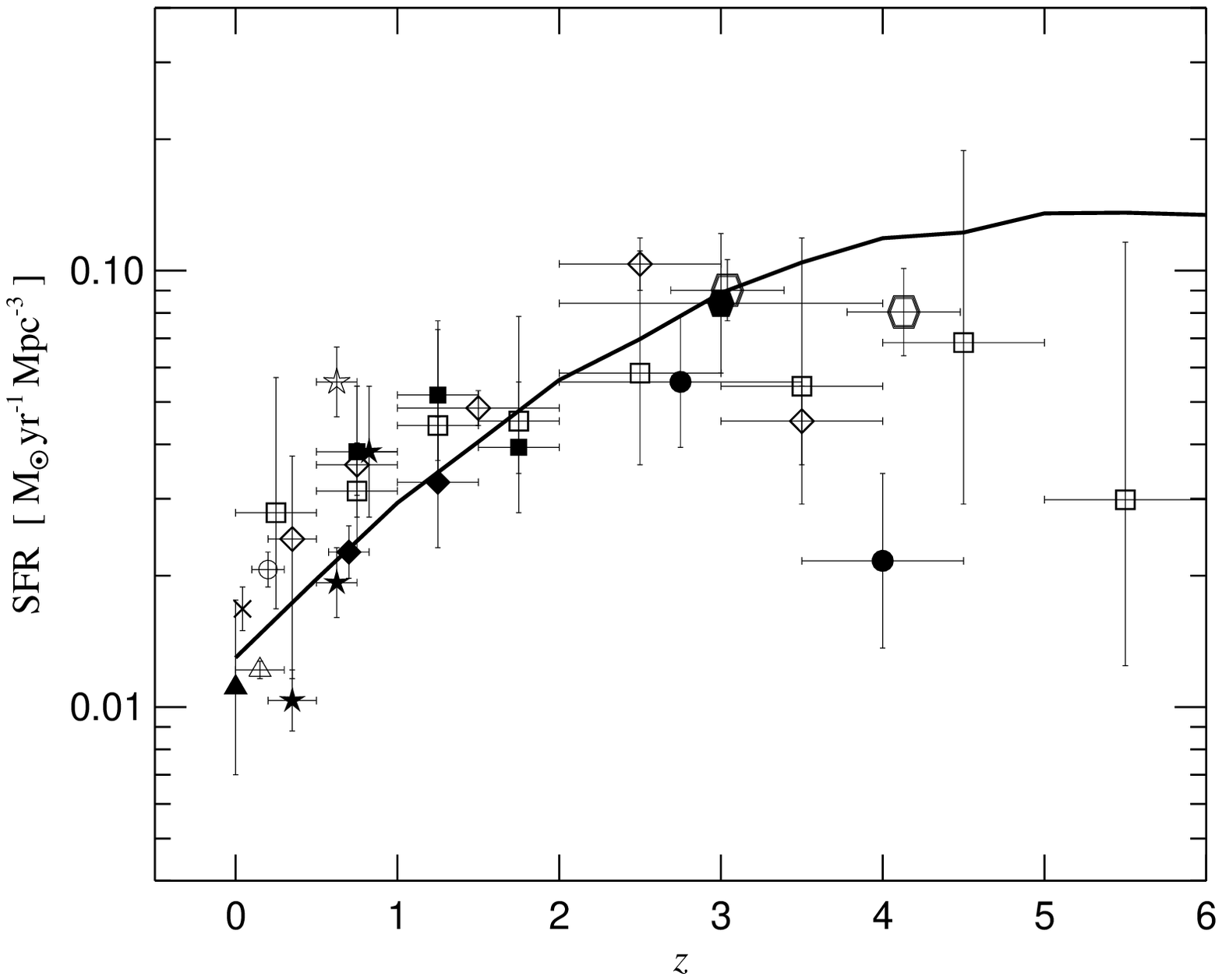}}
\vspace*{-0.5cm}\caption{Cosmic star formation history for our
composite simulation result (solid line) compared to data. Symbols
mark various observational results, here largely adopted from a recent
compilation by \citet{Som00}, who also applied corrections for
incompleteness and converted to a $\Lambda$CDM cosmology, where
appropriate. The data originate from \citet[filled triangle]{Gal95},
\citet[thin diagonal cross]{Gron99}, \citet[open triangle]{Trey98},
\citet[hollow circle]{Tresse98}, \citet[filled stars]{Lil96},
\citet[filled squares]{Con97}, \citet[filled circles]{Mad98},
\citet[empty squares]{Pas98}, \citet[empty stars]{Flo99},
\citet[filled diamonds]{Cow96}, \citet[filled hexagon]{Hug98},
\citet[hollow hexagons]{Steid99}. Following \citet{Som00}, we also
applied a uniform dust correction factor of 0.33 in the log to the
1500 Angstrom UV-points at $z>2$ of \citet{Mad98}, \citet{Steid99} and
\citet{Pas98}. {\em Note:} In the published version of this figure,
all observational data points were erroneously drawn by a factor
$1/h=1.43$ {\em too high}.
\label{figcumsfr}}
\ec
\end{figure*}

In Figure~\ref{figcumsfr}, we compare a variety of direct
observational constraints of the cosmic star formation rate density
with our prediction.  For this comparison, we have used a compilation
of observational data put together by \citet{Som00}, who have also
computed corrections for incompleteness, and for the $\Lambda$CDM
cosmology if appropriate.  In addition, we have applied a uniform dust
correction of 0.33 in the log to UV-data at high redshift.  It should
be kept in mind that observing the cosmic star formation density
is not only intrinsically challenging, but also involves quite
substantial corrections that make it difficult to assess measurement
uncertainty.

In our model, the star formation rate peaks in the redshift range
$z\sim 5-6$.  This is substantially higher than the peak at $z\sim
1-2$ suggested by the early work of \citet{Mad96}, a result which
however was probably severely affected by dust corrections unknown at
the time.  The newer high-redshift points of \citet{Steid99} and
\citet{Hug98} actually appear to be consistent with our prediction.
Also, our simulations are in good agreement with data for the Local
Universe from \citet{Gal95} and \citet{Trey98}, but not
\citet{Gron99}.

However, if we assume that all the data points are unbiased, our
result seems low compared to the ``average'' at redshifts around $z
\sim 1$.  These high observational results at $z\sim 1$ suggest a very
rapid decline of the star formation rate towards the present epoch.
For example, \citet{Hogg01} analysed the diverse set of available data
from the literature and estimates $\dot \rho_\star \propto
(1+z)^\beta$ with $\beta=3.1 \pm 0.7$.  Such a steep evolution was
first suggested by \citet{Lil96}, based on an analysis of the
Canada-France Redshift survey.  Our model prediction clearly evolves
more slowly than this estimate.

On the other hand, the more gradual decline of star formation found by
us is in better agreement with the result of \citet{Cow99}.  Recently,
this group has been able to substantially increase their observational
sample of multicolour data and spectroscopic redshifts from the Hawaii
Survey Fields and the Hubble Deep Field, allowing a selection based on
rest-frame UV up to a redshift of $z=1.5$.  \citet{Wil02} confirm a
shallow evolutionary rate of $(1+z)^{1.7\pm 1.0}$ over this redshift
range from this new data. We think our results for the star formation
history are broadly consistent with the data, given the current level
of uncertainty in the observational determinations.  Note that a
proper treatment of metal-line cooling has the potential to increase
our star formation estimate at low redshift, as discussed by
\citet{HernSpr2002}. This could eliminate a potential deficit of star
formation at low redshift in our model if future observational
improvements confirm this suggestion of the present data.

It is also interesting to compare our prediction with other
theoretical studies of the cosmic star formation history.
\citet{Bau98} have analysed the cosmic star formation history using
semi-analytic models.  They find a rapid decline of star formation at
low redshift, with a peak occurring around $z\sim 2$, clearly quite
different from the history found here.  Their result implies that 50\%
of the stars form after $z\simeq 1$ in their low-density model, with
very little star formation occurring at high redshift.  Fewer than
10\% of the stars are formed beyond redshift $z>3$ according to their
picture.  This tendency for the bulk of stars to be formed at
relatively low redshift is not found in our simulation result, as we
will discuss in more detail in Section~\ref{SecWhenWhere}.

\citet{Som00} used a similar technique to predict the history of star
formation from the present to a redshift of $z=6$.  They actually find
a star formation history resembling ours, but only for their
``collisional starburst'' model.  Their ``constant efficiency
quiescent'' model appears to predict at least an order of magnitude
less star formation for redshifts higher than 5 than found in our
simulations.  This is quite curious.  While our simulations are in
principle capable of following the triggering of intensified star
formation as a result of gas inflow in mergers, we doubt we presently
have sufficient numerical resolution to follow this process
faithfully, particularly at high redshift. It seems more likely that
our simulations have such high star formation rates because our
intrinsic, ``quiescent'' mode of star formation is very efficient at
high redshift.

In the numerical simulation of \citet{Nag01}, the cosmic star
formation rate rises quickly from the present epoch to about $z\sim
1$, and then continues to increase monotonically but slowly towards
high redshift, staying nearly constant between redshifts 2 and 7, the
highest redshift for which they cite results. The peak of the star
formation rate therefore appears to occur beyond redshift 7 in their
model.  In earlier work by \citet{Nag00}, a peak at $z\simeq 2$ was
observed, but they also showed in this study that this first result
was strongly dependent on numerical resolution and had not converged
yet; doubling the box size from $50$ to $100\,h^{-1}{\rm Mpc}$ and
hence lowering the spatial resolution by a factor of two also lowered
the star formation density by about a factor of two.

\citet{Asc02} also investigated the cosmic star formation history by
means of hydrodynamic mesh simulations.  In fact, they used a
multi-phase model quite similar in spirit to the one employed here,
although theirs differs substantially in the details of its
formulation.  Similar to \citet{Nag01}, their $\Lambda$CDM models
predict a nearly constant star formation density over an extended
redshift range, but \citet{Asc02} find a peak at a low redshift of
about $z=1.0-1.5$, with a gradual decline towards higher redshift.

In both of these numerical studies, hydrodynamical mesh-codes have
been used which have severe difficulty in resolving star-forming
regions with the appropriate size and density, despite the high
accuracy these numerical techniques have for the hydrodynamics of gas
at moderate and low overdensity.  We suspect that this might make it
relatively difficult to achieve numerical convergence for the star
formation density using this approach.  We expect our Lagrangian
simulation technique to offer better behaviour in this respect,
although it can suffer from (different) numerical limitations
\citep[e.g.][]{Weinberg99,Pea2000,SprHe01}, depending on details of
its implementation.

Finally, we comment on the behaviour of the cosmic star formation rate
density at the epoch of reionisation.  In our simulations, an
externally imposed UV background radiation of the form used by
\citet{Da99} starts to reionise the Universe at redshift $z=6$.
However, we do not find the distinctive drop of the cosmic star
formation rate at the epoch of reionisation predicted by
\citet{Bar00}. From Figure~\ref{figMultiplicity}, we see that after
the epoch of reionisation, the UV background efficiently suppresses
star formation in halos of virial temperatures up to $\sim5\times
10^4\,{\rm K}$, but the halos in this mass range do not make a very
significant contribution to the total star formation rate density.  At
the epoch of reionisation at redshift $z=6$, the bulk of star
formation has already shifted to higher mass scales, so that
photoionisation reduces the total star formation rate by only 10-20\%.
For a higher reionisation redshift, we would expect to see a stronger
effect, however. Note that a number of semi-analytic studies of
photoionisation ``squelching'' of star formation in dwarf galaxies
\citep{Bullock00,Bullock01,Bar00,Somer01,Benson2001a,Benson2001b} have
assumed a relatively strong depletion of the gas fraction, and a
corresponding reduction of the star formation rate, in halos of virial
temperatures of up to $10^5\,{\rm K}$.  Our numerical results suggest
that this effect may have been overestimated in these studies.

\subsection{The cosmic density in stars}

An immediate corollary of our star formation history is the total
density of stars and stellar remnants we expect to have formed by the
present epoch.  We find $\Omega_\star = 0.004$ for this quantity in
units of the critical density, corresponding to $\sim$10\% of all
baryons being locked up in long-lived stars.  This is substantially
lower than is typically found in hydrodynamic simulations that do not
include strong feedback processes, and this low value brings us into
better agreement with observational estimates of $\Omega_\star$ based
on the local luminosity density.

\citet{Cole2001} used the 2dF and 2MASS catalogues to constrain the K-
and J-band luminosity functions.  For a $\Lambda$CDM cosmology, they
find a total density in stars of $\Omega_\star=(1.6 \pm 0.24)\times
10^{-3}h^{-1}$ for a Kennicutt IMF, and $\Omega_\star =(2.9 \pm 0.43)
\times 10^{-3}h^{-1}$ for a Salpeter IMF \citep[see also][]{Bal01}.
Note that for the baryon density of $\Omega_b=0.04$ we adopted
(motivated by constraints from big bang nucleosynthesis), this means
that only between 5-12\% of the baryons appear to be locked up in
stars.  Similarly low values are found using optical data.  For
example, \citet{Fuk98} estimated the global budget of baryons in all
states, based on a large variety of data, and found $\Omega_\star=
2.45^{+1.7}_{-1.0}\times 10^{-3}h^{-1}$, corresponding to about 17\%
of the sum of all baryons they were able to detect.  This sum amounted
to $\Omega_b=0.021\pm 0.007$ in their study, falling somewhat short of
current nucleosynthesis constraints.  However, this difference can be
attributed to a state of baryons unaccounted for in their study, the
reservoir of warm-hot gas in the IGM that is seen in simulations
\citep{Cen99,Dave2001}, but which is difficult to detect in either
absorption or emission.  This lowers the relative contribution of
stars in the baryon budget, leading to values in agreement with the
K-band results.  Note that our simulation result of $\Omega_\star =
2.8\times 10^{-3}h^{-1}$ is consistent with these constraints.

\begin{figure}
\bc
\resizebox{8cm}{!}{\includegraphics{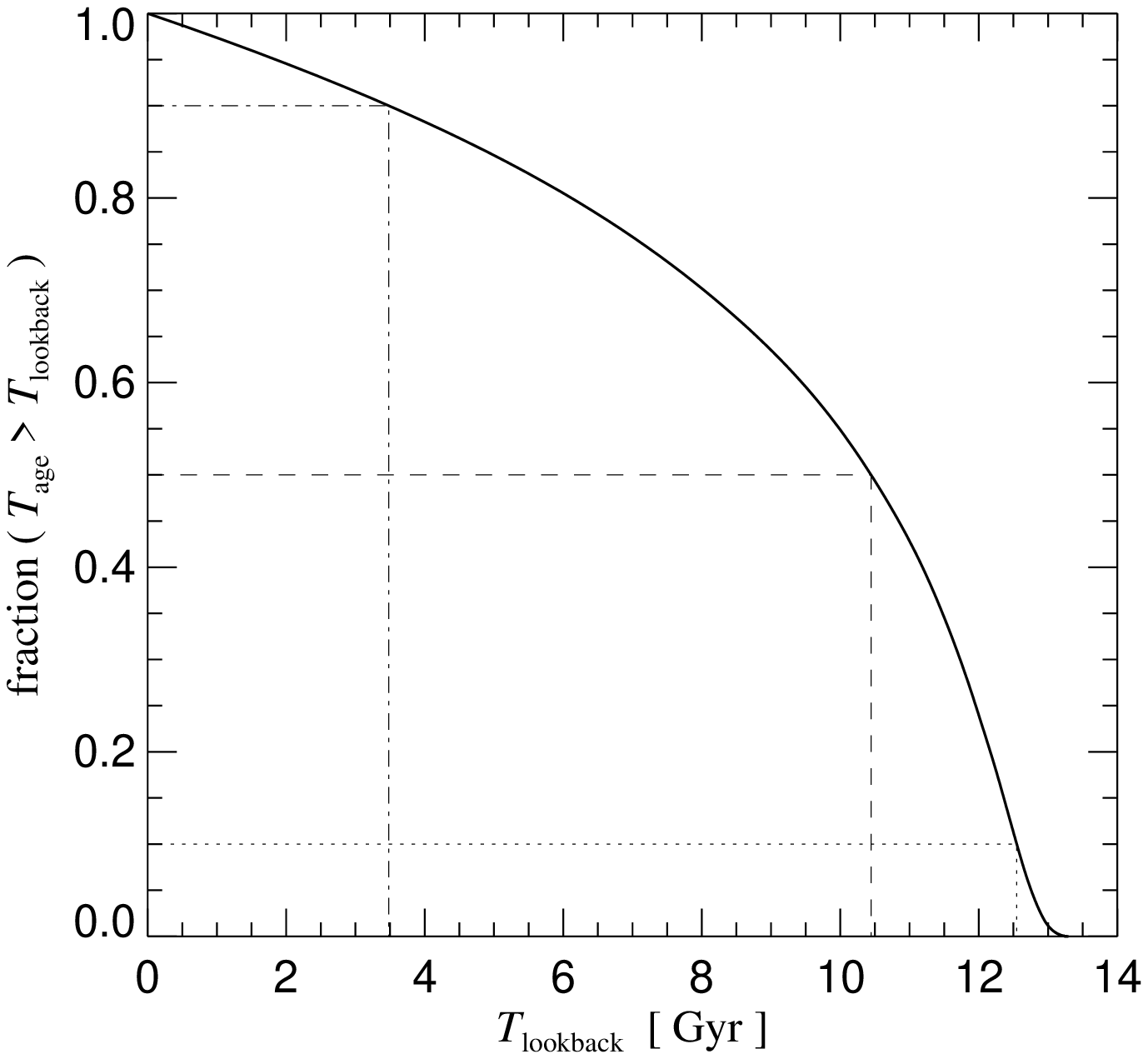}}\\%
\resizebox{8cm}{!}{\includegraphics{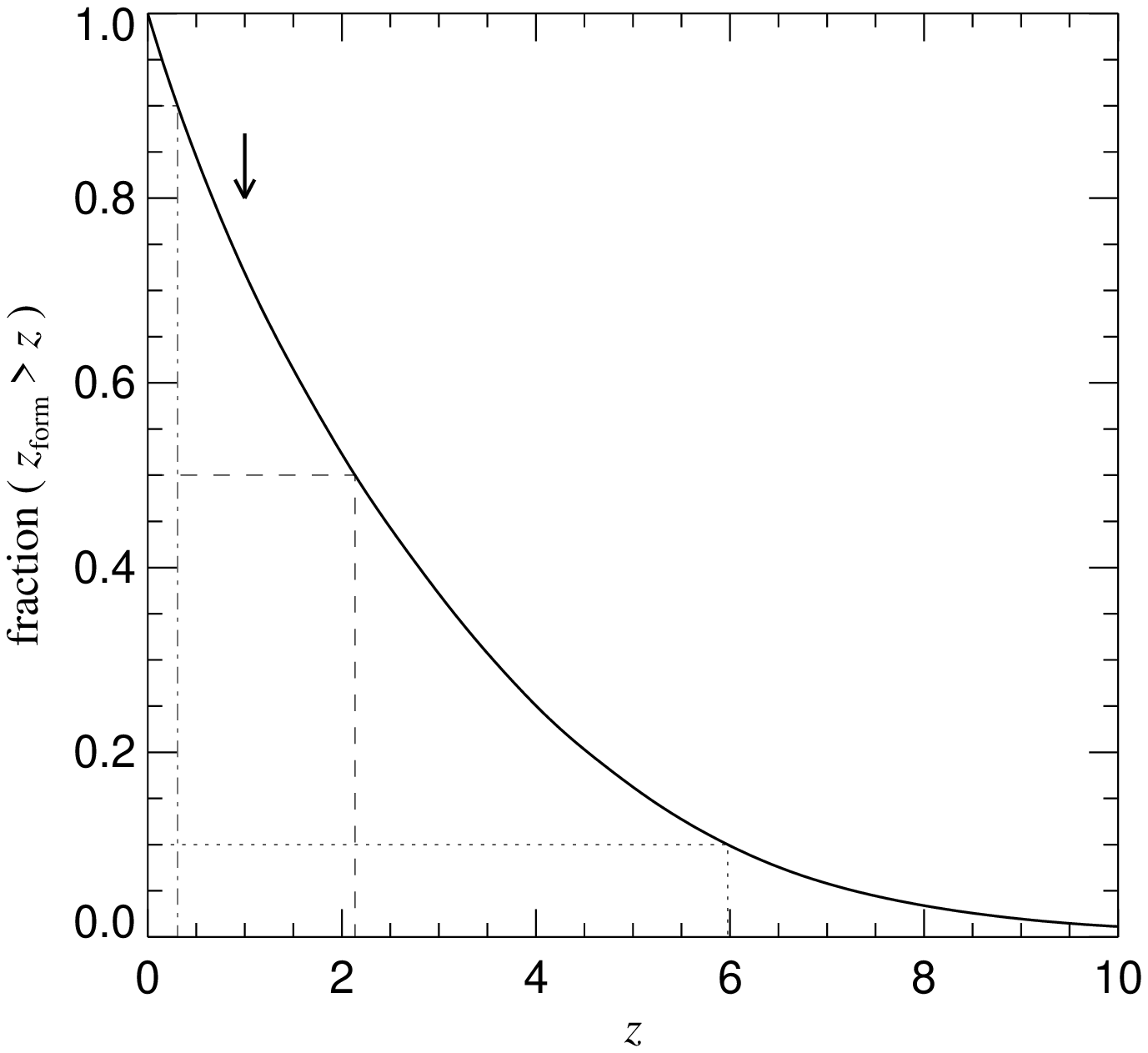}}\\%
\caption{Cumulative star formation as a function of lookback time
(top), or redshift (bottom).  In both panels, we plot the fraction of
stars that have formed by a given lookback time or redshift,
respectively, with the broken lines indicating 10\%, 50\%, and 90\%
percentiles. The arrow marks an upper limit by \citet{Bald02} for the
relative fraction of stars that may at most have formed in the
redshift range $z>1$. We provide a tabulated form of the data in this
figure in Table~\ref{tabSfrCumul}.
\label{figCumulSfr}} 
\ec 
\end{figure}

\begin{figure}
\bc
\resizebox{8cm}{!}{\includegraphics{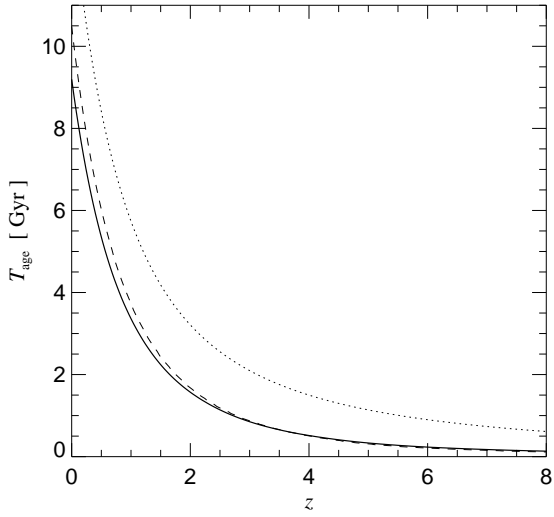}}%
\caption{Mean (solid) and median (dashed) ages of stars that have
formed up to a certain epoch $z$. The dotted line shows the age of the
universe as a function of redshift. For redshifts higher than $z\simeq
2.5$, the global stellar population is quite young, with a mean age
below 1$\,$Gyr.  However, after this epoch, the mean age nearly
increases as rapidly as the age of the Universe itself, because the
star formation rate density is rapidly declining towards low redshift.
As a result, the mean age of all stars is about 9 Gyr at the present
epoch.
\label{figMeanAge}}
\ec
\end{figure}

\section{When and where stars form} \label{SecWhenWhere}

In this section, we examine the cumulative history of cosmic star
formation, as predicted by our composite simulation result.  This is
shown in Figure~\ref{figCumulSfr}, both as a function of redshift and
lookback time.  Interestingly, already half the stars have formed by
redshift $z\sim 2.14$ in our model, or more than 10.4 Gyr ago,
implying that most of the stars at the present epoch should be quite
old, with little recent star formation.  Only slightly more than 10\%
of all stars are younger than 4 Gyr, and only about 25\% of the stars
form at redshifts below 1.

Recently, \citet{Bald02} have constrained the star formation history
based on an averaged ``cosmic spectrum'' of galaxies in the 2dF Galaxy
Redshift Survey.  They argue that one can put {\em upper limits} on
the star formation rate at redshifts higher than $z>1$, requiring that
at most 80\% of all stars formed at redshifts $z>1$. Our result
satisfies this constraint, but not by a wide margin.

The fact that stars should be relatively old at the present time
according to our model becomes clear when we explicitly measure the
mean stellar age versus redshift.  This is shown in
Figure~\ref{figMeanAge}.  Early on, the young age of the Universe and
the high rate of star formation together guarantee that the global
stellar population is quite young, with a mean age below 1$\,$Gyr.
However, due to the rapid decline of star formation at low redshift,
the mean age eventually starts to grow nearly in proportion to the age
of the Universe itself, reaching a value of about 9 Gyr by redshift
$z=0$.  This is substantially larger than what was reported by
\citet{Nag01}, highlighting that our star formation history is
considerably ``older'' than suggested by most previous numerical or
semi-analytic work.  There are large observational uncertainties in
determining the age distribution of stellar populations, so it is
unclear whether this quantity is very constraining for the models.  It
seems clear, however, that our model should have no difficulty in
accommodating the claimed old ages of the stellar populations in most
elliptical galaxies, despite the bona-fide hierarchical formation of
all galaxies in the simulations.

\begin{figure}
\bc
\resizebox{8cm}{!}{\includegraphics{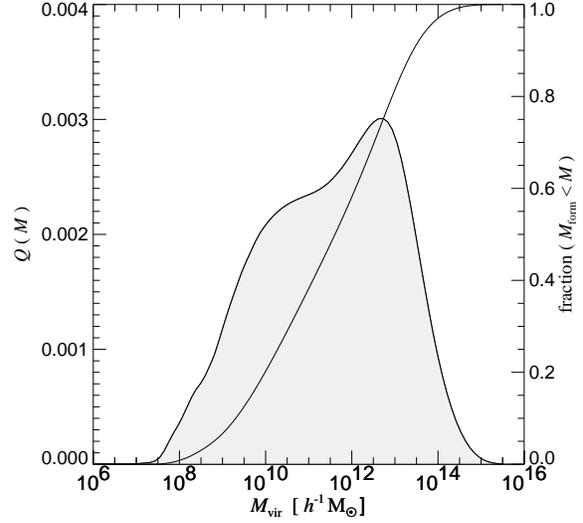}}\\%
\caption{Distribution of the integrated star formation history with
respect to halo mass (shaded).  This function is the time integral of
the multiplicity function of star formation and shows which mass
scales are the dominant {\em formation sites} of stars.
Interestingly, the distribution is quite broad, showing that nearly
the same mass in stars forms per decade of halo mass in the range
$10^{10}-10^{13.5}\,h^{-1}{\rm M}_\odot$, with a slight peak at $\sim
10^{12.5}\,h^{-1}{\rm M}_\odot$. We also include a plot of the
cumulative function of this distribution (right scale).  Roughly half
of the stars are predicted to have formed in halos less massive than
$10^{11.5}\,h^{-1}{\rm M}_\odot$.
\label{figSfrvsHalomass}}
\ec 
\end{figure}

Another interesting question one may ask is {\em where} do most of the
stars form?  By ``where'', we refer to the mass-scale of the halos
that hosted the star formation.  Formally, this distribution is given
by integrating the multiplicity function $S(M,z)$ defined by equation
(\ref{eqm2}) over time:
\be
Q(M)= \int s(M,z)\, g(M,z)\, {\rm d}t.
\ee
In Figure~\ref{figSfrvsHalomass}, we show the
resulting distribution of all star formation with respect to halo
mass.  The distribution is quite broad, with about an equal mass in
stars being formed per decade of halo mass in the range $10^{10} -
10^{13.5} \, h^{-1}{\rm M}_\odot$, with a peak at about $10^{12.5} \,
h^{-1}{\rm M}_\odot$.  Note, however, that this does not imply that
the stars are distributed over halos in this way at the present day.
Most of them will not live in the halo they originally formed in,
either because their parent halo was incorporated into a more massive
halo by merging, or because the halo simply grew in mass through
accretion.

From the cumulative form of this distribution function (also shown in
Figure~\ref{figSfrvsHalomass}), we infer the fraction of stars that
were born in halos less massive than a given threshold value.  This
quantity enables us to assess how much of the star formation might
have been missed by a simulation of a given mass resolution that has
been run to redshift $z=0$.  For example, consider our `G3' run, the
lowest-resolution simulation of the G-Series, which employed $2\times
144^3$ particles in a $100\,h^{-1}{\rm Mpc}$ box.  Assuming that at
least 100 SPH and dark matter particles per halo are necessary to
obtain a converged result for the star formation in the mean (this is
optimistic), then all the star formation in halos below a mass of
$\sim 2.8\times 10^{12}\,h^{-1}{\rm M}_\odot$, amounting to 65\% of
the total, will either have been lost completely, or will have been
computed unreliably.

\begin{table}
\bc
\begin{tabular}{ccc}
\hline
Fraction & $z$ & $T\;[{\rm Gyr}]$ \\
\hline
  0.1 &  5.98 & 12.55  \\
  0.2 &  4.53 & 12.17  \\
  0.3 &  3.56 & 11.73  \\
  0.4 &  2.80 & 11.19  \\
  0.5 &  2.14 & 10.44  \\
  0.6 &  1.58 &  9.45  \\
  0.7 &  1.08 &  8.03  \\
  0.8 &  0.67 &  6.12  \\
  0.9 &  0.31 &  3.48  \\
\hline
\end{tabular}
\caption{Cumulative star formation history as a function of lookback
time $T$ and redshift $z$. This data corresponds to the plots shown in
Figure~\ref{figCumulSfr}.
\label{tabSfrCumul}}
\ec
\end{table}

\section{Early star formation as possible source for hydrogen reionisation} \label{SecReion}

There are both observational and theoretical indications that at least
some of the ionising background at high redshift comes from stars and
that it is not all produced by quasars
\citep[e.g.][]{Steid2001,Haehn2001,Hui2002}.  This is a long debate.
At low redshift, it seems clear that quasars dominate, while at high
redshift, the space density of bright quasars appears to drop so
rapidly that it is difficult to see how they could reionise the
Universe already by redshift $z\sim 6$. Hence, it has been speculated
extensively that hydrogen was mostly reionised by stars, or possibly
harder sources like weak AGN or mini-quasars.  Recently, the first
determinations of the spectrum of Lyman-break galaxies at redshifts
$\left<z\right>=3.4$ by \citet{Steid2001} have revealed a surprisingly
high escape fraction for hydrogen ionising photons, supporting
scenarios where the star forming galaxies reionise hydrogen around
redshift $z\sim 6$, with quasi-stellar sources taking over only at
redshifts $z<3$ \citep{Haehn2001}.  This general picture is consistent
with the analysis of \citet{Sok2002b} who have constrained the
spectral shape of the UV background at $z\sim 2.5 - 5$ from the
opacities of the H{\small I} and He{\small II} Lyman-alpha forests, at
an epoch when He{\small II} reionisation is likely to have occurred
\citep{Sok2002a}.

Here, we roughly estimate whether the star formation rates predicted
at high redshift in our model could be responsible for the
reionisation of hydrogen by redshift $z\sim 6$. The fact that we infer
a star formation rate that peaks around this redshift is certainly
intriguing, although this may just be a coincidence.

\begin{figure}
\bc
\resizebox{8cm}{!}{\includegraphics{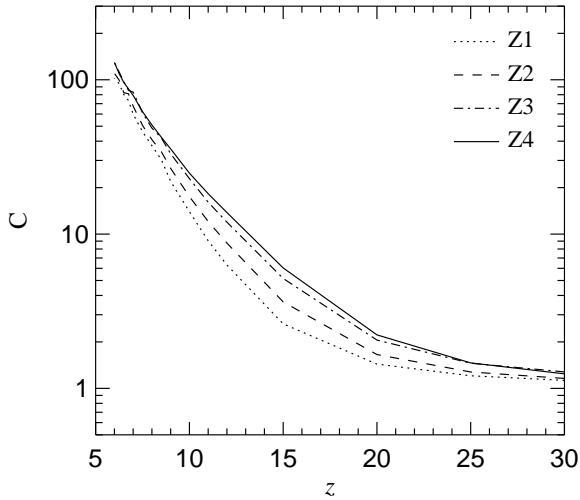}}%
\caption{Clumping factor of the gas in the simulations of the Z-Series
as a function of redshift.  We show results for Z1, Z2, Z3, and
Z4. The clumping factor was estimated for all the gas by means of
equation (\ref{eqnclumping}).
\label{figClumping}}
\ec
\end{figure}

The reionisation of the universe can be described in a statistical
sense by noting that every ionising photon that is emitted is either
absorbed by a newly ionised hydrogen atom, or by a recombining one
\citep{Madau2000}.  The filling factor $Q(t)$ of regions of ionised
hydrogen is then simply given by the total number of ionising photons
emitted per hydrogen atom at earlier times than $t$, minus the total
number of recombinations per atom.  Based on this argument,
\cite{Madau1999} derived a simple differential equation that governs
the transition from a neutral universe to a fully ionised one: \be
\frac{{\rm d}Q}{{\rm d}t} = \frac{\dot n_{\rm ion}}{{\overline{n}_{\rm
H}}} - \frac{Q(t)}{\overline{t}_{\rm rec}}.
\label{eqionize} \ee Here $\dot n_{\rm ion}$
is the comoving emission rate of hydrogen ionising photons,
${\overline{n}_{\rm H}}$ is the average comoving density of hydrogen
atoms, and $\overline{t}_{\rm rec}$ is the volume averaged hydrogen
recombination time.  The latter can be expressed as: \be
\overline{t}_{\rm rec} = \frac{1}{(1+z)^3
(1+2\chi)\,{\overline{n}_{\rm H}}\, \alpha_B\, C}, \ee where \be C
\equiv \frac{\left< n_{\rm H^+}^2 \right>}{\left<n_{\rm H^+}\right>^2}
\ee is the clumping factor of ionised hydrogen, $\chi$ is the helium
to hydrogen abundance ratio, and $\alpha_B$ is the recombination
coefficient.  Assuming a gas temperature of $10^4\,{\rm K}$, the
recombination time can be written as \citep{Madau1999}: \be
\overline{t}_{\rm rec}= 588\frac{a^3}{C(a)}\,{\rm Gyr}. \ee We will
approximate $C$ with the clumping factor of {\em all} the gas,
measured directly from the simulations. This will tend to overestimate
the clumping slightly, since higher density regions will typically tend
to be less ionised.  On the other hand, measurements from the
simulations are expected to be biased low due to their intrinsic
resolution limit.

\begin{figure}
\bc
\resizebox{8cm}{!}{\includegraphics{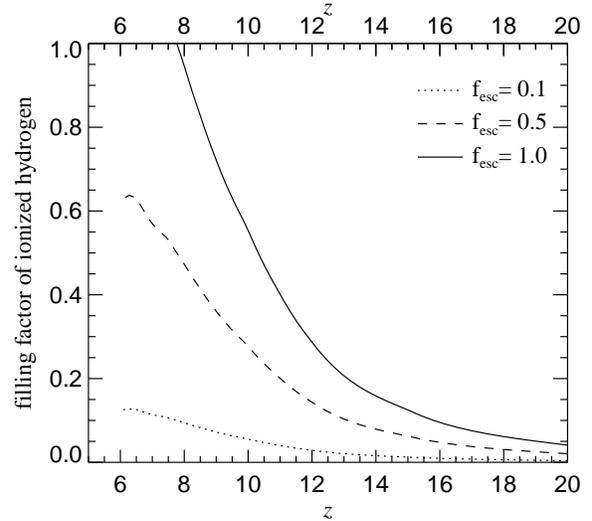}}%
\caption{Filling factor of ionised hydrogen as a function of redshift
when the star-forming galaxies are assumed to be the only ionising
sources.  Results are shown for the different escape fractions of 10\%,
50\%, and 100\%.  Interestingly, the star formation history at high
redshift might be sufficient to ionise hydrogen even without the
proposed population III generation of stars, but only for
high escape fractions close to unity.
\label{figFilling}}
\ec
\end{figure}

For the Lagrangian SPH simulations, we can estimate the clumping
factor conveniently from \be C= \frac{ \sum_i m_i \rho_i^{-1} \sum_j
m_j \rho_j }{\left(\sum_k m_k\right)^2}. \label{eqnclumping} \ee In
Figure~\ref{figClumping}, we show the clumping factors measured in
this way for the four simulations Z1-Z4.  As expected, for higher mass
resolution, there is a tendency to detect an earlier increase of the
clumping due to the first structures that form, but overall there is
gratifying agreement between the simulations, especially in the most
interesting redshift regime of $z \simeq 6-10$. Note that our measured
clumping values are also in good agreement with the semi-analytic
prediction of \citet{Ben2001}.

\citet{Madau1999} estimate that approximately $10^{53}$ ionising
photons per second are released by every star forming region with a
star formation rate of $1\,{\rm M}_\odot\,{\rm yr}^{-1}$.  A large
uncertainty is to what extent these photons are able to escape a
galaxy without being absorbed in the star-forming ISM itself.  The
escape fraction $f_{\rm esc}$ has been estimated to be quite small at
low redshift \citep[e.g.][]{Leit95}, but recent results by
\citet{Steid2001} suggest that it may actually have been quite high in
Lyman-break galaxies at high redshift.  If these results are
confirmed, it could well be above 50\% at these redshifts.  We model
the emission rate as \be \dot n_{\rm ion} = f_{\rm esc}\, \eta \,
\dot\rho_\star, \ee where $\eta =10^{53}\,{\rm
phot\;s^{-1}\,M_\odot^{-1} yr}$.

For a given value of the escape fraction, the measured evolution of
the star formation history and of the clumping factor in the
simulations can be used to integrate equation~(\ref{eqionize}) in
time.  In Figure~\ref{figFilling}, we show our results for escape
fractions of 10\%, 50\%, and 100\%.  It is rather interesting to note
that for the latter case, we obtain a reionisation redshift of
$z\simeq 8$.  However, this result is quite sensitive to the assumed
escape fraction. Lowering the number of available photons by a factor
of 2 will already delay the reionisation redshift to $z\simeq 4$, while an
escape fraction as low as 10\% together with the strong clumping that
develops at low redshift will prevent hydrogen ionisation by stars
altogether. By the same token, the result is sensitive to the assumed
IMF because the latter influences the production rate of ionising photons. For
a Salpeter IMF, \citet{Hui2002} actually estimate $\eta= 1.5\times
10^{53}\,{\rm phot\;s^{-1}\,M_\odot^{-1} yr}$, which is 50\% higher
than the value we adopted in our estimate and would thus give more
room for lower escape fractions.

It thus appears that hydrogen reionisation by stars at redshifts
around $z\sim 6$ is plausible with a cosmic star formation history as
predicted here.  It requires, however, a high escape fraction. Further
observational determinations of the escape fractions, and theoretical
studies of radiative transfer in the star forming regions, are clearly
needed to shine more light on this important issue.  In particular, a
quantitative estimate of the escape fraction in a galaxy experiencing
a strong wind is desirable, as the sweeping of gas in galaxy
halos by these winds should enable high energy photons to escape more
easily.  Promising attempts to include self-consistent treatments of
radiative transfer directly in simulations have already been made
\citep[e.g.][]{Gne97,Ciardi2001,Ciardi2002,Sok2001}, and the recent
progress in developing better algorithmic treatments of radiative
transfer \citep[e.g.][]{GnedinAbel01,AbelWandelt2002,Cen2002} will
soon improve the reliability of these computations.

\section{Discussion}  \label{SecDisc}

We have studied the star formation history of the Universe using
cosmological SPH simulations that employ a sophisticated treatment of
star formation, supernova feedback, and galactic outflows.  Using
present day computational capabilities, we have been able to show that
this model yields numerically converged results for halos of all
relevant mass scales, assuming that the vast majority of stars can
form only in halos in which gas can condense by atomic line cooling.
Using a large set of high-resolution simulations on interlocking
scales and at interlocking redshifts, we have been able to determine
the multiplicity function of cosmic star formation and infer its
evolution from high redshift to the present.  We think that the
numerical methodology used here represents perhaps the most
comprehensive attempt thus far to constrain the cosmic star formation
history in the $\Lambda$CDM cosmogony with simulations.

Interestingly, our predicted star formation density peaks early,
already at redshifts around $z\sim 5 - 6$, and then falls to the
present time by a factor of about 10. The decline at low redshift is
hence shallower than suggested by a subset of observational results,
and our star formation rate also appears low at redshifts around
$z\sim 1$ compared to some of the data. However, there is very large
scatter between different observations, which are partly inconsistent
with each other, emphasising the challenges posed by these
measurements.  Our result matches a number of the observational
points, and does so particularly well at very low and very high
redshift.  We thus conclude that at present our results are consistent
with direct data on the star formation rate density as a function of
epoch.  It will be very interesting to see whether this will remain
true when observational uncertainty is reduced in the future.

One consequence of the star formation history we predict is that the
majority of stars at the present should be quite old, despite the
hierarchical formation of galaxies. Already 10.4 Gyr ago, by redshift
$z=2.14$, half of the stars should have formed, with only $\sim 25\%$
forming at redshifts below unity. The mean stellar age at $z=0$
predicted by our model is 9 Gyr.

Integrated over cosmic time, star formation occurs predominantly in
halos with masses between $10^{8}\,h^{-1}{\rm M}_\odot$ and
$10^{14}\,h^{-1}{\rm M}_\odot$, with 50\% forming in halos below
$10^{11.5}\,h^{-1}{\rm M}_\odot$.  It is thus clear that simulations
of galaxy formation need to be able to resolve at least these mass
scales well in order to have a chance at giving a reasonably accurate
accounting of the formation of the luminous component of the Universe.

The total integrated star formation rate predicts a density in stars
of about $\Omega_\star=0.004$, or expressed differently, 10\% of all
baryons should have been turned into long-lived stars by the present.
This is in comfortable agreement with recent determinations of the
luminosity density of the Universe, while earlier theoretical work was
typically predicting substantially larger numbers of stars, by up to a
factor of three or so.  Our model hence appears to have resolved the
so-called ``over-cooling crisis''.  This was primarily made possible by
the strong feedback we adopted in our simulations in the form of
galactic winds.

We note that our predictions do depend on the model for star formation
and feedback we adopted.  In particular, without the inclusion of
galactic outflows, which have been introduced on a phenomenological
basis in our approach, star formation in the low redshift universe
would clearly have been higher. Our strategy has been to normalise the
free parameters in our star formation law (the consumption timescale
of cold gas) to observations of local disk galaxies, and to select the
parameters for the galactic winds as suggested by observations.  Under
the assumption that the same laws hold roughly at all redshifts, we
have then computed what simulations predict for the $\Lambda$CDM
model.  In this context, one should clearly distinguish between the
computational difficulty of the problem on one hand, and the
uncertainty and complexity of the modeling of the physics on the
other.  We think that we have been able to make great progress on the
computational side of the problem, but we are aware that large
uncertainties remain in our handling of star formation and feedback.
It is possible that the modeling of the physics we adopted could be
incomplete in crucial respects.

Perhaps one of the most important effects that has been neglected in
our simulations is metal line cooling.  It is well known that metals
can substantially boost cooling rates, and hence can potentially have
a very prominent effect on the rates at which gas becomes available
for star formation \citep{Wh91}.  However, the extent to which metal
enrichment can enhance gas cooling is strongly dependent on how
efficiently metals can be dispersed and mixed into gas that has yet to
cool for the first time.  In simulations without galactic winds, we
find that metals are largely confined to the star-forming ISM at high
overdensity.  Metal-line cooling would not significantly increase the
star formation rates in these simulations.  In the present set of
simulations, metals can be transported by winds into the low-density
IGM.  If the corresponding gas is reaccreted at later times into
larger systems, metal-line cooling should then accelerate cooling,
thereby potentially increasing the total star formation density
compared to what we estimated here, particularly at low redshift.  But
recall that metal line cooling has also been neglected when we
normalised the star formation timescale used in our model for the ISM
to match the Kennicutt law. If we had included metal line cooling,
this normalisation would have been somewhat different in order to
compensate for the accelerated cooling, such that the net star
formation rate would have again matched the Kennicutt law. This effect
will alleviate any difference that one naively expects from the
inclusion of metal line cooling.  More work is therefore needed to
quantitatively estimate how important the effect of metal line cooling
would ultimately be in the present model of feedback due to galactic
winds.

The set of simulations we carried out offers rich information on many
aspects of galaxy formation and structure formation, not only on the
cosmic star formation history.  Note in particular that our
simulations are among the first that can self-consistently address the
interaction of winds with the low-density IGM, and the transport of
heavy metals along with them.  We have already investigated effects of
the winds on secondary anisotropies of the cosmic microwave background
\citep{WhiHerSpr02}.  Among the issues we plan to address next, is the
question of whether winds imprint specific signatures in the
Lyman-$\alpha$ forest that can be identified observationally..

\section*{Acknowledgements}

We thank Simon White for instructive discussions and critical comments
that were helpful for the work on this paper.  This work was supported
in part by NSF grants ACI 96-19019, AST 98-02568, AST 99-00877, and
AST 00-71019.  The simulations were performed at the Center for
Parallel Astrophysical Computing at the Harvard-Smithsonian Center for
Astrophysics.

\bibliographystyle{mnras}

\bibliography{paper_sfr}

\begin{thebibliography}{110}
\expandafter\ifx\csname natexlab\endcsname\relax\def\natexlab#1{#1}\fi

\bibitem[Abel et~al.(2002)Abel, Bryan \& Norman]{Abel2002}
Abel T., Bryan G.~L., Norman M.~L., 2002, Science, 295, 93

\bibitem[Abel \& Wandelt(2002)]{AbelWandelt2002}
Abel T., Wandelt B.~D., 2002, MNRAS, 330, 53

\bibitem[Ascasibar et~al.(2002)Ascasibar, Yepes, {Gottl\"{o}ber} \&
  {M\"{u}ller}]{Asc02}
Ascasibar Y., Yepes G., {Gottl\"{o}ber} S., {M\"{u}ller} V., 2002, A\&A, 387,
  396

\bibitem[Baldry et~al.(2002)Baldry, Glazebrook, Baugh et~al.]{Bald02}
Baldry I.~K., Glazebrook K., Baugh C.~M., et~al., 2002, ApJ, 569, 582

\bibitem[Balogh et~al.(2001)Balogh, Pearce, Bower \& Kay]{Bal01}
Balogh M.~L., Pearce F.~R., Bower R.~G., Kay S.~T., 2001, MNRAS, 326, 1228

\bibitem[Barkana \& Loeb(2000)]{Bar00}
Barkana R., Loeb A., 2000, ApJ, 539, 20

\bibitem[Baugh et~al.(1998)Baugh, Cole, Frenk \& Lacey]{Bau98}
Baugh C.~M., Cole S., Frenk C.~S., Lacey C.~S., 1998, ApJ, 498, 504

\bibitem[Benson et~al.(2002{\natexlab{a}})Benson, Lacey, abd S.~Cole \&
  Frenk]{Benson2001a}
Benson A.~J., Lacey C.~G., abd S.~Cole C. M.~B., Frenk C.~S.,
  2002{\natexlab{a}}, MNRAS, 333, 156

\bibitem[Benson et~al.(2002{\natexlab{b}})Benson, Lacey, abd S.~Cole \&
  Frenk]{Benson2001b}
Benson A.~J., Lacey C.~G., abd S.~Cole C. M.~B., Frenk C.~S.,
  2002{\natexlab{b}}, MNRAS, 333, 177

\bibitem[Benson et~al.(2001)Benson, Nusser, Sugiyama \& Lacey]{Ben2001}
Benson A.~J., Nusser A., Sugiyama N., Lacey C.~G., 2001, MNRAS, 320, 153

\bibitem[{Bland-Hawthorn}(1995)]{Bland95}
{Bland-Hawthorn} J., 1995, Proc. Astron. Soc. Australia, 12, 190

\bibitem[Blanton et~al.(1999)Blanton, Cen, Ostriker \& Strauss]{Bl99}
Blanton M., Cen R., Ostriker J.~P., Strauss M., 1999, ApJ, 522, 590

\bibitem[Bromm et~al.(1999)Bromm, Coppi \& Larson]{Bromm99}
Bromm V., Coppi P.~S., Larson R.~B., 1999, ApJ, 527, 5

\bibitem[Bullock et~al.(2000)Bullock, Kravtsov \& Weinberg]{Bullock00}
Bullock J.~S., Kravtsov A.~V., Weinberg D.~H., 2000, ApJ, 539, 517

\bibitem[Bullock et~al.(2001)Bullock, Kravtsov \& Weinberg]{Bullock01}
Bullock J.~S., Kravtsov A.~V., Weinberg D.~H., 2001, ApJ, 548, 33

\bibitem[Carr et~al.(1984)Carr, Bond \& Arnett]{Carr84}
Carr B.~J., Bond J.~R., Arnett W.~D., 1984, ApJ, 277, 445

\bibitem[Cen(2002)]{Cen2002}
Cen R., 2002, ApJS, 141, 211

\bibitem[Cen et~al.(1994)Cen, {Miralda-Escud\'e}, Ostriker \& Rauch]{CMOR94}
Cen R., {Miralda-Escud\'e} J., Ostriker J.~P., Rauch M.~J., 1994, ApJ, 437, L9

\bibitem[Cen \& Ostriker(1993)]{Ce93}
Cen R., Ostriker J.~P., 1993, ApJ, 417, 415

\bibitem[Cen \& Ostriker(1999)]{Cen99}
Cen R., Ostriker J.~P., 1999, ApJ, 519, L109

\bibitem[Cen \& Ostriker(2000)]{Ce00}
Cen R., Ostriker J.~P., 2000, ApJ, 538, 83

\bibitem[Ciardi et~al.(2002)Ciardi, Bianchi \& Ferrara]{Ciardi2002}
Ciardi B., Bianchi S., Ferrara A., 2002, MNRAS, 331, 463

\bibitem[Ciardi et~al.(2001)Ciardi, Ferrara, Marri \& Raimondo]{Ciardi2001}
Ciardi B., Ferrara A., Marri S., Raimondo G., 2001, MNRAS, 324, 381

\bibitem[Cole(1991)]{Col91}
Cole S., 1991, ApJ, 367, 45

\bibitem[Cole et~al.(1994)Cole, {Aragon-Salamanca}, Frenk, Navarro \&
  Zepf]{Col94}
Cole S., {Aragon-Salamanca} A., Frenk C.~S., Navarro J.~F., Zepf S.~E., 1994,
  MNRAS, 271, 781

\bibitem[Cole et~al.(2001)Cole, Norberg, Baugh et~al.]{Cole2001}
Cole S., Norberg P., Baugh C.~M., et~al., 2001, MNRAS, 326, 255

\bibitem[Connolly et~al.(1997)Connolly, Szalay, Dickinson, Subbarao \&
  Brunner]{Con97}
Connolly A.~J., Szalay A.~S., Dickinson M., Subbarao M.~U., Brunner R.~J.,
  1997, ApJ, 486, L11

\bibitem[Cowie et~al.(1999)Cowie, Songaila \& Barger]{Cow99}
Cowie L.~L., Songaila A., Barger A.~J., 1999, ApJ, 118, 603

\bibitem[Cowie et~al.(1996)Cowie, Songaila, Hu \& Cohen]{Cow96}
Cowie L.~L., Songaila A., Hu E.~M., Cohen J.~G., 1996, AJ, 112, 839

\bibitem[Croft et~al.(2001)Croft, {Di~Matteo}, Dav\'{e} et~al.]{Cr00}
Croft R. A.~C., {Di~Matteo} T., Dav\'{e} R., et~al., 2001, ApJ, 557, 67

\bibitem[Dahlem et~al.(1997)Dahlem, Petr, Lehnert, Heckman \& Ehle]{Dahlem97}
Dahlem M., Petr M.~G., Lehnert M.~D., Heckman T.~M., Ehle M., 1997, A\&A, 320,
  731

\bibitem[Dav\'{e} et~al.(2001)Dav\'{e}, Cen, Ostriker et~al.]{Dave2001}
Dav\'{e} R., Cen R., Ostriker J.~P., et~al., 2001, ApJ, 552, 473

\bibitem[Dav\'{e} et~al.(1999)Dav\'{e}, Hernquist, Katz \& Weinberg]{Da99}
Dav\'{e} R., Hernquist L., Katz N., Weinberg D.~H., 1999, ApJ, 511, 521

\bibitem[Flores et~al.(1999)Flores, Hammer, Thuan et~al.]{Flo99}
Flores H., Hammer F., Thuan T.~X., et~al., 1999, ApJ, 517, 148

\bibitem[Frye et~al.(2002)Frye, Broadhurst \& Benitez]{Frye01}
Frye B., Broadhurst T., Benitez N., 2002, ApJ, 568, 558

\bibitem[Fukugita et~al.(1998)Fukugita, Hogan \& Peebles]{Fuk98}
Fukugita M., Hogan C.~J., Peebles P. J.~E., 1998, ApJ, 503, 518

\bibitem[Gallego et~al.(1995)Gallego, Zamorano, Aragon-Salamanca \&
  Rego]{Gal95}
Gallego J., Zamorano J., Aragon-Salamanca A., Rego M., 1995, ApJ, 455, L1

\bibitem[Gnedin \& Ostriker(1997)]{Gne97}
Gnedin N., Ostriker J., 1997, ApJ, 486, 581

\bibitem[Gnedin \& Abel(2001)]{GnedinAbel01}
Gnedin N.~Y., Abel T., 2001, New Astronomy, 6, 437

\bibitem[Gronwall(1999)]{Gron99}
Gronwall C., 1999, in { After the Dark Ages: When galaxies where young\/},
  edited by S.~Holt, E.~Smith,  335, Am. Inst. Phys. Press, Woodbury, NY

\bibitem[Haardt \& Madau(1996)]{Ha96}
Haardt F., Madau P., 1996, ApJ, 461, 20

\bibitem[Haehnelt et~al.(2001)Haehnelt, Madau, Kudritzki \& Haardt]{Haehn2001}
Haehnelt M.~G., Madau P., Kudritzki R., Haardt F., 2001, ApJ, 549, L151

\bibitem[Heckman et~al.(1995)Heckman, Dahlem, Lehnert, Fabbiano, Gilmore \&
  Waller]{Heck95}
Heckman T.~M., Dahlem M., Lehnert M.~D., Fabbiano G., Gilmore D., Waller W.~H.,
  1995, ApJ, 448, 98

\bibitem[Heckman et~al.(2000)Heckman, Lehnert, Strickland \& Armus]{Heck00}
Heckman T.~M., Lehnert M.~D., Strickland D.~K., Armus L., 2000, ApJS, 129, 493

\bibitem[Hernquist et~al.(1996)Hernquist, Katz, Weinberg \& {Miralda-Escud{\'
  e}}]{He96}
Hernquist L., Katz N., Weinberg D.~H., {Miralda-Escud{\' e}} J., 1996, ApJ,
  457, L51

\bibitem[Hernquist \& Springel(2002)]{HernSpr2002}
Hernquist L., Springel V., 2002, preprint, astro-ph/0209183

\bibitem[Hogg(2001)]{Hogg01}
Hogg D.~W., 2001, preprint, astro-ph/0105280

\bibitem[Hughes et~al.(1998)Hughes, Serjeant, Dunlop et~al.]{Hug98}
Hughes D.~H., Serjeant S., Dunlop J., et~al., 1998, Nature, 394, 241

\bibitem[Hui et~al.(2002)Hui, Haiman, Zaldarriaga \& Alexander]{Hui2002}
Hui L., Haiman Z., Zaldarriaga M., Alexander T., 2002, ApJ, 564, 525

\bibitem[{Jang-Condell} \& Hernquist(2001)]{Jang2001}
{Jang-Condell} H., Hernquist L., 2001, ApJ, 548, 68

\bibitem[Jenkins et~al.(2001)Jenkins, Frenk, White et~al.]{Jen01}
Jenkins A., Frenk C.~S., White S. D.~M., et~al., 2001, MNRAS, 321, 372

\bibitem[Katz et~al.(1999)Katz, Hernquist \& Weinberg]{Kat99}
Katz N., Hernquist L., Weinberg D.~H., 1999, ApJ, 523, 463

\bibitem[Katz et~al.(1996)Katz, Weinberg \& Hernquist]{Ka96}
Katz N., Weinberg D.~H., Hernquist L., 1996, ApJS, 105, 19

\bibitem[Kauffmann et~al.(1994)Kauffmann, Guiderdoni \& White]{Kau94}
Kauffmann G., Guiderdoni B., White S. D.~M., 1994, MNRAS, 267, 981

\bibitem[Kauffmann et~al.(1993)Kauffmann, White \& Guiderdoni]{Kau93a}
Kauffmann G., White S. D.~M., Guiderdoni B., 1993, MNRAS, 264, 201

\bibitem[Kennicutt(1989)]{Ke89}
Kennicutt R.~C., 1989, ApJ, 344, 685

\bibitem[Kennicutt(1998)]{Ke98}
Kennicutt R.~C., 1998, ApJ, 498, 541

\bibitem[Keshet et~al.(2002)Keshet, Waxman, Loeb, Springel \&
  Hernquist]{Keshet2002}
Keshet U., Waxman E., Loeb A., Springel V., Hernquist L., 2002, preprint,
  astro-ph/0202318

\bibitem[Lacey et~al.(1993)Lacey, Guiderdoni, {Rocca-Vomerange} \&
  Silk]{Lac93b}
Lacey C., Guiderdoni B., {Rocca-Vomerange} B., Silk J., 1993, ApJ, 402, 15

\bibitem[Lacey \& Silk(1991)]{Lac91}
Lacey C., Silk J., 1991, ApJ, 381, 14

\bibitem[Lanzetta et~al.(2002)Lanzetta, Yahata, Pascarelle, Chen \&
  {Fernandez-Soto}]{Lan2002}
Lanzetta K.~M., Yahata N., Pascarelle S., Chen H., {Fernandez-Soto} A., 2002,
  ApJ, 570, 492

\bibitem[Lehnert \& Heckman(1996)]{Lehn96}
Lehnert M.~D., Heckman T.~M., 1996, ApJ, 462, 651

\bibitem[Leitherer et~al.(1995)Leitherer, Ferguson, Heckman \&
  Lowenthal]{Leit95}
Leitherer C., Ferguson H.~C., Heckman T.~M., Lowenthal J.~D., 1995, ApJ, 454,
  L19

\bibitem[Lilly et~al.(1996)Lilly, Fevre, Hammer \& Crampton]{Lil96}
Lilly S.~J., Fevre O.~L., Hammer F., Crampton D., 1996, ApJ, 460, L1

\bibitem[Madau(2000)]{Madau2000}
Madau P., 2000, Philos. Trans. R. Soc. London A, 358, 2221

\bibitem[Madau et~al.(1996)Madau, Ferguson, Dickinson, Giavalisco, Steidel \&
  Fruchter]{Mad96}
Madau P., Ferguson H.~C., Dickinson M.~E., Giavalisco M., Steidel C.~C.,
  Fruchter A., 1996, MNRAS, 283, 1388

\bibitem[Madau et~al.(1999)Madau, Haardt \& Rees]{Madau1999}
Madau P., Haardt F., Rees M.~J., 1999, ApJ, 514, 648

\bibitem[Madau et~al.(1998)Madau, Pozzetti \& Dickinson]{Mad98}
Madau P., Pozzetti L., Dickinson M., 1998, ApJ, 498, 106

\bibitem[Martin(1999)]{Mar99}
Martin C.~L., 1999, ApJ, 513, 156

\bibitem[McKee \& Ostriker(1977)]{McKee77}
McKee C.~F., Ostriker J.~P., 1977, ApJ, 218, 148

\bibitem[Mo \& White(2002)]{Mo2002}
Mo H.~J., White S. D.~M., 2002, MNRAS, 336, 112

\bibitem[Nagamine et~al.(2000)Nagamine, Cen \& Ostriker]{Nag00}
Nagamine K., Cen R., Ostriker J.~P., 2000, ApJ, 541, 25

\bibitem[Nagamine et~al.(2001)Nagamine, Fukugita, Cen \& Ostriker]{Nag01}
Nagamine K., Fukugita M., Cen R., Ostriker J.~P., 2001, ApJ, 558, 497

\bibitem[Navarro et~al.(1996)Navarro, Frenk \& White]{NFW}
Navarro J.~F., Frenk C.~S., White S. D.~M., 1996, ApJ, 462, 563

\bibitem[Navarro et~al.(1997)Navarro, Frenk \& White]{NFW2}
Navarro J.~F., Frenk C.~S., White S. D.~M., 1997, ApJ, 490, 493

\bibitem[Pascarelle et~al.(1998)Pascarelle, Lanzetta \& Fernandez-Soto]{Pas98}
Pascarelle S.~M., Lanzetta K.~M., Fernandez-Soto A., 1998, ApJ, 508, L1

\bibitem[Pearce et~al.(1999)Pearce, Jenkins, Frenk et~al.]{Pea99}
Pearce F.~R., Jenkins A., Frenk C.~S., et~al., 1999, ApJ, 521, L99

\bibitem[Pearce et~al.(2001)Pearce, Jenkins, Frenk et~al.]{Pea2000}
Pearce F.~R., Jenkins A., Frenk C.~S., et~al., 2001, MNRAS, 326, 649

\bibitem[Pettini et~al.(2001)Pettini, Shapley, Steidel et~al.]{Pett01}
Pettini M., Shapley A.~E., Steidel C.~C., et~al., 2001, ApJ, 554, 981

\bibitem[Pettini et~al.(2000)Pettini, Steidel, Adelberger, Dickinson \&
  Giavalisco]{Pett00}
Pettini M., Steidel C.~C., Adelberger K.~L., Dickinson M., Giavalisco M., 2000,
  ApJ, 528, 96

\bibitem[Press \& Schechter(1974)]{Pre74}
Press W.~H., Schechter P., 1974, ApJ, 187, 425

\bibitem[Scannapieco et~al.(2001)Scannapieco, Thacker \& Davis]{Scan01c}
Scannapieco E., Thacker R.~J., Davis M., 2001, ApJ, 557, 605

\bibitem[Sheth et~al.(2001)Sheth, Mo \& Tormen]{SMT01}
Sheth R.~K., Mo H.~J., Tormen G., 2001, MNRAS, 323, 1

\bibitem[Sheth \& Tormen(1999)]{ShTo99}
Sheth R.~K., Tormen G., 1999, MNRAS, 308, 119

\bibitem[Sheth \& Tormen(2002)]{ShTo02}
Sheth R.~K., Tormen G., 2002, MNRAS, 329, 61

\bibitem[Sokasian et~al.(2002{\natexlab{a}})Sokasian, Abel \&
  Hernquist]{Sok2002a}
Sokasian A., Abel T., Hernquist L., 2002{\natexlab{a}}, MNRAS, 332, 601

\bibitem[Sokasian et~al.(2002{\natexlab{b}})Sokasian, Abel \&
  Hernquist]{Sok2002b}
Sokasian A., Abel T., Hernquist L., 2002{\natexlab{b}}, preprint,
  astro-ph/0206428

\bibitem[Sokasian et~al.(2001)Sokasian, Abel \& Hernquist]{Sok2001}
Sokasian A., Abel T., Hernquist L.~E., 2001, New Astronomy, 6, 359

\bibitem[Somerville(2002)]{Somer01}
Somerville R.~S., 2002, ApJ, 572, 23

\bibitem[Somerville et~al.(2001)Somerville, Primack \& Faber]{Som00}
Somerville R.~S., Primack J.~R., Faber S.~M., 2001, MNRAS, 320, 504

\bibitem[Springel \& Hernquist(2002{\natexlab{a}})]{SprHerMultiPhase}
Springel V., Hernquist L., 2002{\natexlab{a}}, preprint, astro-ph/0206393

\bibitem[Springel \& Hernquist(2002{\natexlab{b}})]{SprHe01}
Springel V., Hernquist L., 2002{\natexlab{b}}, MNRAS, 333, 649

\bibitem[Springel et~al.(2001)Springel, Yoshida \& White]{SprGadget2000}
Springel V., Yoshida N., White S. D.~M., 2001, New Astronomy, 6, 79

\bibitem[Steidel et~al.(1999)Steidel, Adelberger, Giavalisco, Dickinson \&
  Pettini]{Steid99}
Steidel C.~C., Adelberger K.~L., Giavalisco M., Dickinson M., Pettini M., 1999,
  ApJ, 519, 1

\bibitem[Steidel et~al.(2001)Steidel, Pettini \& Adelberger]{Steid2001}
Steidel C.~C., Pettini M., Adelberger K.~L., 2001, ApJ, 546, 665

\bibitem[Steinmetz \& M\"{u}ller(1995)]{St95}
Steinmetz M., M\"{u}ller E., 1995, MNRAS, 276, 549

\bibitem[Thacker \& Couchman(2000)]{Th2000}
Thacker R.~J., Couchman H. M.~P., 2000, ApJ, 545, 728

\bibitem[Thacker et~al.(2000)Thacker, Tittley, Pearce, Couchman \&
  Thomas]{Th98}
Thacker R.~J., Tittley E.~R., Pearce F.~R., Couchman H. M.~P., Thomas P.~A.,
  2000, MNRAS, 319, 619

\bibitem[Tresse \& Maddox(1998)]{Tresse98}
Tresse L., Maddox S.~J., 1998, ApJ, 495, 691

\bibitem[Treyer et~al.(1998)Treyer, Ellis, Milliard, Donas \& Bridges]{Trey98}
Treyer M.~A., Ellis R.~S., Milliard B., Donas J., Bridges T.~J., 1998, MNRAS,
  300, 303

\bibitem[Weinberg et~al.(1999)Weinberg, {Dav\'{e}}, Gardner, Hernquist \&
  Katz]{Weinberg99}
Weinberg D.~H., {Dav\'{e}} R., Gardner J.~P., Hernquist L., Katz N., 1999, in {
  Photometric Redshifts and the Detection of High Redshift Galaxies\/}, edited
  by R.~Weymann, L.~{Storrie-Lombardi}, M.~Sawicki, R.~Brunner, vol. 191, ASP
  Conference Series,  341

\bibitem[Weinberg et~al.(1997)Weinberg, Hernquist \& Katz]{We97}
Weinberg D.~H., Hernquist L., Katz N., 1997, ApJ, 477, 8

\bibitem[Weinberg et~al.(2002)Weinberg, Hernquist \& Katz]{Wein2000}
Weinberg D.~H., Hernquist L., Katz N., 2002, ApJ, 571, 15

\bibitem[White et~al.(2002)White, Hernquist \& Springel]{WhiHerSpr02}
White M., Hernquist L., Springel V., 2002, preprint, astro-ph/0205437

\bibitem[White \& Frenk(1991)]{Wh91}
White S. D.~M., Frenk C.~S., 1991, ApJ, 379, 52

\bibitem[White \& Rees(1978)]{Whi78}
White S. D.~M., Rees M.~J., 1978, MNRAS, 183, 341

\bibitem[Wilson et~al.(2002)Wilson, Cowie, Barger \& Burke]{Wil02}
Wilson G., Cowie L.~L., Barger A.~J., Burke D.~J., 2002, AJ, 124, 1258

\bibitem[Yepes et~al.(1997)Yepes, Kates, Khokhlov \& Klypin]{Ye97}
Yepes G., Kates R., Khokhlov A., Klypin A., 1997, MNRAS, 284, 235

\bibitem[Yoshida et~al.(2002)Yoshida, St\"{o}hr, Springel \&
  White]{Yoshida2001}
Yoshida N., St\"{o}hr F., Springel V., White S. D.~M., 2002, MNRAS, 335, 762

\bibitem[Zhang et~al.(1995)Zhang, Anninos \& Norman]{Zh95}
Zhang Y., Anninos P., Norman M.~L., 1995, ApJ, 453, L57

\end{thebibliography}

\end{document}